\documentclass{article}
\usepackage[utf8]{inputenc}
\usepackage{amsmath,amssymb,indentfirst,epsfig,url}
\usepackage{makeidx} % Diagrama
\usepackage{graphicx}
\usepackage{subcaption} % Subfigure
\usepackage{multicol} % Criar 2 ou mais colunas em tabelas
\usepackage{multirow} % Criar 2 ou mais linhas em tabelas
\usepackage{hyperref} % Suporte para hipertexto, links para referências e figuras
\hypersetup{colorlinks=true, linkcolor=red, citecolor=blue, filecolor=black, urlcolor=green} % Configurações dos links 
\usepackage{verbatim} % Inclui caracteres externos de sem alteração gerado
\usepackage[affil-it]{authblk} % afiliação com affil
\usepackage{natbib} % Referências bibliográficas e afins
\bibpunct[; ]{(}{)}{,}{a}{,}{;} % Formatar as citações no texto e a lista de referências
\usepackage{pifont}
\usepackage{xcolor}

\newtheorem{definition}{Definition}
\providecommand{\keywords}[1]{\textbf{\textit{Keywords: }} #1}

\begin{document}
	\baselineskip = 5.7mm  % Espaço entre linhas;
	%\pagestyle{myheadings} % Estilo da página;
	
	%%Título
	\title{Transmission of Macroeconomic Shocks to Risk Parameters:\\
		Their uses in Stress Testing}
	
	\author[1]{Helder Rojas\thanks{Electronic address: \texttt{hmolina@santander.com.br}}}
	\affil{\small{Institute of Mathematics and Statistics, University of S\~ao Paulo -- IME-USP, Brazil}}
	
	\author[2]{David Dias\thanks{Electronic address: \texttt{davdias@santander.com.br}}}
	\affil{\small{Institute of Mathematical and Computer Sciences, University of S\~ao Paulo -- ICMC-USP, Brazil}}
	\affil[1,2]{\small{Santander Bank, Brazil}}
	
	\date{} 

	\maketitle
	
%%--------------------------------------------------
	\begin{abstract}
	%We are interested in evaluating the resilience of the portfolios of financial institutions in extreme economic conditions. Therefore, we propose empirical measures that characterize the transmission process of macroeconomic shocks to risk parameters. We propose the use of a extensive family of models, called General Transfer Function Models, that condense well the characteristics of the transmission described by the impact measurements. The procedure to estimate the parameters of these models is described through the Bayesian approach, using the prior information contained in the impact measurements. In addition, we illustrate the use of the model estimated from credit risk data of a portfolio.	
	In this paper, we are interested in evaluating the resilience of financial portfolios under extreme economic conditions. Therefore, we use empirical measures to characterize the transmission process of macroeconomic shocks to risk parameters. We propose the use of an extensive family of models, called General Transfer Function Models, which condense well the characteristics of the transmission described by the impact measures. The procedure for estimating the parameters of these models is described employing the Bayesian approach and using the prior information provided by the impact measures. In addition, we illustrate the use of the estimated models from the credit risk data of a portfolio.
\\
\\	
\end{abstract}

%%--------------------------------------------------
\section{Introduction}

	Stress testing resurfaced as a key tool for financial supervision after the 2007-2009 crisis; before the crisis these tests were rarely exhaustive and rigorous and financial institutions often considered them only as a regulatory exercise with no impact at capital. The lack of scope and rigor in these tests played a decisive role in the fact that financial institutions were not prepared for the financial crisis. Currently, the regulatory demands on stress tests are strict and the academic interest for related issues has increased considerably. Stress tests were designed to assess the resilience of financial institutions to possible future risks and, in some cases, to help establish policies to promote resilience. Stress tests generally begin with the specification of stress scenarios in macroeconomic terms, such as severe recessions or financial crisis, which are expected to have an adverse impact on banks. A variety of different models are used to estimate the impact of scenarios on banks’ profits and balance sheets. Such impacts are measured through the risk parameters corresponding to the portfolios evaluated. For a more detailed discussion of the use, future and limitations of stress tests, see \cite{dent2016stress}, \cite{SiddiqueHasan2012}, \cite{ScheuleRoesch2008} and \cite{aymanns2018models}. For an optimal evaluation of the resilience, it is necessary to implement models that condense the relevant empirical characteristics of the transmission process of the macroeconomic shocks to the risk parameters.
    
    In this work, we study empirical characteristics of the transmission process of macroeconomic shocks to financial risk parameters, and suggest measures that describe the propagation and persistence of impacts in the transmission. We propose the use of an extensive family of models, called General Transfer Function Models, that condense well the characteristics of the transmission described by the impact measurements. In addition, we incorporated to the model the possibility of including a learning or deterioration process of resilience, through a flexible structure which implies a stochastic transfer of shocks. We describe the procedure for estimating model parameters, where the prior information provided by the impact measures is incorporated into the procedure of  estimation through Bayesian approach. We also illustrate the use of the impact measures and model, estimated from the credit risk data of a portfolio.

	\subsection*{Outline} 
	
    The paper is organized as follows. In Section \ref{Sec2} we state the objective of the models in stress testing and present a case study. In Section \ref{Transmission} we describe the characteristics of the shock transmission process and propose impact measures. Section \ref{Sec4} describes the General Transfer Function Models. In section \ref{Sec5} we describe how to incorporate a flexible behavior to the resilience of the risk parameter. Section \ref{Sec6} describes the procedure for estimating model parameters. In the section \ref{Simulation} we present the simulation study to evaluate the Response Function dacay and the parameters estimation. Finally in section \ref{Sec7} we present results of the models applied to a case study.
%%--------------------------------------------------
\section{Stress Test Modeling}\label{Sec2}
	
	Stress testing models are equations that express quantitatively how macroeconomic shocks impact the different risk dimensions of financial institutions. Each risk dimension (for example, credit risk, interest rate risk, market risk, among others) is monitored by parameters, called risk parameters, which quantify the exposure of the portfolio to possible losses.
	As we mentioned before, the objective in stress tests is to assess the resilience of the portfolios and this is carried out through the risk parameters corresponding to each portfolio. Consequently, a specific objective is to model the behavior of risk parameters in terms of macroeconomic variables and to use these dependency relationships to extrapolate the behavior to the risk parameters in hypothetical downturn scenarios. Stress tests are also done in parameters that monitor the profitability and performance of banks, which is the case of stress tests for Pre-Provision Net Revenue known as PPNR models; all the concepts and tools discussed in this paper are also applicable to this type of models.	Portfolios used in stress tests are usually organized with each line of business or following some criterion of homogeneity with commercial utility.
	
    The complex relationship between series, the restricted availability of observations (usually between 20 and 30 quarters of observation), the diffuse behavior of all series (Random Walk) and the need to maintain a simple economic narrative are relevant issues to take into account when proposing models for stress testing. 
	
	\subsection*{Case Study: Credit Risk Data}
	
	Credit risk has great potential to generate losses on its assets and, therefore, has significant effects on capital adequacy. In addition, the credit risk is, possibly, the dimension of risk with the biggest bank regulation regarding stress tests. The most relevant credit risk parameters to assess resilience are: Probability of Default (PD) and Loss Given Default (LGD). Other risk parameters can be considered, but these have a definition superimposed with the parameters already mentioned. Furthermore, PD and LGD are used explicitly in the calculation of capital for a financial institution.

The definition of default and of the parameters mentioned above is given below:
\begin{definition}
	In Stress Testing, it is considered to be the default the borrower who does not fulfill the obligations for determined period of time.
\end{definition}

\begin{definition}
Probability of default (PD) is a financial term describing the likelihood of a default over a particular time horizon (this time horizon depends of the financial institution). It provides an estimate of the likelihood that a borrower will be unable to meet its debt obligations. The most intuitive way to estimate PD is through the Frequency of Observed Default (ODF), which is given by number of bad borrower (or borrower in default) in time t and under total number borrower (number of bad and good borrower) in time t.
%	\begin{equation}
%		ODF_{t} = \frac{\#BAD_{t}}{\#BAD_{t} + \#GOOD_{t}},
%	\end{equation}
%where $\#BAD_{t}$ means number of bad borrower or borrower in default in time t and $\#GOOD_{t}$ number of good borrower in time t.
\end{definition}

\begin{definition}
Loss Given Default or LGD is the share of an asset that is lost if a borrower defaults. Theoretically, LGD is calculated in different ways, but the most popular is 'Gross' LGD, where total losses are divided by Exposure at Default (EAD). Thus, the LGD is the total debt\% minus recuperate of the debt\%,  i.e., the percentage of debt that was recovered by the financial institution with the borrower's payments.
\end{definition}
For a better understanding of the different parameters of credit risk and their use to calculate the capital, see \cite{HenryKok2013macro}.
	
	\begin{figure}[h!]
		\centering
		\begin{subfigure}[b]{0.495\linewidth}
			\includegraphics[width=\linewidth]{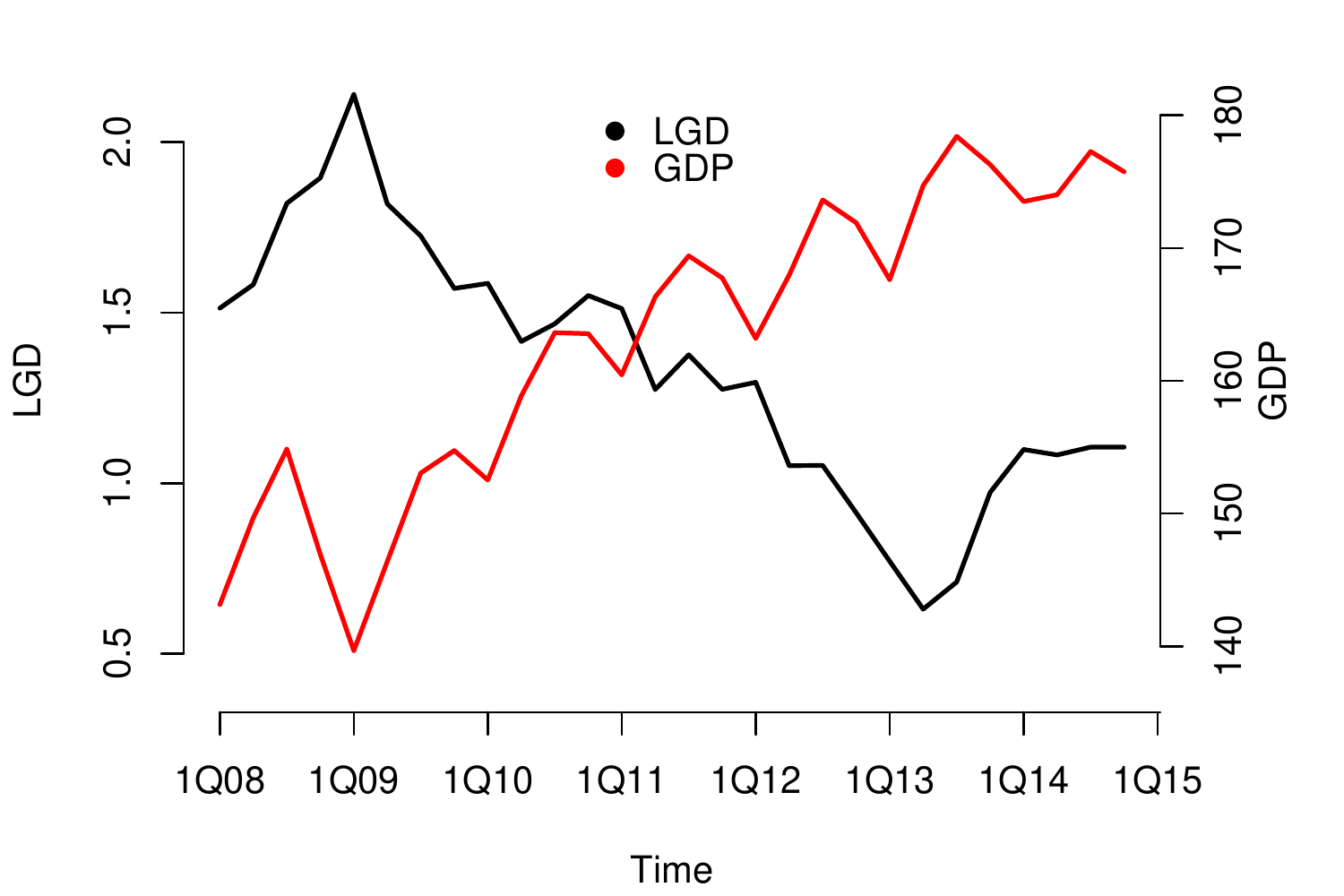}
		\end{subfigure}
		\hfill
		\begin{subfigure}[b]{0.495\linewidth}
			\includegraphics[width=\linewidth]{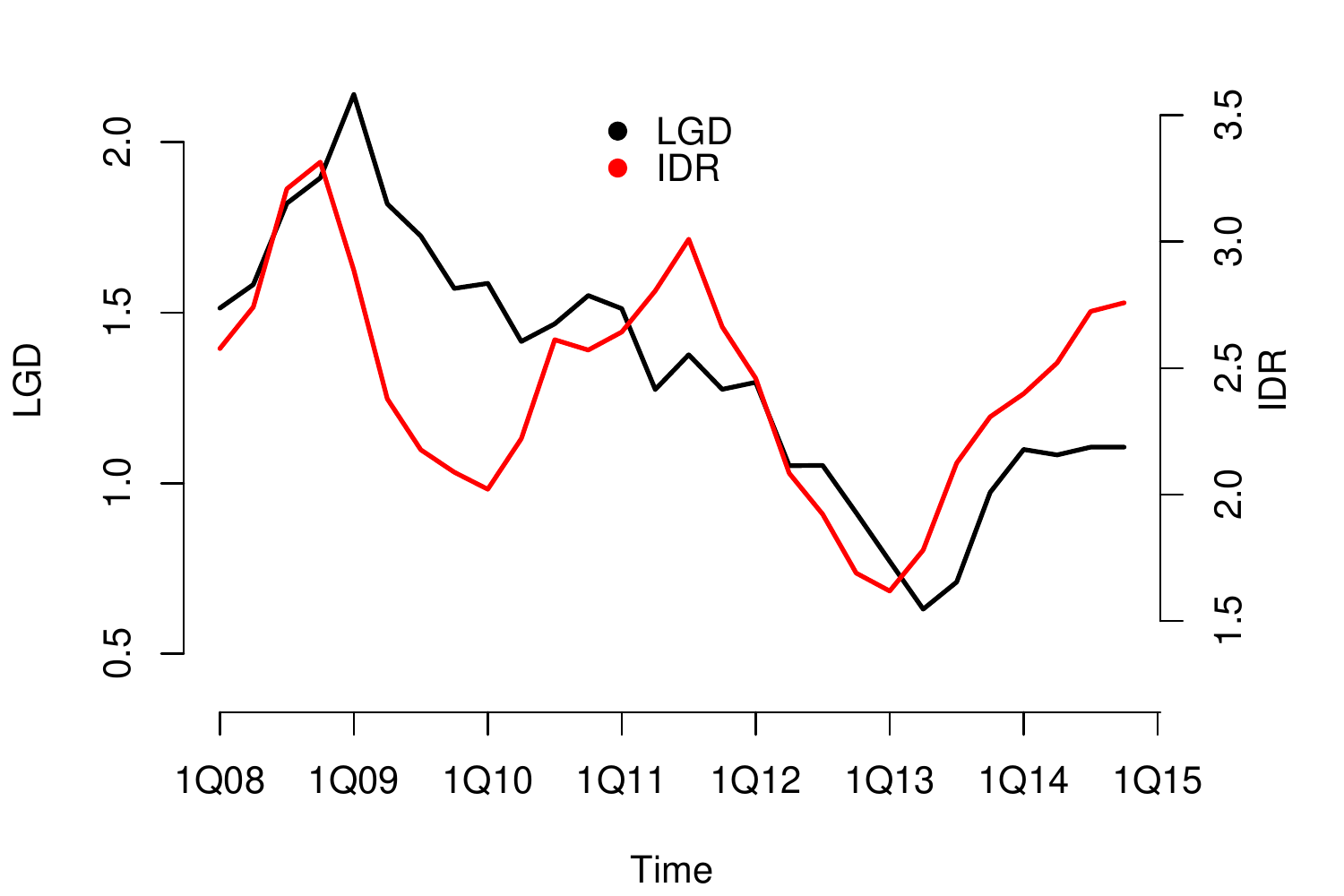}
		\end{subfigure}
		
		\begin{subfigure}[b]{0.495\linewidth}
			\includegraphics[width=\linewidth]{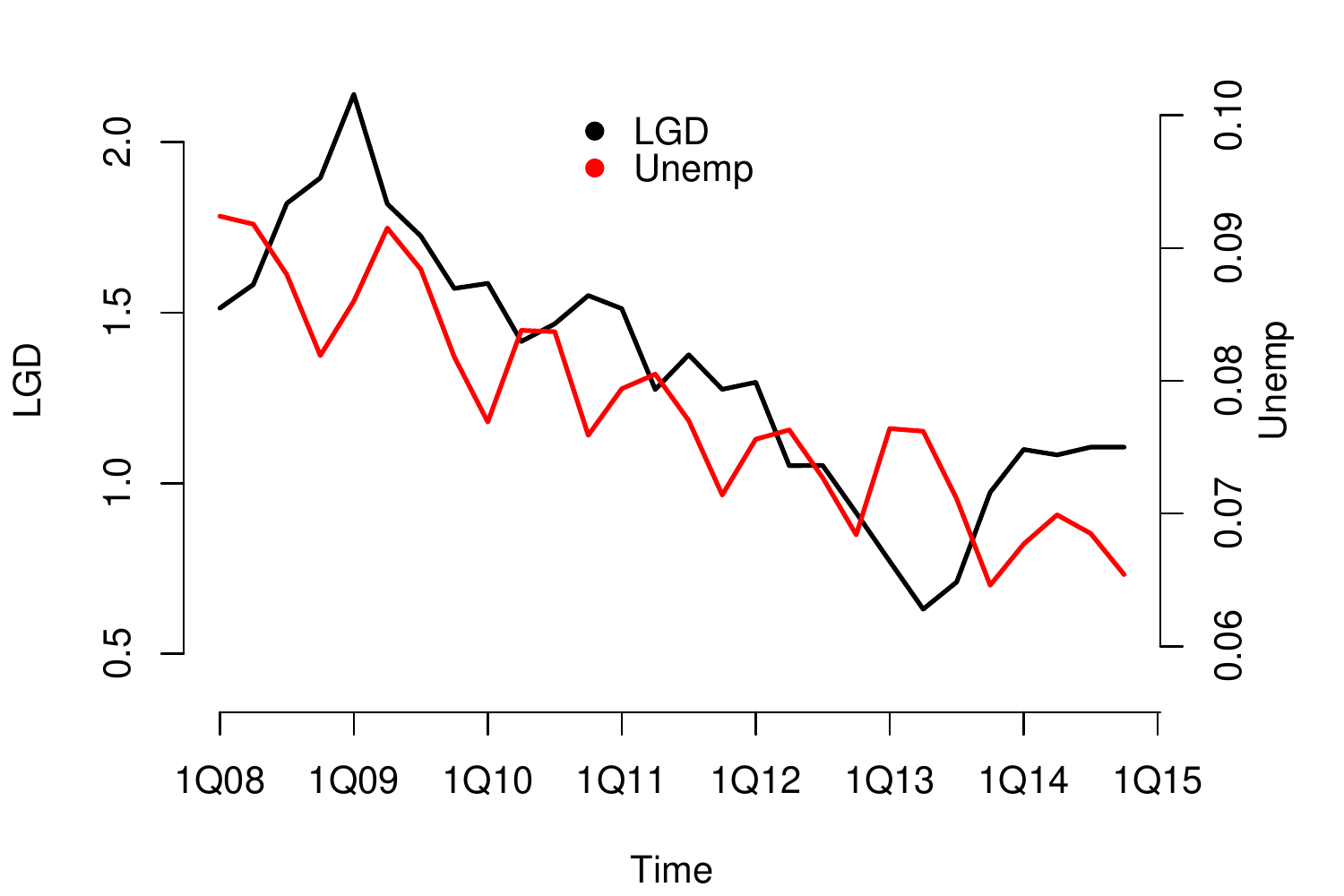}
		\end{subfigure}
		\caption{Risk parameter vs individual macroeconomic series, covering the period from 1Q2008 to 4Q2014.}
		\label{series}\label{fig1}
	\end{figure}
	
Without loss of generality, since they can be applied to other risk dimensions, we analyzed the credit risk data of a portfolio from the first quarter of 2008 to the fourth quarter of 2014, observed quarterly. . We considered three macroeconomic variables that represent the macroeconomic environment for this specific portfolio: Gross Domestic Product (GDP), Interbank Deposit Rate (IDR) and Unemployment (Unemp). The macroeconomic variables were observed in the same period as the risk parameters, LGD\footnote{For information security reasons, we used monotonic transformations in the original series.} in this case (see Figure \ref{fig1}). We considered four macroeconomic scenarios Optimistic, Baseline, Global and Local, where the Local is more severe than Global, Global is more severe than Baseline and Baseline is more severe than Optimistic (see Appendix \ref{AnexoProjectionScenarios}). The severity of the scenarios depends of the regulatory exercise of supervisory authorities and central banks. In our study, the scenarios were generated from the first quarter of 2016 to the fourth quarter of 2021.

%%--------------------------------------------------
 \section*{Modeling on Efficient Environment}
	
Due to its simplicity, possibly, a first stress test model that could be considered is linear regression for time series. A linear regression will link the macroeconomic variables with the evaluated risk parameter, where the coefficients of the regression will represent the impact of the macroeconomic shocks on the risk parameter. A linear regression establishes a static relationship between the variables, which would imply an immediate transfer of the shocks to the risk parameters. An immediate transfer of shocks would only be possible in a fully efficient economic environment, that is, perfectly rational and well informed. The assumption of an efficient environment, where the agents that determine the behavior of the risk parameter are fully informed and rational agents, is generally far from what is observed in the stress tests data. On the contrary, agents are very focused in the short term and are blind in the long term, which generates collective irrationality and a slow process of adaptation to shocks.

Using a static relationship, a linear regression in this case, and without taking into account the inefficiency of the environment; it would generate non-coherent forecasts of scenarios (overlapping scenarios forecasts) (see Appendix \ref{AnexoRegressionModel} for more results). Overlapping scenarios would constitute evidences that the model may not be adequate, given that the model would be making predictions inconsistent with stress levels, which would generate a relevant source of risk. It is worth mentioning that, in some cases, in environments that do not deviate much from the ideal conditions of an efficient environment, some transformations of the variables can be applied to comply with the assumptions of a linear regression model, but in general, these transformations eliminate relevant information at transmission process of shocks that a good model should contain.

%%--------------------------------------------------
    \section{Transmission of Shocks}\label{Transmission}
	
	\subsection{Propagation and Persistence}
	
To understand the transmission process of macroeconomic shocks to risk parameters, it is necessary to consider the characteristics and limitations of the agents that determine the dynamic behavior of the risk parameters. The series of risk parameters observed, derive from the relationship established between imperfect agents (in terms of rationality), which interact within an imperfect economic environment (in terms of information). For example, the agents that determine the dynamic behavior of the credit risk parameters are: the financial institution that grants the credits, the clients whom the credit is granted to and the financial regulator. All these agents have relevant limitations of rationality and information.

The imperfection of the economic environment conditions determines a slow processing of information by agents (behavioral or institutional inertia) and at the same time causes an unequal distribution of information among agents (asymmetry of information). The behavioral inertia and information asymmetry cause that macroeconomic shocks dynamically impact the risk parameters, that is, the impacts are propagated in time. The decay in the propagation of the impacts can be fast or slow; when the decay is slow we say that the propagation is of high persistence; and when the decay is fast, then the propagation is low. Thus, in inefficient environments the transmission process is characterized by persistence in the propagation. The nature and intensity of the persistence in the data of a specific portfolio need to be evaluated empirically and corroborated with some coherent economic criteria.

%%--------------------------------------------------
	\subsection{Impact Measures}\label{im}

To empirically contrast the persistent nature of the impacts, we propose the use of two simple functions (Response and Diffusion):
	
 \subsubsection*{i) Response Function } 

A simple way to assess the impact of macroeconomic shocks on the average level of risk parameters is through what we call Response Function $\mathfrak{R}(j)$ that is defined as:
	
	\begin{equation}{\label{RF}}
	\mathfrak{R}(j):= \mathbb{E}\big[(Y_{t+j}-\mu_{Y})(X_{t}-\mu_{X})\big],\quad \forall \quad j\geq 0
	\end{equation}
	where $Y_{t}$ is the stochastic process generator of risk parameter with mean function $\mu_{Y}=\mathbb{E}[Y_{t}]$ and $X_{t}$ is the stochastic process generator of a univariate macroeconomic series with mean function $\mu_{X}=\mathbb{E}[X_{t}]$. To each $j$, the quantity $\mathfrak{R}(j)$ measures the mean impact on $Y$, a time $j$ later,  of a shock in $X$ at time $t$. The time window used to obtain the estimation of $\mathfrak{R}(j)$ generally comprises the entire length of the series, in this case, in statistics and probability, $\mathfrak{R}(j)$ it is referred to as Cross-Covariance Function (CCF). In some specific cases, it may happen that the macroeconomic series and the risk parameter present divergent relationships from those expected economically. These divergences occur only temporarily due to political changes in the portfolio. In the presence of these divergences, the temporal window to estimate $\mathfrak{R}(j)$ has to be chosen trying to maintain the economic relationships that we know priori that make sense. To be more recurrent, in this work we will use the whole length of the series.
	
	Given the bivariate series $\{Y_{t},X_{t}\}_{t=1}^{T}$, the consistent estimator of $\mathfrak{R}(j)$ (denoted by $\widehat{\mathfrak{R}}$(j)), is given by
	
	\begin{equation}
	\widehat{\mathfrak{R}}(j)= \frac{1}{T}\sum_{t=1}^{T-j}(Y_{t+j}-\bar{Y})(X_{t}-\bar{X})
	\end{equation}
	where $\bar{Y}$ and $\bar{X}$ are the sample mean values of both series observed respectively (see \cite{box2008time}). 

	It is known from the literature that the CCF is not recommended to be used in the model specification when the series are not stationary (which is the case of the series used in Stress Testing). On the other hand, it is worth mentioning that $\widehat{\mathfrak{R}}(j)$, in this work, is not used to specify the model, but to remove qualitative information from the decay of shocks, that is, the functional form of the decays that representing the endogenous behavior of the parameter to macroeconomic shocks.
	
\clearpage 
	\subsubsection*{ii) Diffusion Function}
	
    The Diffusion Function $\mathfrak{D}(j)$ is defined as:
	
	\begin{equation}{\label{DF}}
	\mathfrak{D}(j):= \mathbb{E}\big[(Y_{t+j}-\mu_{Y})^2(X_{t}-\mu_{X})\big],\quad \forall j\geq 0
	\end{equation}
	
	where $Y_{t}$ is the stochastic process generator of risk parameter with mean function $\mu_{Y}=\mathbb{E}[Y_{t}]$ and $X_{t}$ is the stochastic process generator of macroeconomic series with mean function $\mu_{X}=\mathbb{E}[X_{t}]$. To each $j$, the quantity $\mathfrak{D}(j)$  measures the average fluctuation of $Y$ between time $t$ and $t+j$, associated to shocks on $X$ at time $t$. Note that $\mathfrak{D}(j)$ is used to analyze the shocks impact on the variance of the risk parameter.  If there is a persistent pattern in this measure, it should be considered  in construction from candidates models. 
	
	Given the series $\{Y_{t},X_{t}\}_{t=1}^{T}$, the consistent estimator of $\mathfrak{D}(j)$ (denoted by $\widehat{\mathfrak{D}}$(j)), is given by
	
	\begin{equation}
	\widehat{\mathfrak{D}}(j)= \frac{1}{T}\sum_{t=1}^{T-j}(Y_{t+j}-\bar{Y})^2(X_{t}-\bar{X})
	\end{equation}
	where $\bar{Y}$ and $\bar{X}$ are sample mean values. 
	
\subsection{Interpretation of Impact Measures}\label{iim}	

\begin{figure}[h!]
		\centering
		\begin{subfigure}[b]{0.4\linewidth}
			\includegraphics[width=\linewidth]{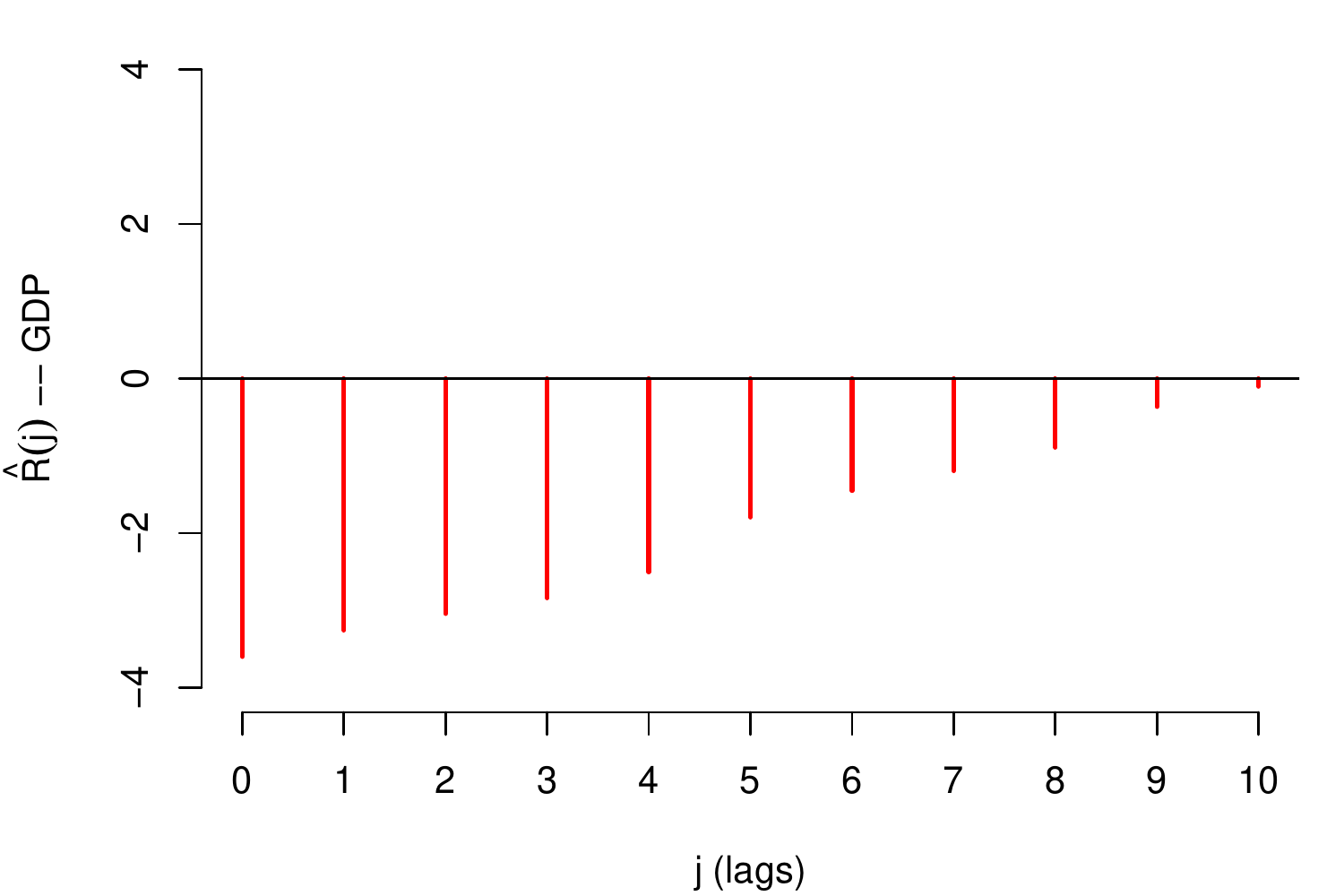}
		\end{subfigure}
		 \begin{subfigure}[b]{0.4\linewidth}
			\includegraphics[width=\linewidth]{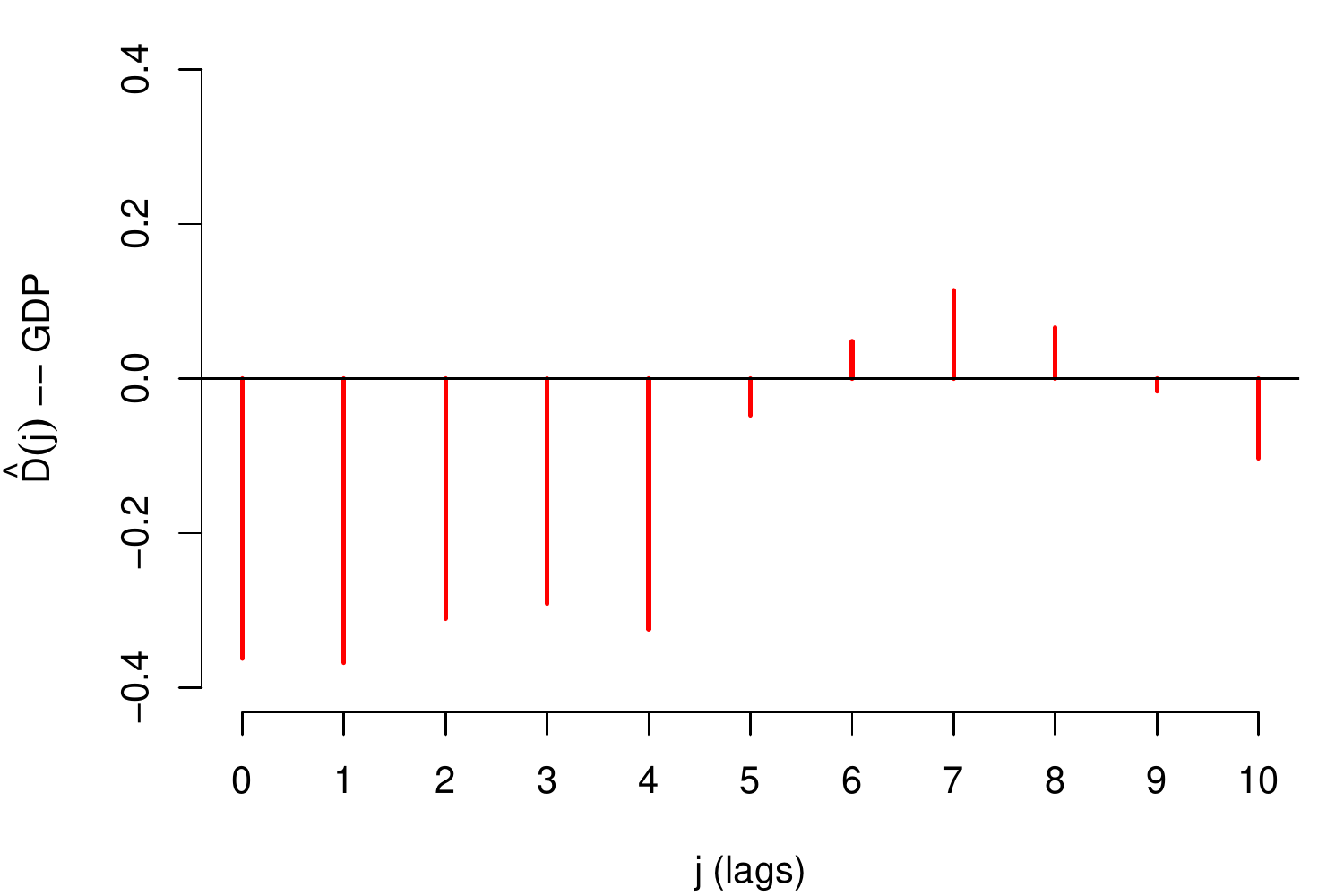}
		\end{subfigure}
		\begin{subfigure}[b]{0.4\linewidth}
			\includegraphics[width=\linewidth]{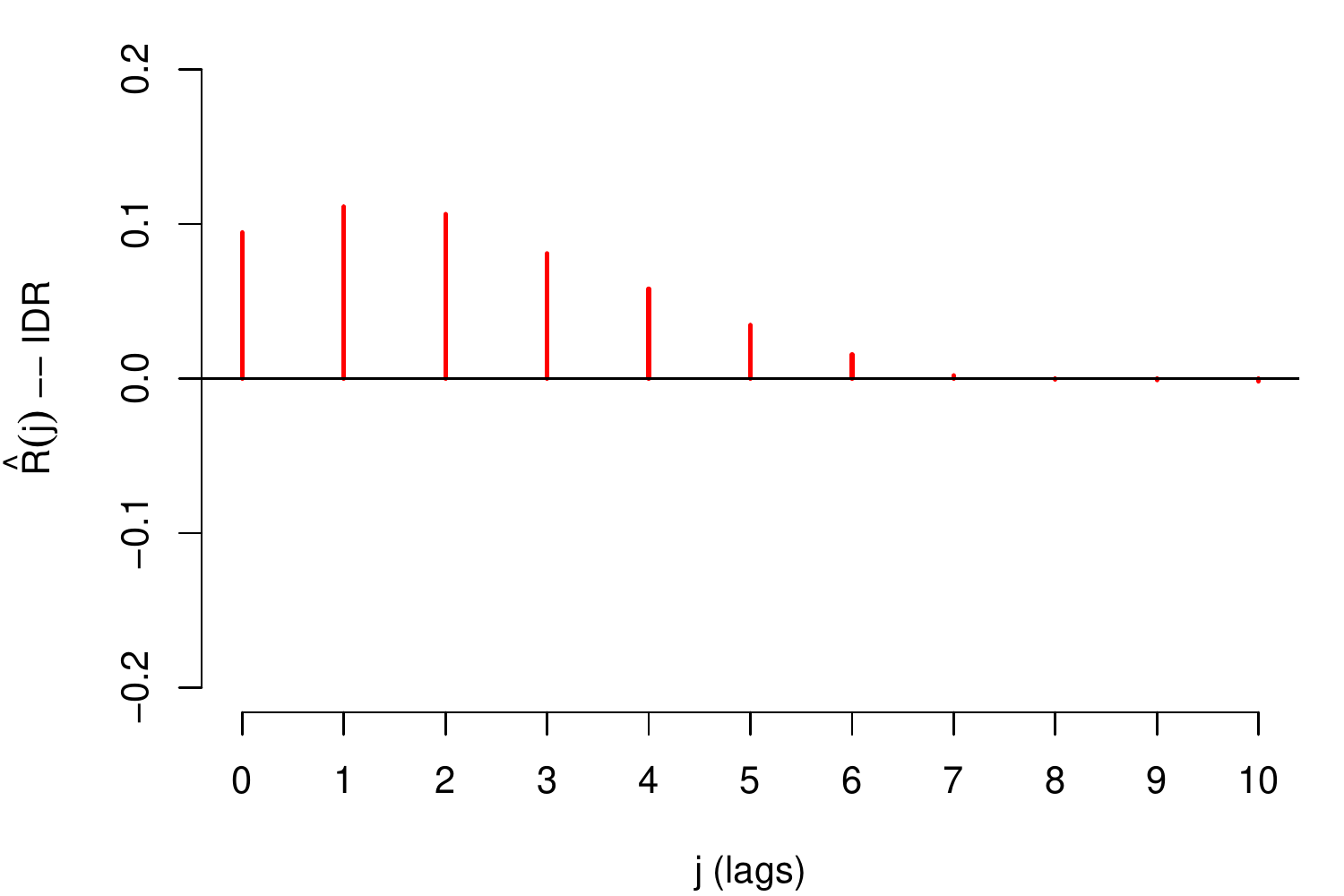}
		\end{subfigure}
		   		\begin{subfigure}[b]{0.4\linewidth}
			\includegraphics[width=\linewidth]{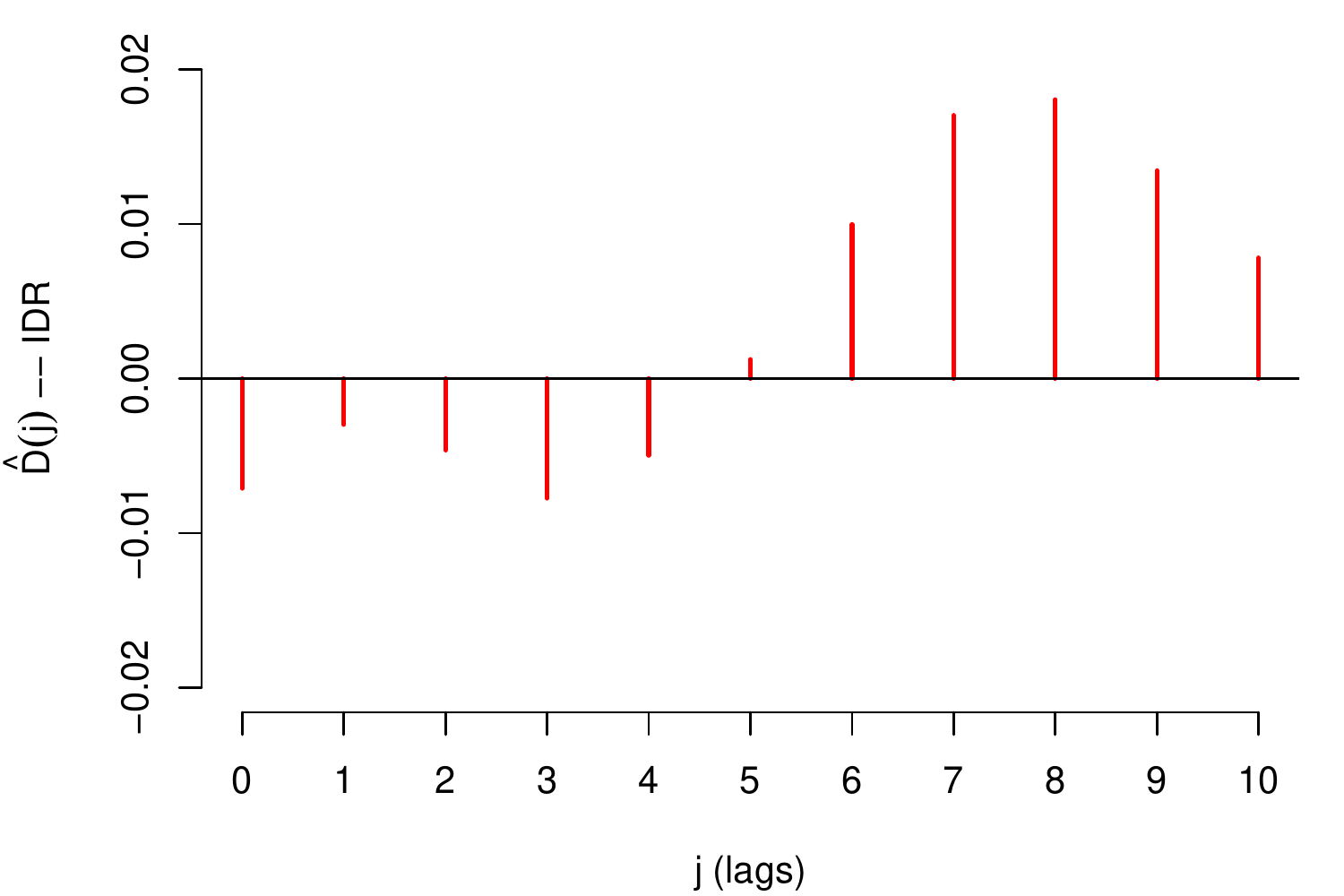}
		\end{subfigure}
		\begin{subfigure}[b]{0.4\linewidth}
			\includegraphics[width=\linewidth]{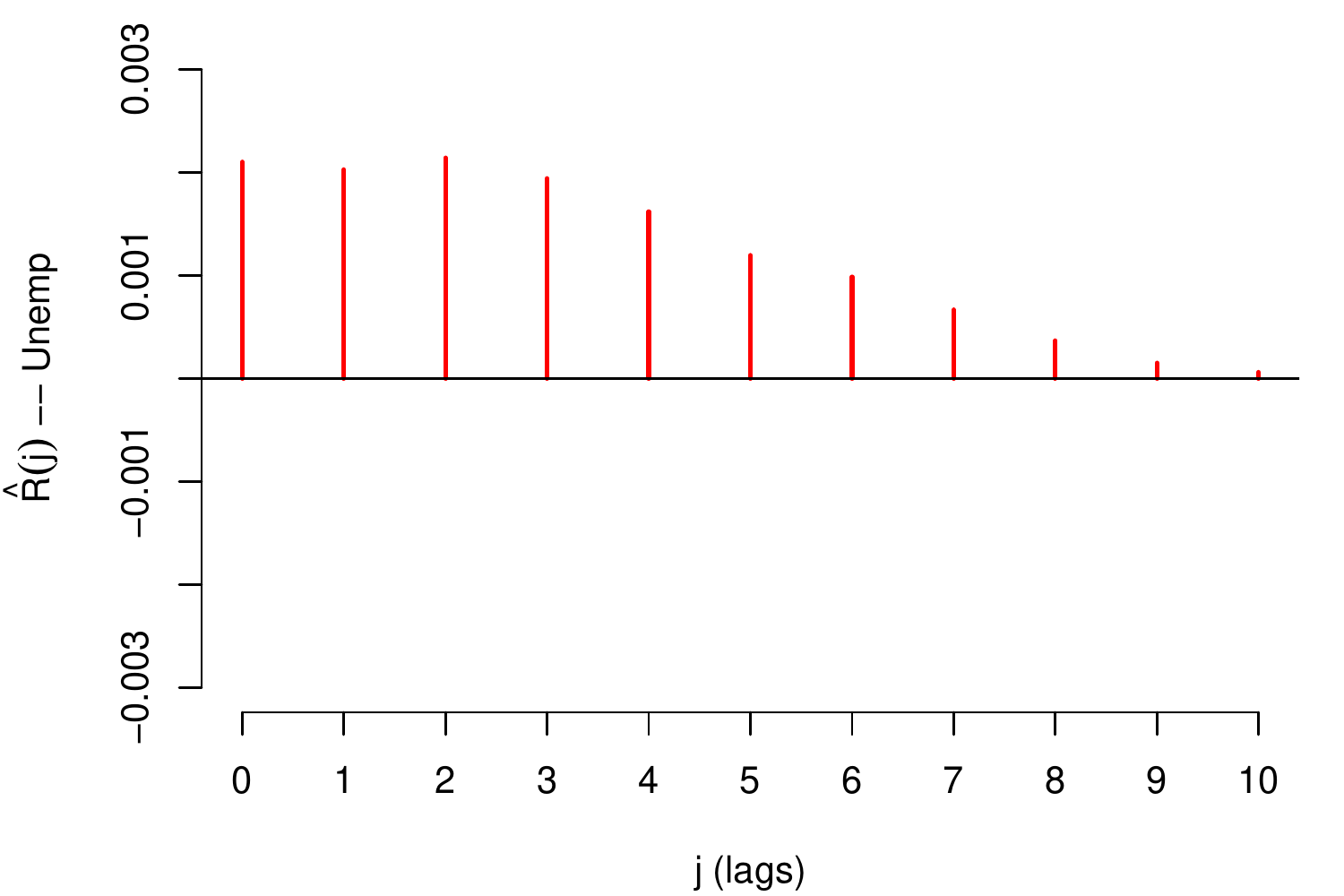}
		\end{subfigure}
		\begin{subfigure}[b]{0.4\linewidth}
			\includegraphics[width=\linewidth]{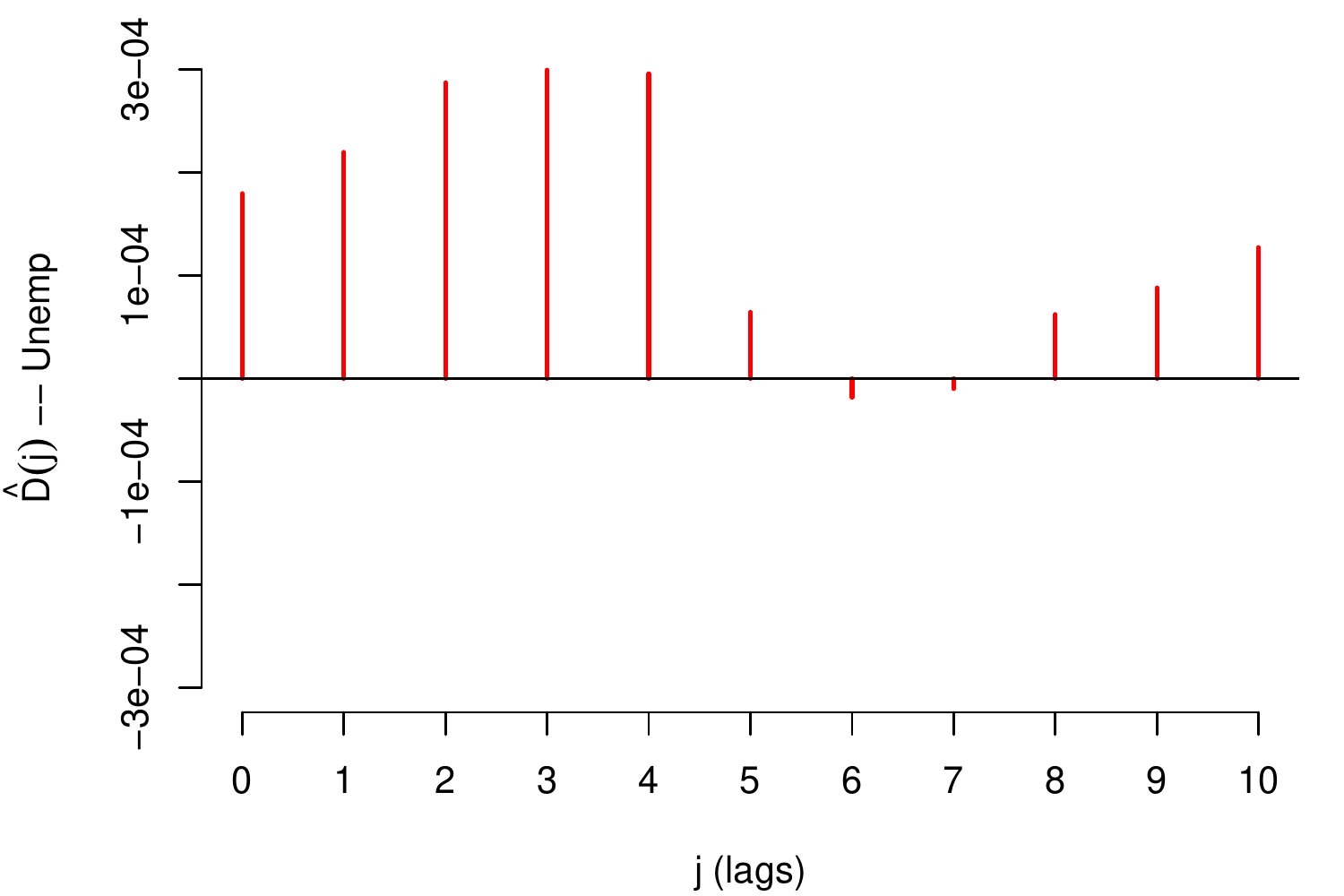}
		\end{subfigure}
		\caption{Impact measures in credit risk parameter with 10 lags}\label{Fig:ImpactMeasures}
	\end{figure}

The impact measurements presented in (\ref{RF}) and (\ref{DF}) describe the dynamic response of the risk parameter $Y$ to shocks in the macroeconomic variable $X$. In other words, they describe the way risk parameter absorbs the shocks, which is determined by the reaction of the agents to return to equilibrium. A slow decrease of the impact measures implies a high persistence at the absorption of shocks. Through these impact measures, two types of impact can be identified: Impacts with permanent effects and impacts with temporary effects. Shocks with permanent effects impact the stochastic trend of the risk parameter and their effects are measured by $\mathfrak{R}(j)$. Shocks with temporary effects impact the stationary fluctuations around the stochastic trend and their effects are measured by $\widehat{\mathfrak{D}}(j)$. For example, we can empirically identify these two types of impacts with the impact measurements referring to the credit risk data, see Figure \ref{Fig:ImpactMeasures}.
	
    The persistent nature of the shocks is observed in the decay rate of the impacts measured by $\widehat{\mathfrak{R}}(j)$. It is important to mention that due to the aggregate nature of the macroeconomic variables, the type of persistence needs to be contrasted with economic criteria that justify its nature. For example in the case of GDP and IDR, observing their corresponding Response Functions, see Figure \ref{Fig:ImpactMeasures}, and considering that both macroeconomic variables are directly involved in the credit recovery policies and LGD calculation respectively, we can consider that the impact of the shocks of these two macroeconomic variables on the LGD of this portfolio are of low persistence. In the case of the macroeconomic variable Unemployment, its response function shows slow decay in the first lags, besides, it is a variable associated with labor laws and institutional policies that justify the high persistence of shocks in this macroeconomic variable. In this case, $ \widehat{\mathfrak{D}}(j) $ not present a persistent pattern, that is, the shocks do not affect the variance of the risk parameter of persistence way, the only effect that the macroeconomic shocks do is to balance around their averages.
	 
%%--------------------------------------------------
	\section{Transfer Function Models}\label{Sec4}
	We are interested in the transmission of shocks in macroeconomic variables, for example $ X$, to risk parameters, for example $Y$. In the previous section, we showed that the $ X $ effect on $ Y $ persists for a period and decays to zero as time passes. A simplified and didactic way to represent the cumulative impact of $X_{t}$ on $Y_{t}$ is
	
	\begin{equation} \label{TFM}
	Y_{t}=\sum_{j=0}^{\infty}\beta (j)X_{t-j}+\xi_{t}
	\end{equation}
	where any shock in $X_{t}$ will impact $\mathbb{E}[Y_{t}]$ $(E_{t})$ in all the later periods. The term $\beta(j)$ in ($\ref{TFM}$) is a function of $j$ that represents the $j$-th impact coefficient. A condition generally assumed for dynamic stability of system (all finite shock has a finite cumulative impact), is that $\sum_{j=0}^{\infty}\beta(j)<\infty$, which implies $\lim_{j\to\infty}\beta(j)=0$. It is important to mention that in the last condition nor $X_{t}$ neither $Y_{t}$ must to be stationary. But the condition establish is the stability of the responses of the parameter risk to the macroeconomic shocks, which is a recurring characteristic in ergodic systems.
	
	It is common to establish specific restrictions on the impact coefficients functionally related, often known as Distributed Lag Models \citep{Zellner1971introduction}. The principal Distributed Lag Models are described in \cite{Koyck1954distributed}, \cite{Solow1960family} and \cite{Almon1965distributed}. 

    In this work we adopted a more general perspective described by \cite{BoxJenkins2015time}. Alternatively, the impact of $X_{t}$ can be written as a linear filter:
	
	\begin{equation}\label{fl}
	\begin{split}
	E_{t}&=\beta(0)X_{t}+\beta(1)X_{t-1}+\beta(2)X_{t-2}+ \ldots\\
	&=\mathcal{T}(L)X_{t}
	\end{split}
	\end{equation}
	
	The polynomial on the lag operator $\mathcal{T}(L)=\beta(0)+\beta(1)L+\beta(2)L^{2}+\ldots$ is called Transfer Function and represents the cumulative impact of $X$ on $Y$. The sequence $\{\beta(j)\}_{j=0}^{\infty}$ called Dynamic Multiplier is the mechanism of propagation of impacts underlying the Transfer Function and expresses the instantaneous impact of $X$ on $Y$ at present and further times. 
	
	\subsection{General Transfer Function Model}
	
    Let $X_{t}$ be the value of an macroeconomic, scalar variable $X$  and $Y_{t}$ the value of risk parameter at time $t$. A General Transfer Function Model (GTFM) for the dynamic effect of $X$ on the risk parameter $Y$, consistent with the empirical facts discussed in section  (\ref{Transmission}), is defined as
	\begin{subequations}\label{gm}
		\begin{align}
		\label{gm1}
		Y_{t} &=\mathbf{F}_{t}^{\top}\mathbf{\theta}_{t}+v_{t}\\ 
		\label{gm2}
		\theta_{t} &=\mathbf{G}_{t}\theta_{t-1}+\psi_{t}X_{t}+\partial\theta_{t}\\
		\label{gm3}
		\psi_{t} &=\psi_{t-1}+\partial\psi_{t}
		\end{align}
	\end{subequations}
	with terms described as follows: $\theta_{t}$ is an $\mathit{n}$-dimensional states vector (unobserved), $\mathbf{F}_{t}$ is a known $\mathit{n}$-vector (design vector), $\mathbf{G}_{t}$ a known transition matrix (evolution matrix) which determines the evolution of the states vector and, $v_{t}$ and $\partial\theta_{t}$ are observation and evolution noise terms. All these terms are precisely as in the standard Dynamic Linear Models (DLM), with the usual independence assumptions for the noise terms hold here. The term $\psi_{t}$ is an $\mathit{n}$-vector of parameters, evolving via addition of a noise term $\partial\psi_{t}$, assumed to be zero-mean normally distributed independently of $v_{t}$ (though not necessarily of 
	$\partial\theta_{t}$).
	
    In accordance with the  empirical facts discussed in section (\ref{Transmission}), the GTFM defined in (\ref{gm}) establishes that $X$ impacts the stochastic trend of $Y$ (\ref{gm2}) and, at the same time, generates stationary fluctuations around that stochastic trend (\ref{gm1}). The parameters that determine the transfer can vary over time, allowing greater flexibility to the transmission process of shocks (\ref{gm3}). This model constitutes a small variation of the first model presented by \cite{HarissonWest1999}.
	
	The states vector $\theta_{t}$ carries the effect of current and past values of the $X$ series to $Y_{t}$ in equation (\ref{gm1}); this is formed in (\ref{gm2}) as the sum of a linear function of past effects; $\theta_{t-1}$, and the current effect $\psi_{t}X_{t}$, plus a noise term. Notice that the parameterization (\ref{gm}) is more flexible than that of equation (\ref{TFM}), allowing different stochastic interpretations for the effect of $X$ on $Y$.
	
	The general model (\ref{gm}) can be rewritten in the standard DLM as follows. Define a 2$\mathit{n}$-dimensional state parameters vector $\tilde{\theta}_{t}$ by concatenating $\theta_{t}$ and $\psi_{t}$, giving $\tilde{\theta}_{t}=(\theta_{t}^{\top},  \psi_{t}^{\top})$. Similarly, extend the $\mathbf{F}_{t}$ vector by concatenating an $\mathit{n}$-vector of zeros, giving a new $\tilde{\mathbf{F}}_{t}$ such that $\tilde{\mathbf{F}}_{t}=(\mathbf{F}_{t}, 0, \cdots, 0)$. For the evolution matrix, define by
	\begin{align*}
	\tilde{\mathbf{G}}_{t}=
	\begin{bmatrix}
	\mathbf{G}_{t}&X_{t}\mathbf{I}_{\mathit{n}}\\
	\mathbf{0}&\mathbf{I}_{\mathit{n}}
	\end{bmatrix}
	\end{align*}
	where $\mathbf{I}_{\mathit{n}}$ is the $\mathit{n}$ x $\mathit{n}$ identity matrix. Finally, let $\mathbf{\omega}_{t}$ be the noise vector defined by $\omega_{t}^{\top}=(\partial\theta_{t}^{\top}+X_{t}\partial\psi_{t}^{\top}, \partial\psi_{t}^{\top})$. Then the model (\ref{gm}) can be written as
	\begin{equation}\label{DLM}
	\begin{split}
	Y_{t}&=\tilde{\mathbf{F}}^{\top}\tilde{\theta}_{t}+v_{t},\\
	\tilde{\theta}_{t}&=\tilde{\mathbf{G}}\tilde{\theta}_{t-1}+\mathbf{\omega}_{t}.
	\end{split}
	\end{equation}
	
	Thus the GTFM (\ref{gm}) is written in the standard DLM form (\ref{DLM}) and the usual inferential analysis corresponding to State Space Models can be applied (see \cite{DurbinKoopman2012time} and \cite{WestHarrison2006bayesian}). Note that the generalization to include several macroeconomic variables in the model (\ref{gm}) is trivial.
	
	\subsection{Some Transfer Functions }	
In this section, we present some Transfer Functions underlying models that are specific cases of the GTFM.
	\subsubsection{Geometric Propagation}
	
 Assuming a Dynamic Multiplier with geometric decay in (\ref{TFM}), this is:
	\begin{equation}\label{koyck}
	\beta(j)=\beta\phi^j,\quad \forall\quad j\geq 0, \quad \textrm{where}\quad |\phi|<1 \quad\textrm{and}\quad \beta \in \mathbb{R}
	\end{equation}
	the mechanism of propagation (\ref{koyck}) was first associated to \cite{Koyck1954distributed}, and postulates the gradual and rapid decay of the impacts and generates the following Transfer Function
	
	\begin{equation}\label{tkoyck}
	\mathcal{T}(L)=\frac{\beta}{1-\phi L}
	\end{equation}
	replacing  (\ref{tkoyck}) in (\ref{fl}), the Transfer Function model would be
	
	\begin{subequations}\label{koyck_dlm}
		\begin{align}
		Y_{t}&=E_{t}+v_{t}\\
		E_{t}&=\phi E_{t-1}+\beta X_{t}+\partial E_{t},
		\end{align}
	\end{subequations}
	where 
	\begin{align*}
	E_{t}=\sum_{j=0}^{\infty}\beta\phi^j X_{t-j}+\sum_{j=0}^{\infty}\phi^j\partial E_{t-j},
	\end{align*}
	with $v_{t}$ ($\partial E_{t}$) normally distributed with zero mean and variance $\sigma_{v}^{2}$ ($\sigma_{E}^{2}$), $\eta=\frac{\sigma_{v}}{\sigma_{E}}$ is signal-to-noise ratio. In this model, the dynamic response of $Y_{t}$ is $\beta \phi^j X_{t-j}$. Based on the representation in (\ref{gm}), we have $\mathit{m}=1$ (model with only one latent state), $\theta_{t}=E_{t}$, the cumulative impact until $t$, $\psi_{t}=\beta$, the current impact for all $t$, $\mathbf{F}_{t}=1$, $\mathbf{G}_{t}=\phi$ and $\partial \theta_{t}=\partial E_{t}$. Two simplifications of the equation (\ref{koyck_dlm}) can be obtained assuming $\eta=0$ or $\sigma_{E}=0$, these simplifications are often known as Autoregressive Distributed Lag (ADL) model. For a detailed review of ADL models from a classical approach, see \cite{PesaranShin1998ARDL}, and from a Bayesian approach, see \cite{BauwensLubrano1999bayesian}, \cite{Zellner1971introduction}.
	
\begin{figure}[h!]
		\centering
		\begin{subfigure}[b]{0.42\linewidth}
			\includegraphics[width=\linewidth]{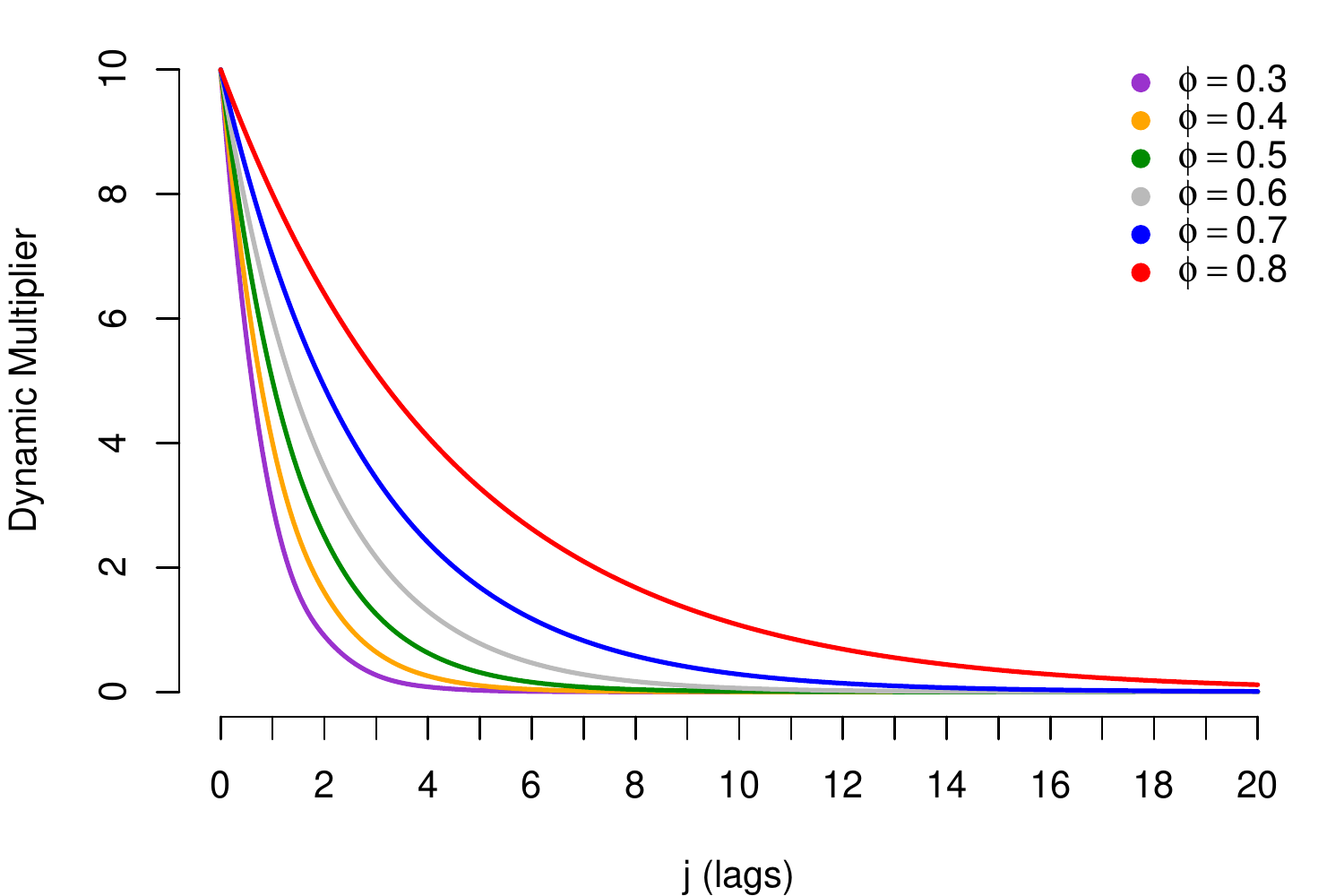}
	\caption{Dynamic Multiplier $\beta=10$ and $0< \phi < 1$}
		\end{subfigure}
		\begin{subfigure}[b]{0.42\linewidth}
			\includegraphics[width=\linewidth]{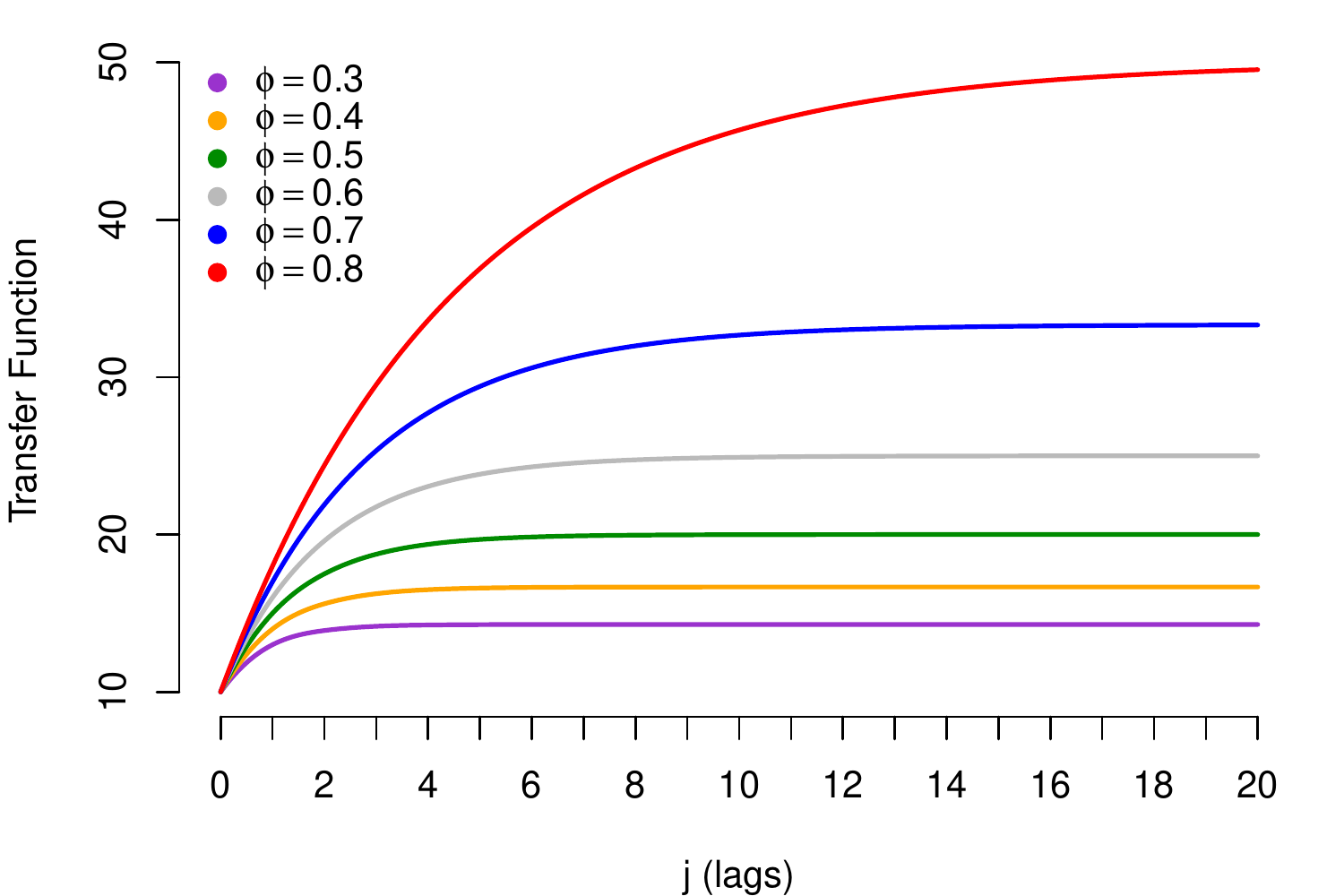}
\caption{Transfer Function $\beta=10$ and $0 < \phi < 1$}	
		\end{subfigure}   
		
		\begin{subfigure}[b]{0.42\linewidth}
			\includegraphics[width=\linewidth]{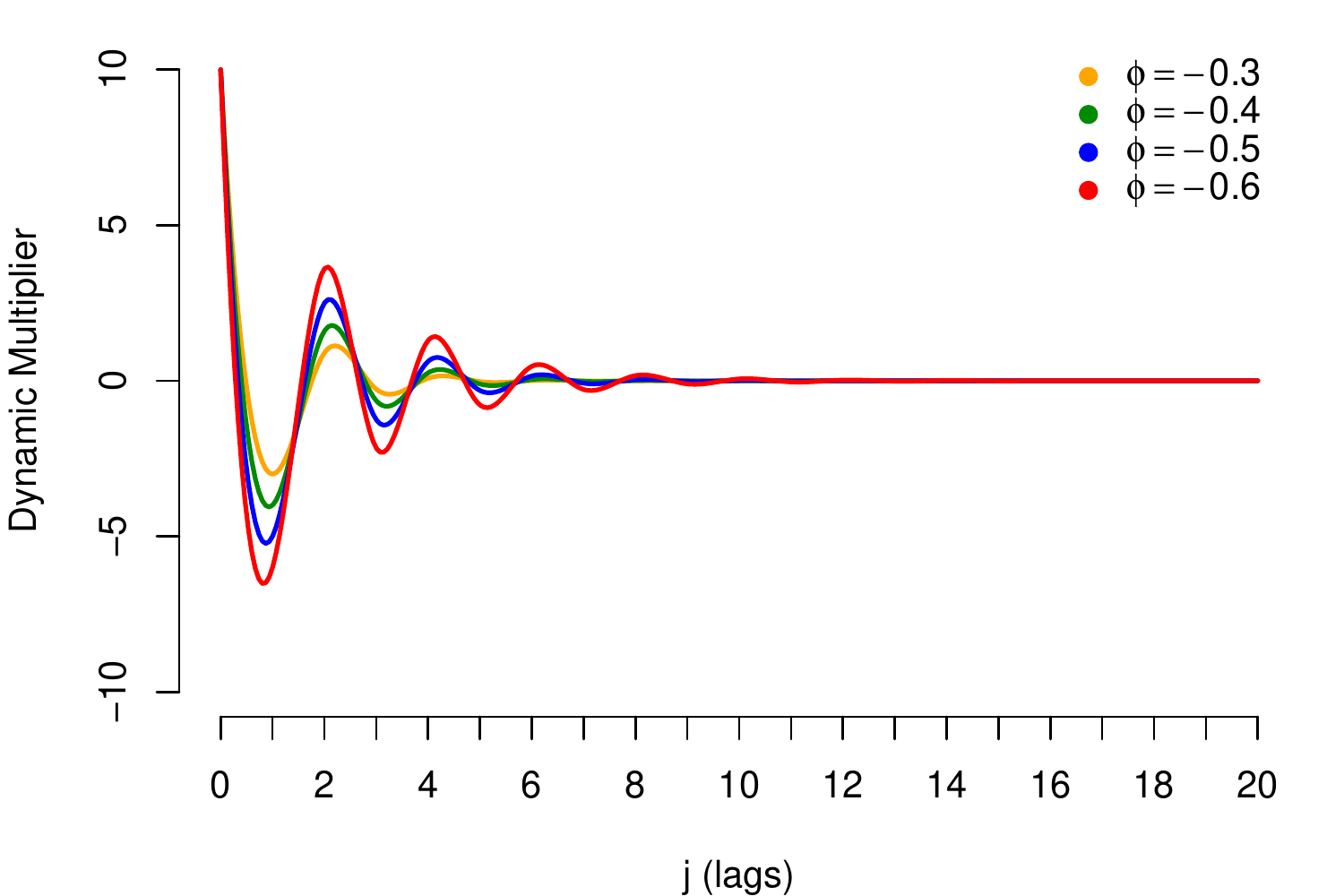}
	\caption{Dynamic Multiplier $\beta=10$ and $-1<\phi < 0$}
		\end{subfigure}
		\begin{subfigure}[b]{0.42\linewidth}
			\includegraphics[width=\linewidth]{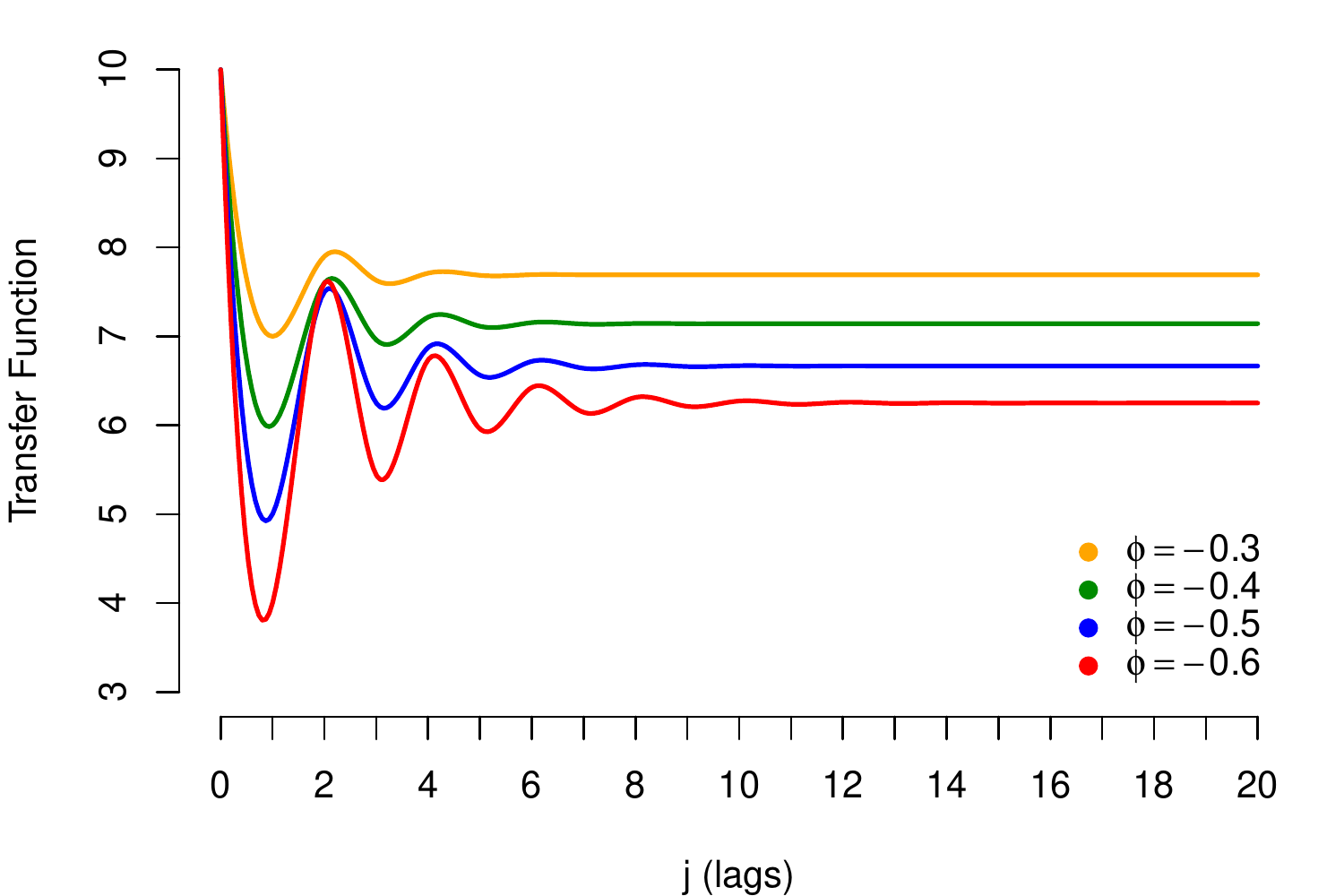}
\caption{Transfer Function $\beta=10$ and $-1 < \phi < 0$}	
		\end{subfigure}    
		\caption{Examples of the Dynamic Multiplier and Transfer Function with different values of $\phi$}
		\label{Fig: koyck_examples}
	\end{figure}  	
	
	Note that the condition $|\phi|<1$ in (\ref{koyck}), corresponds to the stability condition mentioned in (\ref{TFM}), in which case the Dynamic Multiplier may exhibit the following patterns: (i) If $0<\phi<1$, shows monotone geometric decay  and (ii) If $-1<\phi<0$, shows geometric decay with alternating signs. The parameter $\phi$ (resilience parameter) controls the velocity of the decay; absolute values next to 1 imply high persistent of impacts. Such patterns are exhibited in the Figure \ref{Fig: koyck_examples}, with $\beta >0$ (positive relationship) and different values for $\phi$, for $\beta<0$ (negative relationship) the patterns are similar but the decay has inverted sign.

	An extension of Koyck Distributed Lag (\ref{koyck}) was proposed by \cite{Solow1960family}, generally known as Solow Distributed Lag. Solow's assumption about the $\beta(j)$ is that they are generated by a Pascal distribution, which generates a more flexible transfer. An even more flexible proposal is known as Almon Distributed Lag, proposed by \cite{Almon1965distributed}. Almon's assumption is that the coefficients are well approximated by a low order polynomial on the lags, this polynomial approximation provides a wide variety of shapes for the lag function $\beta(j)$. With an adequate reparameterization and identification of the corresponding Transfer Function, both the Koyck, Solow and the Almon proposals can be expressed as a particular case of the GTFM (see, \cite{RavinesSchmidtMingon2006revisiting}).
	
	\subsubsection{Propagation of High Persistence}\label{php}
	
	The simplification proposed by Koyck implies a rapid decay of the impacts. However, many times the shock impact of some macroeconomic variables may present a high persistence, for example, the Unemployment variable referring to the credit risk data in section (\ref{iim}). An alternative to model highly persistent shocks is to choose Transfer Functions corresponding to either the Solow's or Almon's simplifications presented in the previous section. A disadvantage of these alternatives is the need to observe high persistence in all the shocks of the macroeconomic variables considered. But in general, as seen in credit risk data, high persistence is empirically observed only in some of the total macroeconomic variables considered, and in many cases only in one of them. Then, it is necessary to generate a flexible propagation mechanism only for the variables identified with high persistence. A simple way to approach this problem is to propose a general geometric decay for all the variables and generate a superposition effect of lag decays for the variables identified with high persistence.
	
\begin{figure}[h!]
		\centering
		\begin{subfigure}[b]{0.42\linewidth}
			\includegraphics[width=\linewidth]{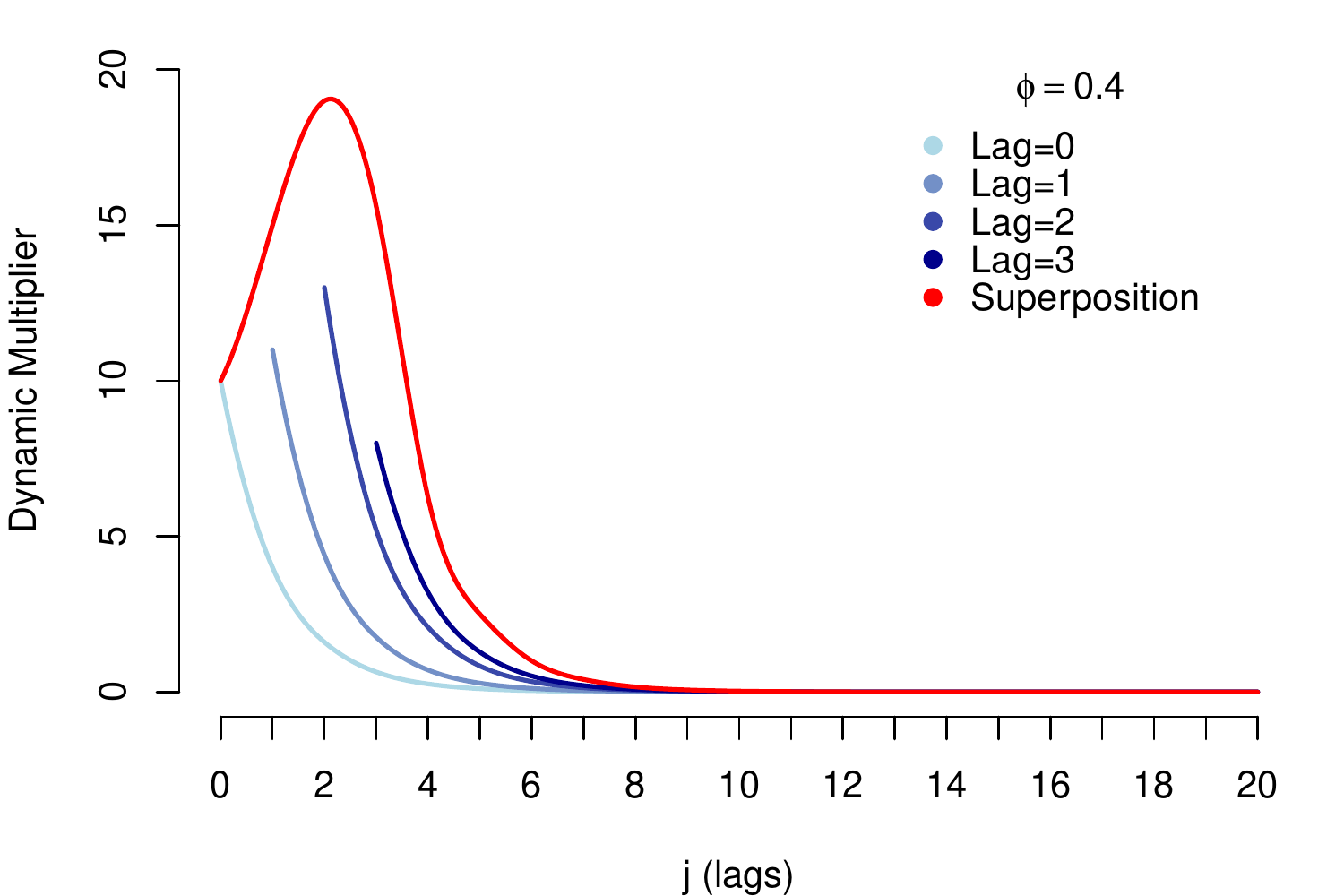}
		\end{subfigure}
		\begin{subfigure}[b]{0.42\linewidth}
			\includegraphics[width=\linewidth]{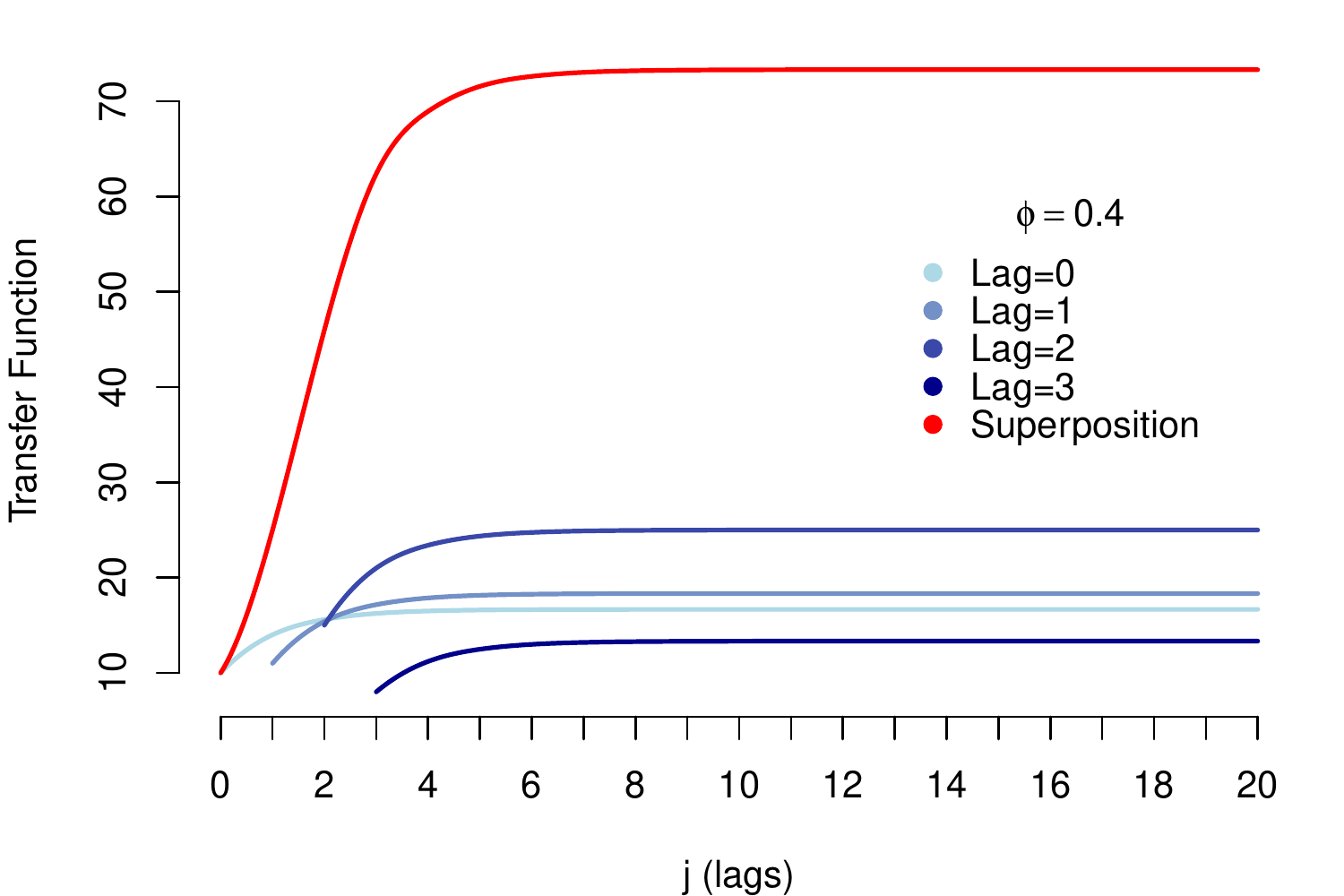}
		\end{subfigure}
    \caption{Dynamic Multiplier and Transfer Function with superposition of lags. Each of the blue curves represents a model with only one of the lags of the explanatory variable, and the red curve represents a model considering all the lags of this variable}
     \label{Fig:superpositionlags}
	\end{figure}

	For example, suppose that $X$ is a variable identified with high persistence; then we assume the following superposition of decays as Dynamic Multiplier
	
\begin{equation}\label{super}
	\beta(j)=\beta_{0}\phi^{j}+\beta_{1}\phi^{j-1}\mathbb{I}_{\{j\geq 1\}}+\ldots +\beta_{s}\phi^{j-s}\mathbb{I}_{\{j\geq s\}},\quad \forall \quad  j\geq 0
\end{equation}
where $|\phi|<1$ and $\beta_{k} \in \mathbb{R}$ for all $k \in \{0,1,\ldots,s\}$ and $\mathbb{I}_{\{.\}}$ is the indicator function. Note that $s$ is an arbitrary number that determines the number of lags used in the superposition (superposition-order), which is calibrated in function of intensity of the persistence in the data. Every lag has geometric decay but generates interference in the decay of their predecessor lags, see Figure \ref{Fig:superpositionlags}.
	
	The propagation defined in (\ref{super}) generates the following Transfer Function
	\begin{equation}
	\mathcal{T}(L)=\sum_{k=0}^{s}\frac{\beta_{k}L^k}{1-\phi L}
	\end{equation} 
	then, the Transfer Function model is
	
	\begin{subequations}\label{super_model}
		\begin{align}
		Y_{t}&=E_{t}+v_{t}\\
		E_{t}&=\phi E_{t-1}+\beta_{0} X_{t}+\beta_{1} X_{t-1}+\ldots + \beta_{s} X_{t-s}+\partial E_{t}
		\end{align}
	\end{subequations}
		with $v_{t}$ ($\partial E_{t}$) normally distributed with zero mean and variance $\sigma_{v}^{2}$ ($\sigma_{E}^{2}$), $\eta=\frac{\sigma_{v}}{\sigma_{E}}$ is signal-to-noise ratio. Same as in the model (\ref{koyck_dlm}), the model (\ref{super_model}) can be easily written using the general representation (\ref{gm}).
	
	\subsection{Choice of Transfer Function}
    Impact measures discussed in section \ref{im} represent the empirical propagation of shocks. We will to use this qualitative information embedded in the  impact measures to choose the Transfer Function. The functional form of the theoretical decay of the impacts (Dynamic Multiplier) must be consistent with the empirical propagation of the permanent impacts. Then, a criterion that we employed to choose the functional form of the Dynamic Multiplier of a macroeconomic variable, was that the functional form be consistent with the rapidity of decay of its Response Function $\widehat{\mathfrak{R}}(j)$. For example, if the Response Function $\widehat{\mathfrak{R}}(j)$ of a macroeconomic variable shows a high persistence or a behavior with wave decay, then the functional form proposed for its Dynamic Multiplier has to imitate that pattern, because it condenses the empirical shocks transmission.

    It is important, as seen above,  to point out that these measures are not used to specify the model with precision, as in the case of the autocorrelation function in the modeling of stationary series. Rather, due to the aggregate nature and the diffuse behavior (random walk) of the macroeconomic variables and the risk parameter, these measures are used to have a general approximation of the functional form of the impact decay that characterizes the transmission process.
%%--------------------------------------------------

\section{Time-varying Resilience and Stochastic Transfer }\label{Sec5}
	In the transfer functions discussed in the previous section the Resilience Parameter is constant over time, this implies a state of equilibrium between the shocks and the response of the risk parameter to these shocks, which remains constant. A constant resilience does not consider the learning process or deterioration of the agents' response, and therefore nor of the response of the risk parameter to recurrent macroeconomic shocks. It may be important to identify and incorporate this learning process or deterioration of the resilience of the parameter into the modeling, in order not to overestimate or underestimate the projections in the stress scenarios.

    To incorporate a dynamic behavior to resilience, we assume a mechanism of stochastic propagation:
	
	\begin{subequations}\label{STF}
		\begin{align}
		\label{STF1}
		\beta_{t}(j)&=\beta\prod_{i=0}^{j-1}\phi_{t-i}\\
		\label{STF2}
		\phi_{t}&=\phi_{t-1}+\partial\phi_{t}
		\end{align}
	\end{subequations}
	where $\beta \in \mathbb{R}$ and $\partial \phi_{t}$ are normally distributed with zero mean and variance $\sigma_{\phi}^{2}$. Note that the resilience changes slowly over time according to the simple random walk (\ref{STF2}). 
	
    The Stochastic Dynamic Multiplier in (\ref{STF}) generates the following Stochastic Transfer Function
	
	\begin{subequations}\label{EQ:STF2}
		\begin{align}
		\mathcal{T}_{t}(L)&=\beta\sum_{j=0}^{\infty}\prod_{i=0}^{j-1}\phi_{t-i}L^{j} \\
		\phi_{t}&=\phi_{t-1}+\partial\phi_{t}
		\end{align}
	\end{subequations}
	
\begin{figure}[h!]
		\centering
		\begin{subfigure}[b]{0.55\linewidth}
			\includegraphics[width=\linewidth]{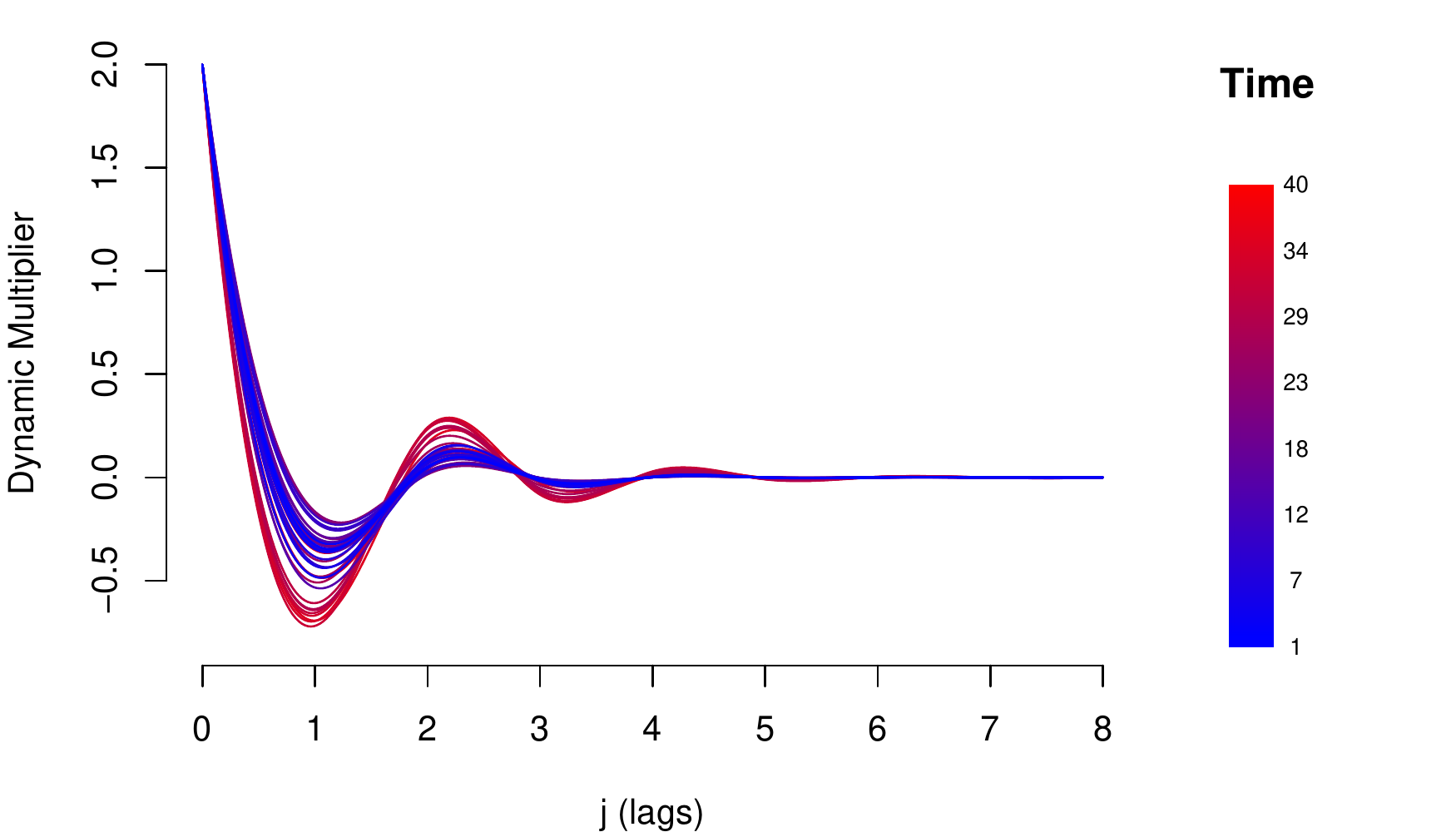}
		\end{subfigure}
		\begin{subfigure}[b]{0.55\linewidth}
			\includegraphics[width=\linewidth]{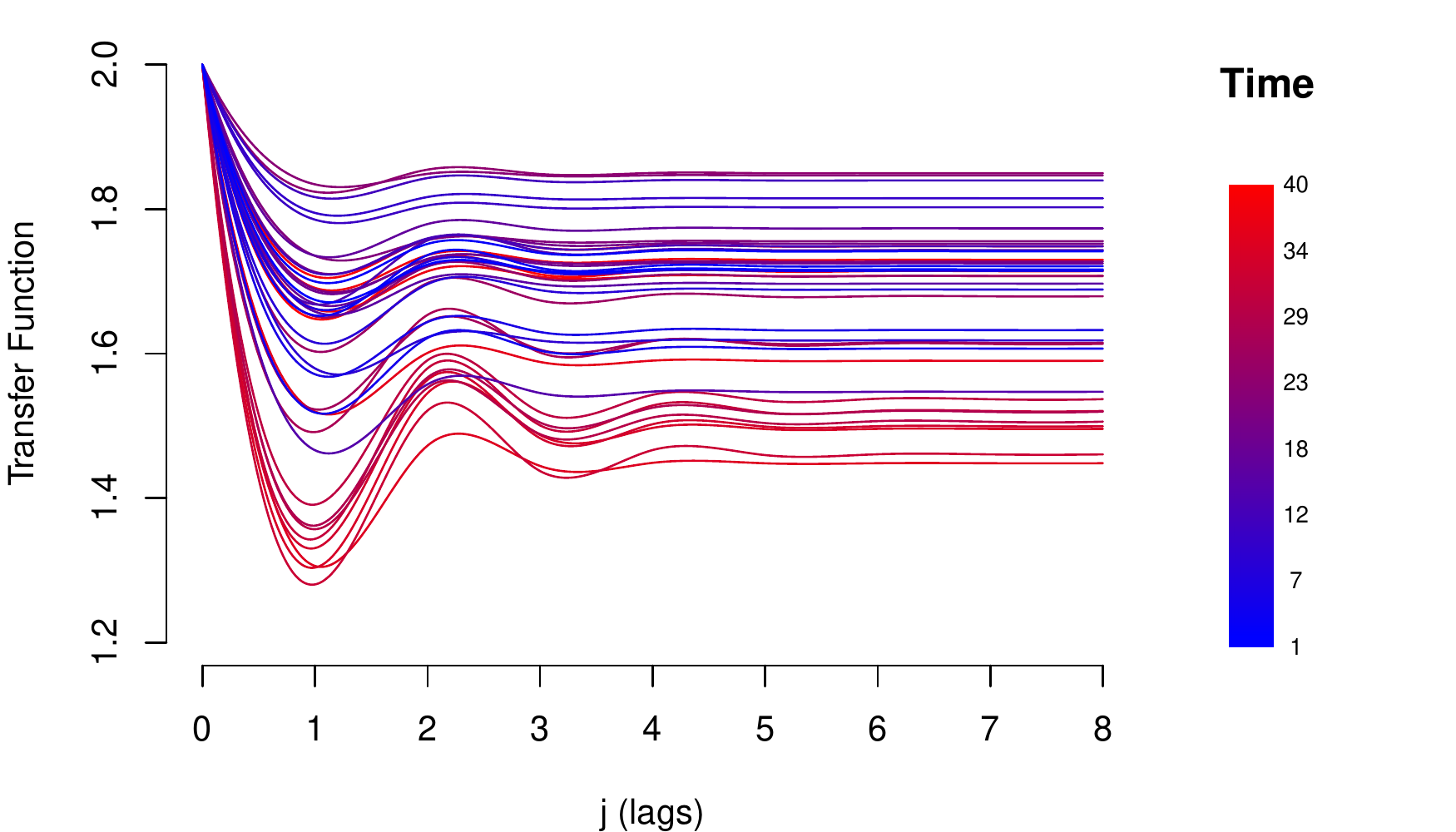}
		\end{subfigure}  
		\caption{Example of the Dynamic Multiplier with time-varying resilience and Stochastic Transfer Function. In these charts we considered $\beta=2$ and windows of 8 lags; using the equations (\ref{STF}) and (\ref{EQ:STF2}), observed for 40 periods. } 
		\label{Fig:stochastic_transfer}
	\end{figure}

	Then, the Transfer Function model is
	\begin{subequations}\label{tvr}
		\begin{align}
		Y_{t}&=E_{t}+v_{t}\\
		E_{t}&=\phi_{t} E_{t-1}+\beta X_{t}+\partial E_{t}\\
		\phi_{t}&=\phi_{t-1}+\partial\phi_{t}
		\end{align}
	\end{subequations}
	with $v_{t}$ ($\partial E_{t}$) normally distributed with zero mean and variance $\sigma_{v}^{2}$ ($\sigma_{E}^{2}$), $\eta=\frac{\sigma_{v}}{\sigma_{E}}$ signal-to-noise ratio. Like all models previously presented, the model (\ref{tvr}) can be easily written using the general representation (\ref{gm}).
	
    Equation (\ref{tvr}) also establishes a relationship of equilibrium between shocks and the response of the risk parameter, but that equilibrium is sensitive and changes before variations occur in the macroeconomic environment, recurrent characteristic of metastable systems. The flexible structure for the risk parameter makes it possible to identify a learning effect or deterioration of resilience over time. In Figure \ref{Fig:stochastic_transfer} is shown an example of deterioration of resilience.

%%--------------------------------------------------
	\section{Inference Procedure}\label{Sec6}
	
    Macroeconomic variables are usually correlated, this could significantly compromise the estimation of the parameters of the models described in this paper. Another problem emerges when it is necessary to model shocks with high persistence through the method of superposition of lags proposed in the section (\ref{php}), implying moderate to large values of $s$ (superposition-order), in which case the estimation of the parameters of the model, when we use the classic inference, is compromised by the autocorrelation in $X_{t}, \ldots ,X_{t-s}$. For these reasons we recommend the use of Bayesian inference whenever possible. Note that both problems could cause that the Response Functions and the Dynamic Multipliers do not have the same decay patterns. We propose to estimate the parameters of these models using the qualitative information contained in the Impact Measures (decay pattern of Response Functions), incorporating them into the prior using estimation procedure through Bayesian approach.
	
	\subsection{Prior Specification}\label{PriorDist}
    Following the Bayesian paradigm, the specification of a model is complete after specifying the prior distribution of all parameters of interest. To complete the specification of the models described, we defined the prior distribution according to the declination pattern of the Response Function. 
    
    For the Resilience Parameter $\phi \in (0,1)$, that corresponds to Response functions with monotone decay pattern, we specified a $\mathcal{B}eta(a,b)$ distribution; and  for $\phi \in (-1,0)$, that corresponds to Response functions with wave decay pattern, it is specified by setting $\phi=-\phi^{*}$ where $\phi^{*}\sim\mathcal{B}eta(c,d)$. We chose to work with the normal distribution for the regressors parameters, however, we decided to restrict the support according to the Response Function observed at the data, that is, considering the dynamic relationship between the risk parameter and the macroeconomic series.
	\begin{itemize}
		\item If the relation present by the Response Function of the $j$-th macroeconomic variable is positive, 
		\begin{align*}
		\beta_{j}\sim\textrm{Half-Normal}(\mu_{0}, \sigma_{0}^{2}) \hspace{1cm} \textrm{with} \hspace{1cm} \beta_{j} \in [0,\infty).
		\end{align*}
		\item If the relation of the Response Function of the $j$-th macroeconomic variable is negative, 
		\begin{align*}
		\beta_{j}\sim\textrm{Half-Normal}(\mu_{0}, \sigma_{0}^{2}) \hspace{1cm} \textrm{with} \hspace{1cm} \beta_{j} \in (-\infty,0].
		\end{align*}
	\end{itemize}
	
	Finally, the prior distribution for $\sigma_{E}$, $\eta$ and $\sigma_{\phi}$ are specified as,
	\begin{align*}
	\sigma_{E}&\sim \textrm{Inv-Gamma}(p_{1}, q_{1}) \\
	\sigma_{\phi}&\sim \textrm{Inv-Gamma}(p_{2}, q_{2})             \\
	\eta&\sim \textrm{Gamma}(\tau, \varpi). \\
	\end{align*}  
%where $\alpha_{1}=\alpha_{2}=2$, $\beta_{1}=\beta_{2}=0.1$, $\alpha_{3}=10$ and $=\beta_{3}=1$

	\subsection{MCMC-based Computation}
	
	There are several approximation methods based on simulations to obtain samples of the posterior distribution. The most widely used is the Markov Chain Monte Carlo Methods (MCMC), described in \cite{RobertCasella2010MCMC}. Within the MCMC family stand out the algorithms: Metropolis \citep{metropolis1953equation}, Metropolis-Hastings \citep{Hastings1970monte}, Gibbs Sampling \citep{GelfandSmith1990Gibbs} and Hamiltonian Monte Carlo \citep{Neal2011hmc}. Great advances have been made recently with Hamiltonian Monte Carlo algorithms (HMC, see for example \cite{Girolami2011AdvanceHMC}) for the estimation Bayesian models. The HMC has an advantage over the other algorithms mentioned, since it avoids the random walk behavior, presents smaller correlations structure and converges with less simulations. In this paper we used the HMC to obtain samples of the posterior distribution.
	
	\subsection{Model Comparison}
	
	One of the challenges in the modeling process is choosing the model that best represents the data structure and that is parsimonious. A widely used metric, in the case of Bayesian approach, is the Deviance Information Criterion (DIC) proposed by \cite{spiegelhalter2002dic}, which is defined as, 
	\begin{align*}
	DIC = -2\log p\big(\mathbf{Y}|\hat{\theta}_{Bayes}\big) + 2\textrm{p}_{DIC},
	\end{align*}
	where $\mathbf{Y}=(Y_{1},\dots ,Y_{T})$ are the data, $\theta$ is the vector of parameters of the model, $\hat{\theta}_{Bayes} = E[\theta|\mathbf{Y}]$ and $\textrm{p}_{DIC}$ is a penalizing term given by,
	\begin{align*}
	\textrm{p}_{DIC} = 2\Big(\log p(\mathbf{Y}|\hat{\theta}_{Bayes}) - E_{\theta|\mathbf{Y}}\big[\log p(\mathbf{Y}|\theta) \big] \Big).
	\end{align*} 
	Then, given M simulations from the posterior distribution of $\theta$, $p_{DIC}$ can be approximated as, 
	\begin{align*}
	\hat{\textrm{p}}_{DIC} = 2\Bigg(\log p\big(\mathbf{Y}|\hat{\theta}_{Bayes}\big) - \frac{1}{M}\sum_{m=1}^{M}\log p\big(\mathbf{Y}|\theta^{m}\big)  \Bigg)
	\end{align*}
	
	Additionally, in this paper, we used the Watanabe Information Criterion (WAIC) described by \cite{watanabe2010waic}. The WAIC does not depend on Fisher's asymptotic theory; therefore, it is not responsible for later displacement to a single point, as well as it is an alternative to more advanced models (see \cite{watanabe2013widely}, \cite{vehtari2015efficient} and \cite{vehtari2017practical}). It is defined as,
	\begin{equation}\label{elpdwaic}
	\textrm{elpd}_{WAIC} = \textrm{lpd} - \textrm{p}_{WAIC}
	\end{equation}
	where $\textrm{p}_{WAIC} = \sum_{i=1}^{T}V_{\theta|\mathbf{Y}}\big(\log p(Y_{i}|\theta)\big)$ penalizes for the effective number of parameters and $\textrm{lpd} = \sum_{i=1}^{T}\log p(Y_{i}|\mathbf{Y})$ is the log pointwise predictive density, as defined in \cite{watanabe2010waic}, where,
	\begin{align*}
	\log p(Y_{i}|\mathbf{Y}) = \log \int_{\Theta} p(Y_{i}|\theta)p(\theta|\mathbf{Y})d\theta.
	\end{align*}
	Similar to DIC, given M simulated from posterior distribution, the two terms in (\ref{elpdwaic}) are estimated as,
	\begin{align*}
	\textrm{lpd} &= \sum_{i=1}^{T}\log \Bigg(\frac{1}{M}\sum_{m=1}^{M}p(Y_{i}|\theta^{m})\Bigg) 
	\end{align*}
	and
	\begin{align*}
	\textrm{p}_{WAIC} &= \sum_{i=1}^{T}V_{m=1}^{M}\Big(\log p(Y_{i}|\theta)\Big).
	\end{align*}
	where $V_{m=1}^{M}$ denotes the sample variances of log $p(Y_{i}|\theta^{(1)}), \ldots, p(Y_{i}|\theta^{(M)})$. Thus, WAIC is defined as,
	\begin{align*}
	WAIC = -2\textrm{elpd}_{WAIC}.
	\end{align*} 
	
	The log pointwise predictive density can also be estimated  using approximate leave-one-out cross-validation (LOOIC) as,
	\begin{align*}
	\textrm{lpd}_{LOOIC} &= \sum_{i=1}^{T}\log p(Y_{i}|Y_{-i}) = \sum_{i=1}^{T}\log\int_{\Theta} p(Y_{i}|\theta)p(\theta|Y_{-i})d\theta, 
	\end{align*}
	where $Y_{-i}$ denotes the data vector with the $i$th observation deleted. \cite{vehtari2015efficient} introduced an efficient approach to compute LOOIC using Pareto-Smoothed Importance Sampling (PSIS). The lower the value of the selection criteria, the better the model is considered. 
	It is worth mentioning that the estimates for WAIC and LOOIC are obtained as the sum of independent components, so it is possible to calculate the approximate standard errors for the estimated predictive errors.

	\subsection{Forecasting with Posterior Distribution}
    Without loss of generality, we will assume a univariate case (only a macroeconomic variable). We want to apply our model analysis to a new set of data, where we observed the new covariate $X_{T+1}$ corresponding to one of the stress scenarios, and we wish to predict the corresponding outcome $Y_{T+1}$. From a Bayesian perspective, this process can be performed considering the posterior predictive distribution.
\begin{equation}
p(Y_{T+1}|D_{T}) = \int p(Y_{T+1}|\gamma, D_{T})p(\gamma|Y_{T},D_{T})d\gamma,
\end{equation}
where $D_{T}$ is all the information until the instant $T$ and $\gamma = (E_{T+1}, \sigma_{E})^{\top}$. However, since there is no analytical determination of this distribution, we adopted the following procedure.  
\begin{align*}
\phi_{T+1}^{(m)}&\sim\mathcal{N}\Big(\phi_{T}^{(m)}, \sigma_{\phi}^{2(m)}\Big) \\
E_{T+1}^{(m)}&\sim\mathcal{N}\Big(\phi_{T}^{(m)}E_{T}^{(m)} + X_{T+1}^{\top}\beta^{(m)}, \sigma_{E}^{2(m)}\Big) \\
Y_{T+1}^{(m)}&\sim\mathcal{N}\Big(E_{T+1}^{(m)}, \sigma_{v}^{2(m)}\Big), 
\end{align*} 
where $m = 1, \ldots, M$ are MCMC samples of the posterior distribution. Then, $\Big(Y_{t+1}^{(1)}, \ldots, Y_{t+1}^{(M)}\Big)$ are an i.i.d. sample from $p(Y_{T+1}|D_{T})$. The process can be repeated until $h$-steps-ahead, considering $X_{T+2},\ldots,X_{T+h}$, to obtain the forecast to $Y_{T+2},\ldots,Y_{T+h}$.  

%%--------------------------------------------------
	\section{Simulations Study}\label{Simulation}
	
	In this section , the theoretical results of the previous sections are illustrated through the simulations study. First we will provide the details of the process of simulating of the data series. We will explain how evaluating the simulations and to comment about the results obtained.

	\textbf{Data Generating Process}. The simulation process of data series was done follow the steps below:
	
	\begin{itemize}
		\item Set the number of observations ($T$) in the series you will simulated;
		\item Generate two different random walk processes $X_{1}$ and $X_{2}$ of size $T$, which in this case represent the macroeconomic variables;
		\item Set the values of the parameters that will generate the response variable $Y$, which in this case represents the risk parameter;
		\item Finally $Y_{t}\sim\mathcal{N}(\alpha+\phi y_{t-1}+\beta_{1}x_{1t}+\beta_{2}x_{2t}, \sigma_{v}^{2})$ of size $T$.
	\end{itemize}	

	As in general the series used in the Stress Test exercises are short, we consider using $T = 40$ and we generated $M = 500$ replicates of series of this size. 
	We divided the simulation study into two steps. In the first stage we check the functional form of the decays presented by the simulated series and calculate the hit rate of the response functions. For the first step we used several configurations, as shown in Table \ref{sims1}. In the second step, we used two of the configurations presented in step 1, performed the estimation of the N series and evaluated some metrics in relation to the estimates, they are: General mean, general deviation, Mean Squared Error (MSE) and Median Absolute Error (MAE).

	To calculate the hit rate of the response function, we consider the following process. Be $\Delta\widehat{\mathfrak{R}}(j) = \widehat{\mathfrak{R}}(j) - \widehat{\mathfrak{R}}(j-1)$, where $j=0,\ldots, L$, consider the mean dacay below:
	\begin{align*}
		\bar{D}_{m} = \frac{1}{L-1}\sum_{j=0}^{L-1}\Delta\widehat{\mathfrak{R}}(j)^{(m)}, 
	\end{align*}
	where $\widehat{\mathfrak{R}}(j)^{(m)}$ is the estimative of the response function of $m$-th iteration of the $M$ simulations. The Hit Rate of the Response Function is given by:
	\begin{align*}
		\textrm{Hit Rate} = \frac{1}{M}\sum_{m=1}^{M}\mathbb{I}_ {\{\bar{D}_{m}, \beta<0\}}.
	\end{align*}
	We used numbers of lags $L=10$.	

	\subsection{Results}
		
	In this section we present the results of the two steps of the simulation study.
 
		\begin{table}[!h]
		\centering
		\caption{Configuration of the parameters and Hit Rate for the Response Function Simulations}\label{sims1}
		\vspace{-0.2cm}
		\begin{tabular}{cccc|cc}
			\hline
			$\phi$	&	$\sigma$	&	$\beta_{1}$	&	$\beta_{2}$	&	\multicolumn{2}{c}{Hit Rate} \\ \hline
\multirow{9}{*}{0.7}	&	1	&	0.4	&	-0.4	&	0.882	&	0.852	\\
				&	1	&	0.4	&	-0.8	&	0.752	&	0.966	\\
				&	1	&	0.8	&	-0.4	&	0.948	&	0.724	\\
				&	0.1	&	0.4	&	-0.4	&	0.874	&	0.844	\\
				&	0.1	&	0.4	&	-0.8	&	0.726	&	0.954	\\
				&	0.1	&	0.8	&	-0.4	&	0.968	&	0.758	\\
				&	0.01	&	0.4	&	-0.4	&	0.874	&	0.882	\\
				&	0.01	&	0.4	&	-0.8	&	0.724	&	0.972	\\
				&	0.01	&	0.8	&	-0.4	&	0.972	&	0.74	\\
\multirow{9}{*}{0.4}	&	1	&	0.4	&	-0.4	&	0.926	&	0.938	\\
				&	1	&	0.4	&	-0.8	&	0.810	&	0.988	\\
				&	1	&	0.8	&	-0.4	&	0.984	&	0.794	\\
				&	0.1	&	0.4	&	-0.4	&	0.932	&	0.916	\\
				&	0.1	&	0.4	&	-0.8	&	0.784	&	0.994	\\
				&	0.1	&	0.8	&	-0.4	&	0.996	&	0.806	\\
				&	0.01	&	0.4	&	-0.4	&	0.934	&	0.918	\\
				&	0.01	&	0.4	&	-0.8	&	0.802	&	0.992	\\
				&	0.01	&	0.8	&	-0.4	&	0.994	&	0.806	\\
			\hline
		\end{tabular}
	\end{table}		

	According the Table \ref{sims1}, the hit rate of the response function is satisfactory. With this we can be use the response function it to evaluate the decay of the shocks of the macroeconomic variables. Note that when the value of one of the betas is double the other, its decay hit rate is higher than the which one with of the lowest weight, but yet is satisfactory to evaluate the decay. 

	To the step two we evaluate the estimations considering two configurations and compare the Response Function Mean of the estimates with the Expected decay from of each variables. 

	The configuration 1 we considered is:  $\alpha=3$, $\phi=0.4$, $\beta_{1}=-0.4$, $\beta_{2}=0.4$ and $\sigma_{v}=0.1$.
	
	\begin{table}[ht!]
		\centering
		\caption{Estimative of simulations of the configuration 1}\label{sims1Est}
		\vspace{-0.2cm}
		\begin{tabular}{cccccc}
			\hline
		Measures	&	$\alpha$	&	$\phi$	          &	$\beta_{1}$	&	$\beta_{2}$	&	$\sigma_{v}$	\\ \hline
			MEAN	&	2.9756	&	0.4032	&	-0.3970	&	0.3981	&	0.1158		\\
			SD	&	0.1053	&	0.0196	&	0.0142	&	0.0141	&	0.0157		\\
			MSE	&	0.0117	&	0.0004	&	0.0002	&	0.0002	&	0.0005		\\
			MAE	&	0.0632	&	0.0129	&	0.0084	&	0.0092	&	0.0164		\\
			\hline
		\end{tabular}
	\end{table}

	As present in the Table \ref{sims1Est}, the estimates obtained are close to the true values of the parameters. Also we observated that the standard deviation of $\alpha$ and the difference in scale of the $\sigma$ parameter is slightly higher than the others presented.

		\begin{figure}[ht!]
		\centering
		\begin{subfigure}[t]{0.35\textwidth}
			\centering
			\includegraphics[width=\textwidth]{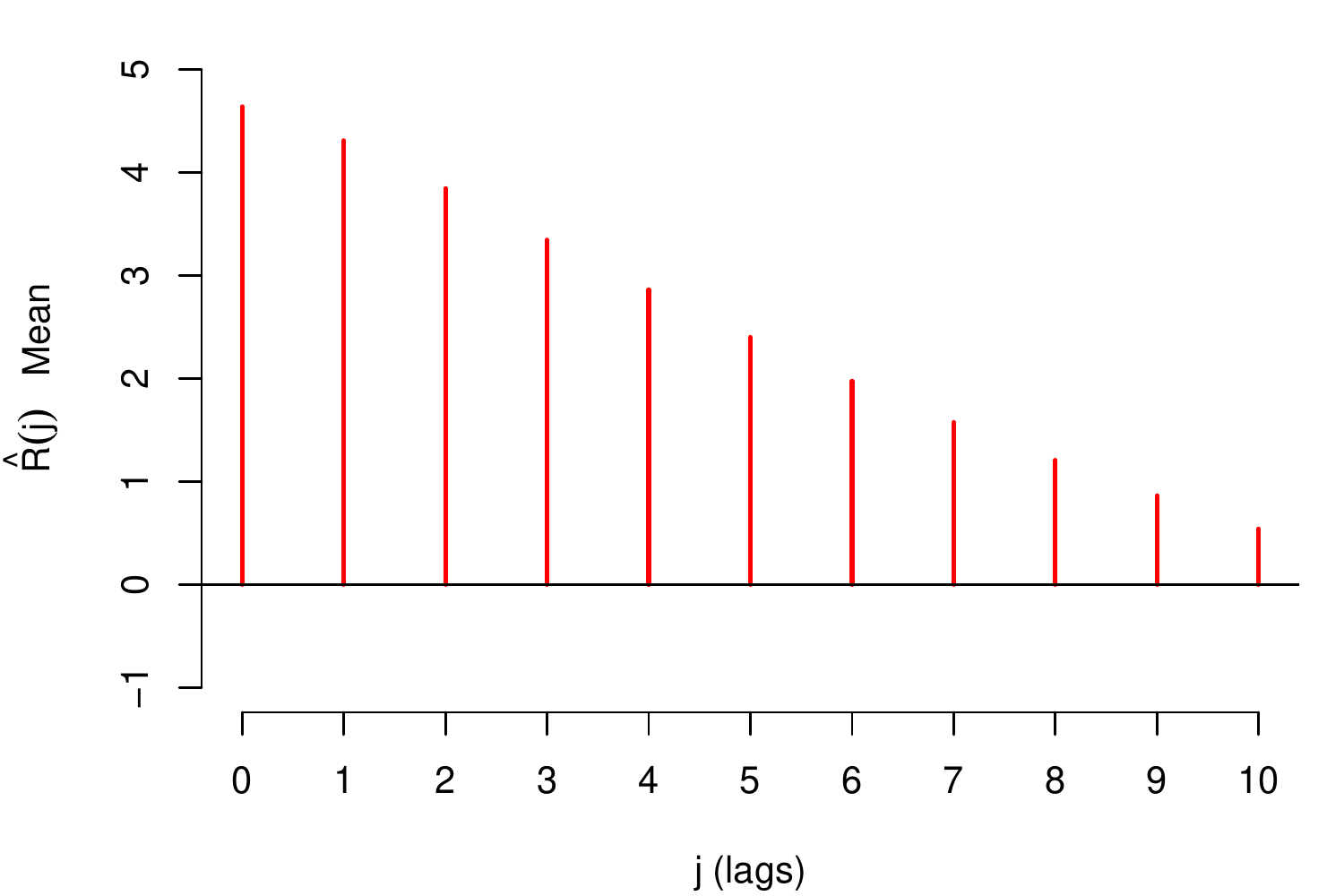}  
			\caption{Response Functions Mean variable 1}
		\end{subfigure}
		\hspace{1cm} %\hfill
		\begin{subfigure}[t]{0.35\textwidth}
			\centering
			\includegraphics[width=\textwidth]{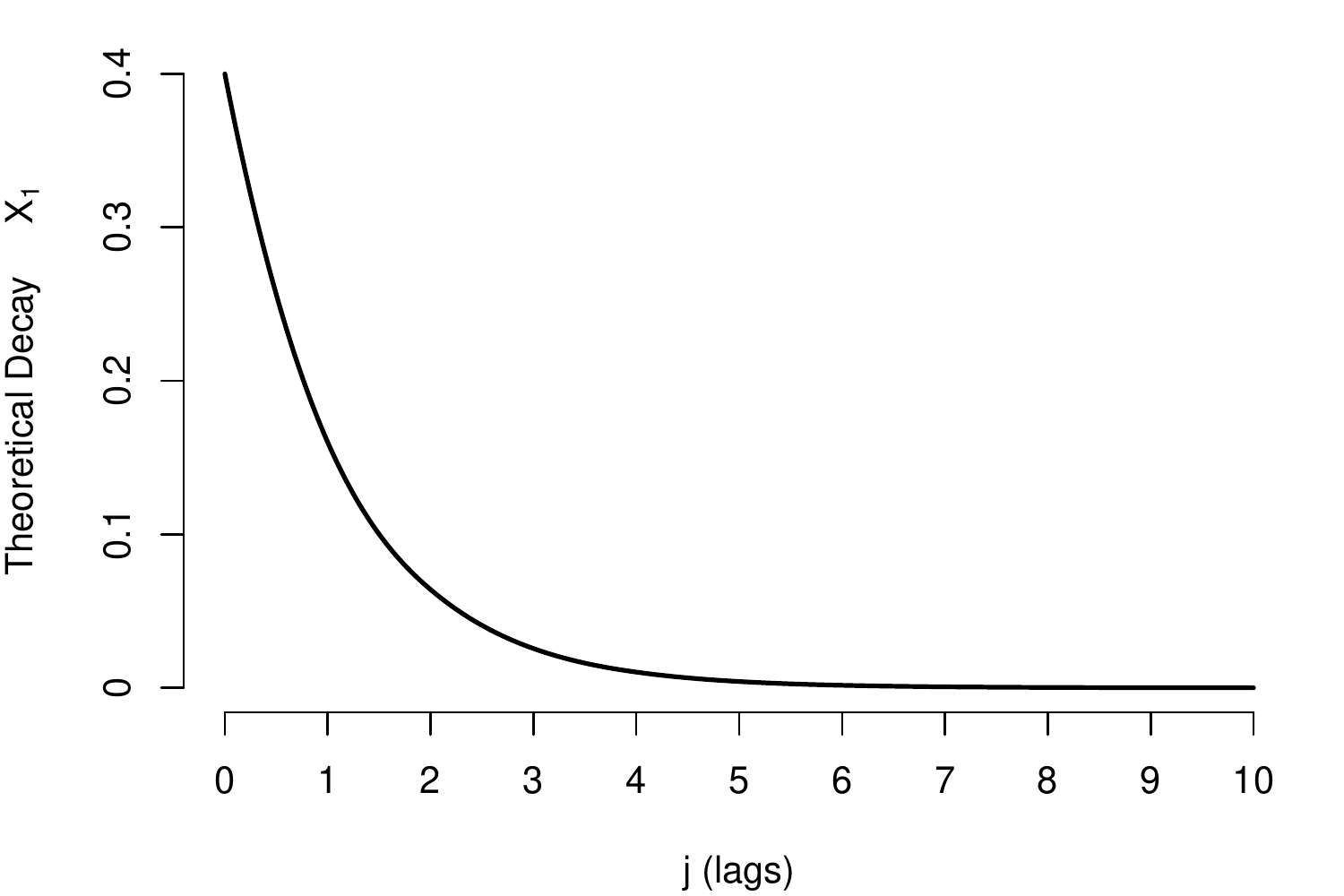}  
			\caption{Expected decay of the variable 1} 
		\end{subfigure}

		\begin{subfigure}[t]{0.35\textwidth}
			\centering
			\includegraphics[width=\textwidth]{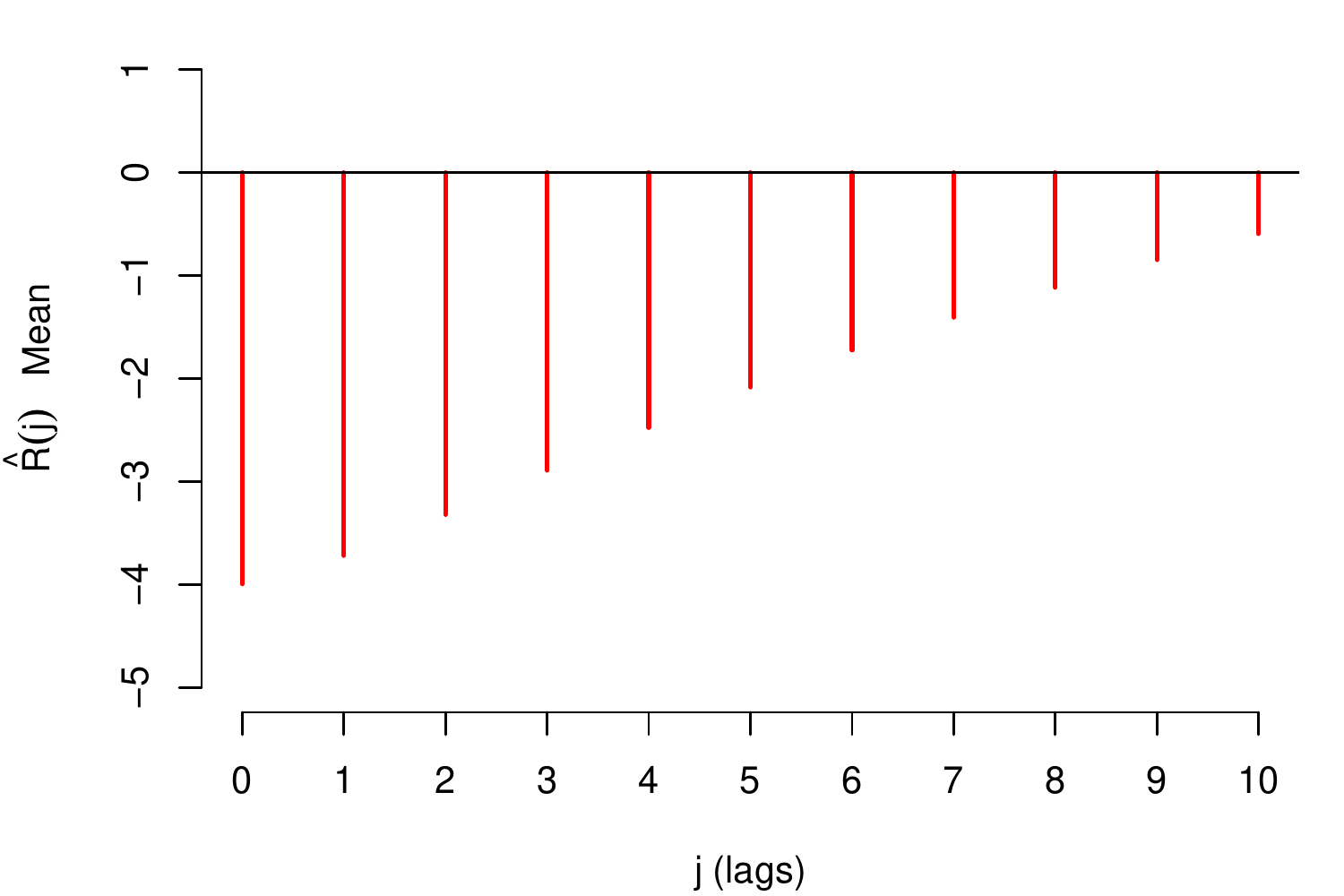}  
			\caption{Response Functions Mean variable 2}
		\end{subfigure}
		\hspace{1cm} %\hfill
		\begin{subfigure}[t]{0.35\textwidth}
			\centering
			\includegraphics[width=\textwidth]{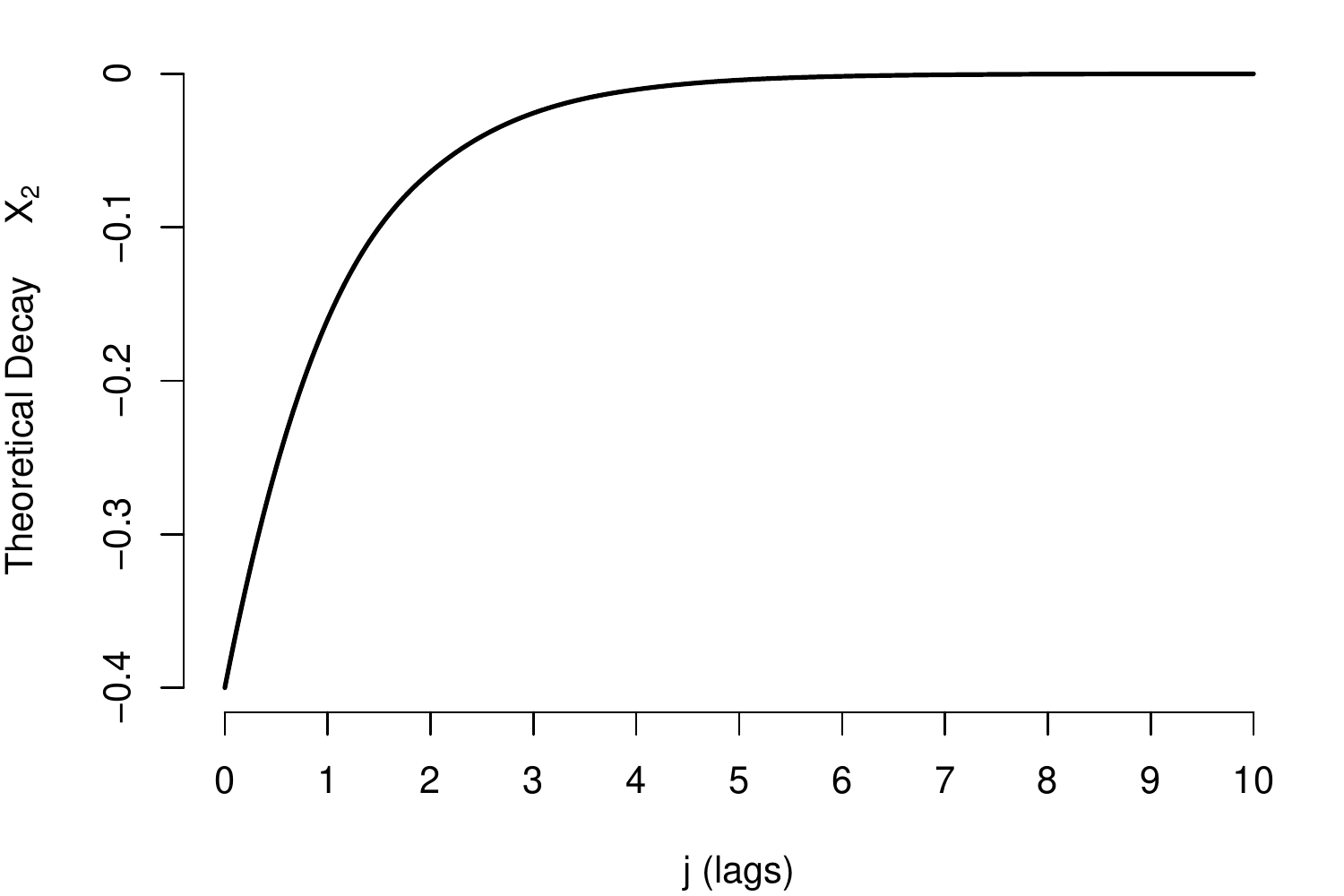}  
			\caption{Expected decay of the variable 2} 
		\end{subfigure}
		\caption{Theoretical transmission of model for the configuration 1}\label{sims1Fig}
	\end{figure}
	
	In the Figure \ref{sims1Fig} we presented the Reponse Function Mean of the $M=500$ simulations and present the Expected decay of the variables used to construct the data series. As shown the average empirical functional form of decay is in accordance with the expected form of theoretical decay of the variables.

	The configuration 1 we considered is:  $\alpha=3$, $\phi=0.7$, $\beta_{1}=-0.8$, $\beta_{2}=0.4$ and $\sigma_{v}=0.01$.
	
	\begin{table}[ht!]
		\centering
		\caption{Estimative of simulations of the configuration 2}\label{sims2Est}
		\vspace{-0.2cm}
		\begin{tabular}{cccccc}
			\hline
		Measures	&	$\alpha$	&	$\phi$	          &	$\beta_{1}$	&	$\beta_{2}$	&	$\sigma_{v}$	\\ \hline
			MEAN	&	2.9995	&	0.7000	&	-0.7999	&	0.4000	&	0.0088	\\
			SD	&	0.0077	&	0.0006	&	0.0015	&	0.0012	&	0.0012	\\
			MSE	&	0.0001	&	0.0000	&	0.0000	&	0.0000	&	0.0000	\\
			MAE	&	0.0045	&	0.0004	&	0.0010	&	0.0007	&	0.0013	\\
			\hline
		\end{tabular}
	\end{table}

	Like in configuration 1, the the estimates obtained in configuration 2, showing in the Table \ref{sims2Est}, are close to the true values of the parameters. We also note that the standard deviation of all estimates has decreased considerably. We believe it is due to the resilience parameter.

		\begin{figure}[ht!]
		\centering
		\begin{subfigure}[t]{0.35\textwidth}
			\centering
			\includegraphics[width=\textwidth]{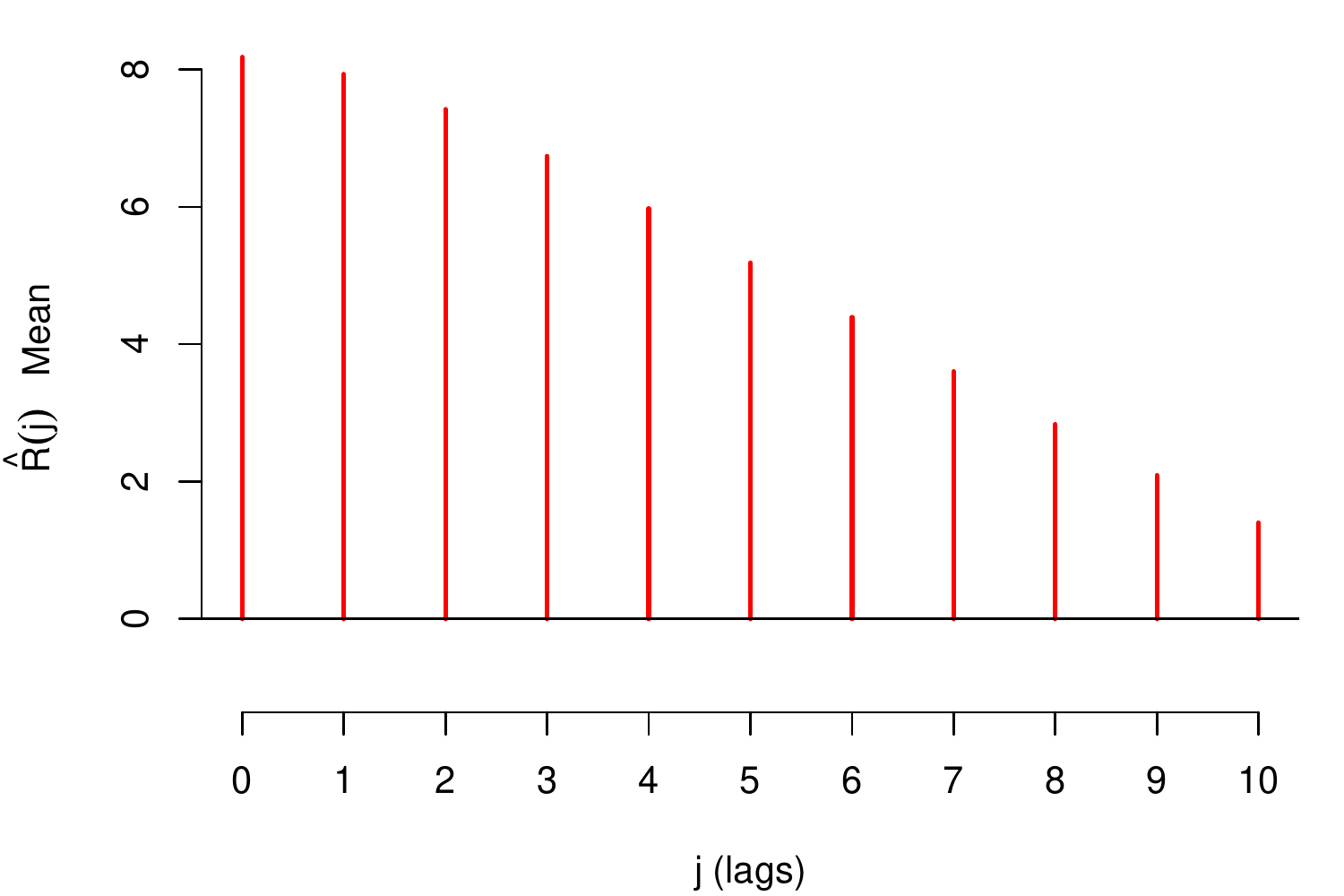}  
			\caption{Response Functions Mean variable 1}
		\end{subfigure}
		\hspace{1cm} %\hfill
		\begin{subfigure}[t]{0.35\textwidth}
			\centering
			\includegraphics[width=\textwidth]{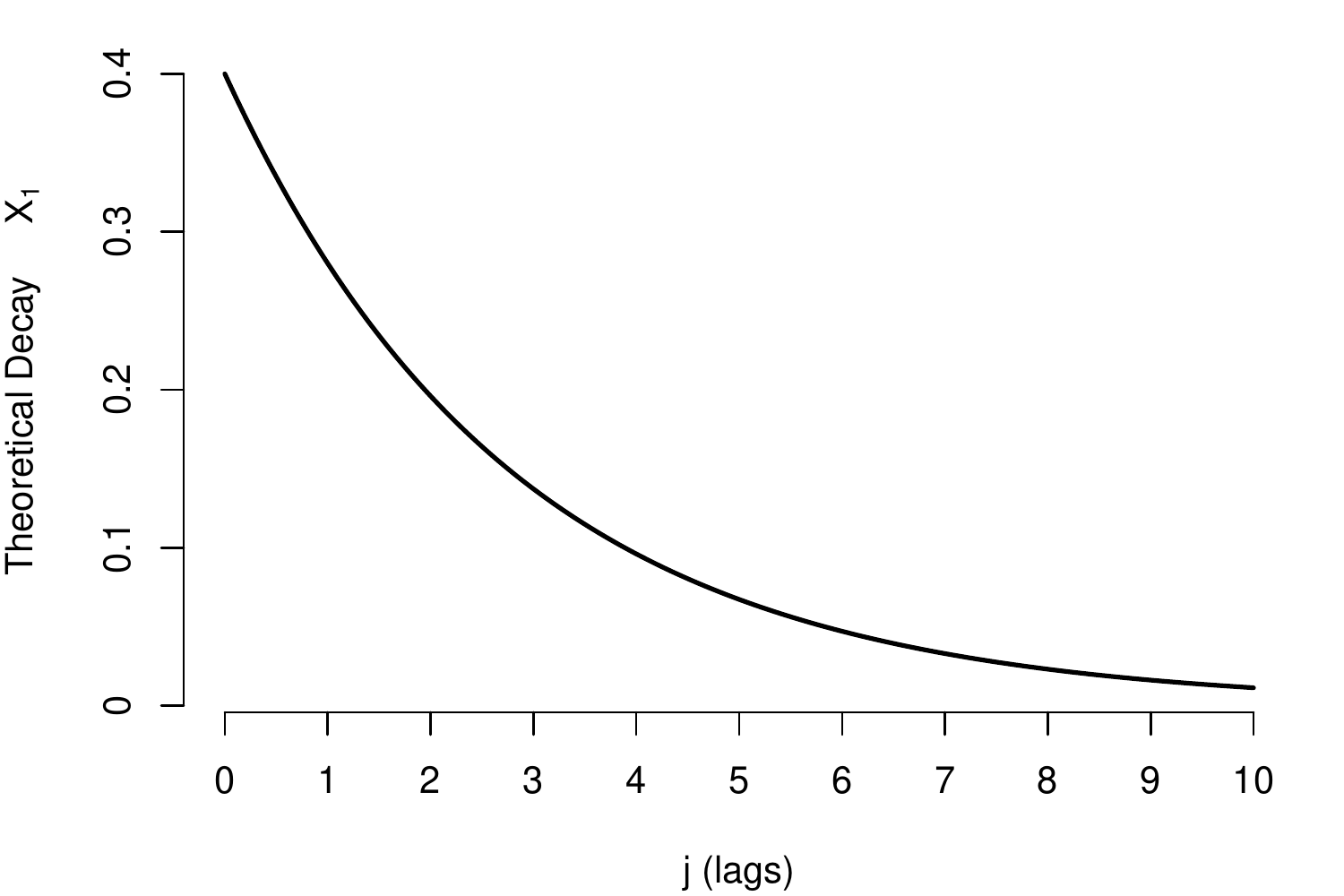}  
			\caption{Expected decay of the variable 1} 
		\end{subfigure}

		\begin{subfigure}[t]{0.35\textwidth}
			\centering
			\includegraphics[width=\textwidth]{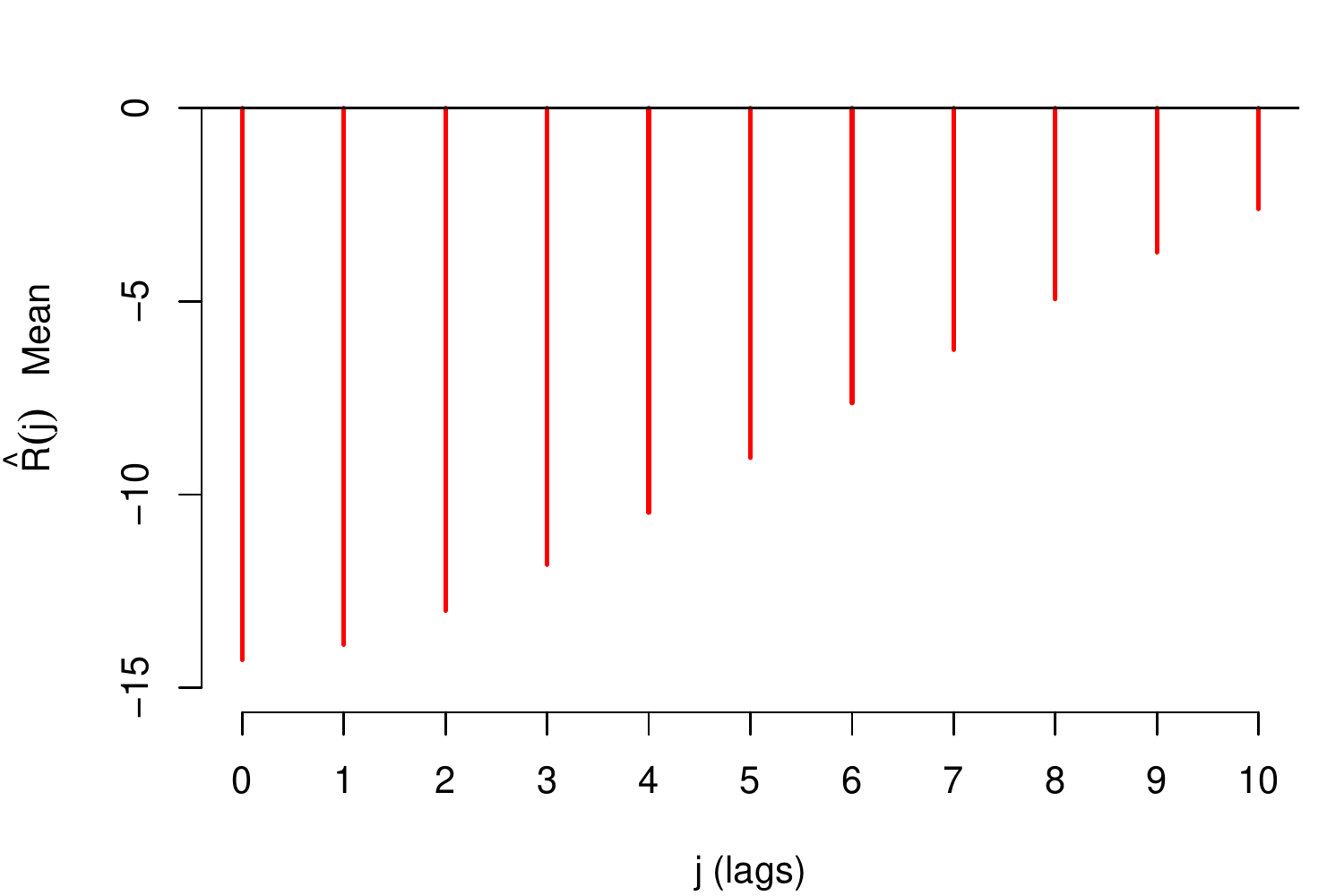}  
			\caption{Response Functions Mean variable 2}
		\end{subfigure}
		\hspace{1cm} %\hfill
		\begin{subfigure}[t]{0.35\textwidth}
			\centering
			\includegraphics[width=\textwidth]{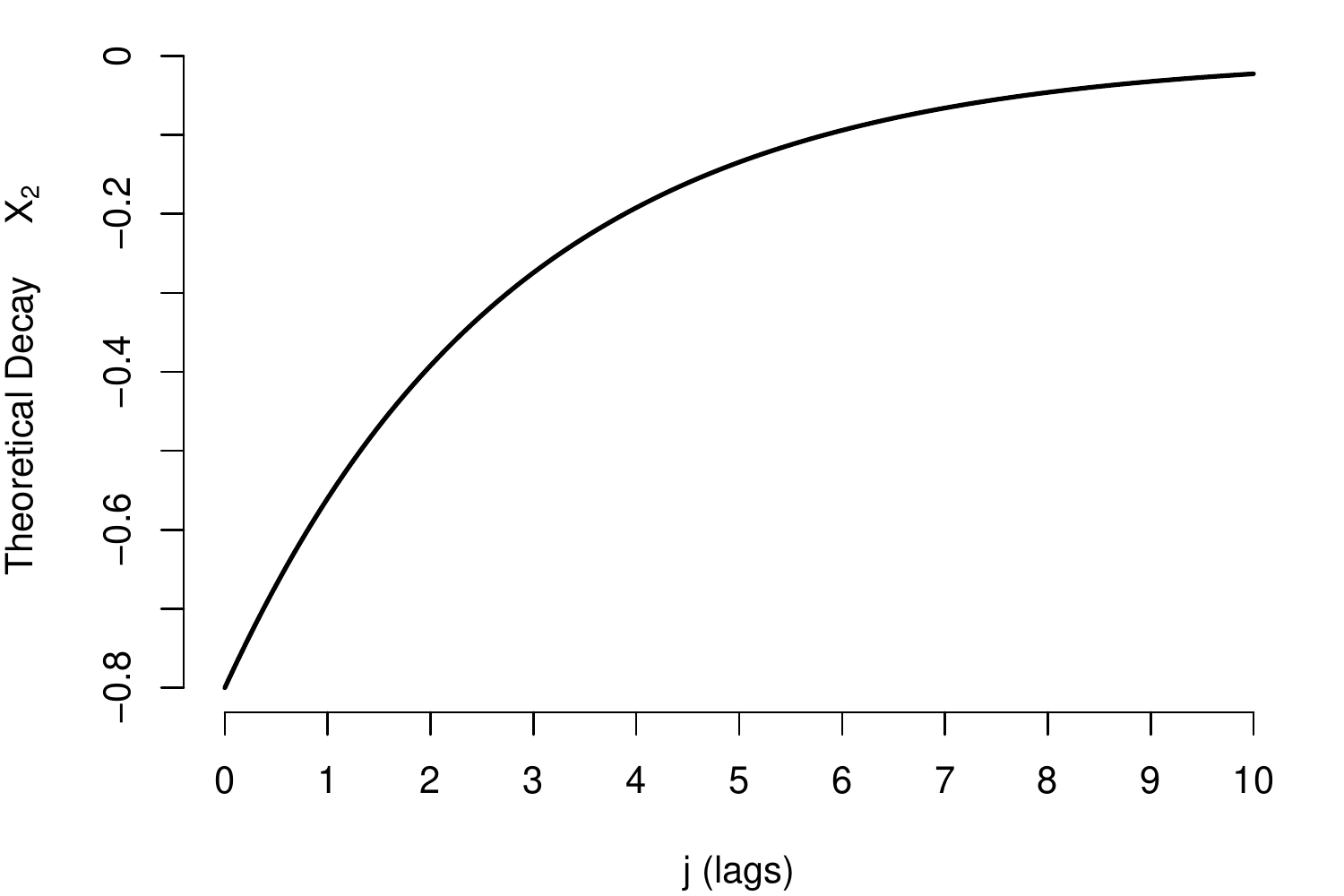}  
			\caption{Expected decay of the variable 2} 
		\end{subfigure}
		\caption{Theoretical transmission of model for the configuration 2}\label{sims2Fig}
	\end{figure}
	
	As observed in both configuration the functional form of average decay that is the qualitative information that will be used to build the priori, present in Figures \ref{sims1Fig} and \ref{sims2Fig} , are in agreement with the form of theoretical decay already defined by construction of the data series. It is worth mentioning that, since Bayesian inference was used, the estimates presented are based on the mean and standard deviation of a posteriori means. 

%%--------------------------------------------------
	\section{Application: Case study}\label{Sec7}
	
	In this section, we illustrate the use of the proposed model in credit risk data that was presented in section \ref{Sec2}. We consider the same macroeconomic series: GDP, IDR and Unemployment. It is noteworthy that the Unemployment series shocks on risk parameter LGD are of persistent nature, since the empirical measures of impact show a slow decay, as presented in section \ref{im} and there are economic arguments to corroborate this fact; therefore, we considered a superposition with three lags (order 3) for the Unemployment variable and geometric decay for GDP and IDR. We also considered the use of the four scenarios mentioned before, being:  Baseline, Optimistic, Local and Global.
	
	In order to demonstrate the flexibility of general model, we tested some variations of the model; these were denominated models I, II, III and IV. For each model we drew 10,000 MCMC samples for the parameters using HMC with Stan and NUTS where 5,000 were discarded as warmup, as detailed below.
	
	\subsection{Model I}
	The Model I can be represented as,
	
\begin{align}
	Y_{t} &=E_{t} + v_{t}, \hspace{4.2cm}             v_{t}\sim\mathcal{N}(0,\sigma_{Y}^{2}) \nonumber \\
	E_{t} &= \phi E_{t-1} + X_{t}^{\top}\beta+\partial E_{t}, \hspace{2cm} \partial E_{t}\sim\mathcal{N}(0,\sigma_{E}^{2}), \\
	\textrm{Initial State:} \hspace{2cm} E_{1} &\sim \mathcal{N}(Y_{1}, \sigma_{E}^{2}), \nonumber
\end{align}
where, $\eta=0$, $X_{t}^{\top}=(1, GDP_{t}, IDR_{t}, Unemp_{t}, Unemp_{t-1}, Unemp_{t-2}, Unemp_{t-3} )$ and $\beta=(\alpha, \beta_{1}, \ldots, \beta_{6})^{\top}$. We defined the following prior distributions,
	\begin{align*}
	\alpha&\sim \mathcal{N}(1.5, 0.5), \hspace{0.5cm} \textrm{with} \hspace{0.5cm} \alpha \in \mathbb{R} \\
	\beta_{j}&\sim\textrm{Half-Normal}(0, 1), \hspace{0.5cm} \textrm{with} \hspace{0.5cm} \beta_{j} \in (-\infty,0] \,\, \textrm{and} \,\, j = 1           \\
	\beta_{j}&\sim\textrm{Half-Normal}(0, 1), \hspace{0.5cm} \textrm{with} \hspace{0.5cm} \beta_{j} \in [0,\infty) \,\, \textrm{and} \,\, j \in \{2,3,4,5,6\}           \\
	\phi&\sim\mathcal{B}eta(2,2), \hspace{0.5cm} \textrm{with} \hspace{0.5cm} \phi \in (0,1) \\
	\sigma_{E}&\sim \textrm{Inv-Gamma}(2, 0.1), \hspace{0.5cm} \textrm{with} \hspace{0.5cm} \sigma_{E} \in [0,\infty). 
	\end{align*} 
The results of this model are presented below,
	
	\begin{table}[ht]
		\centering
		\caption{Posterior measures of the parameters of model I}
		\vspace{-0.2cm}
		\begin{tabular}{cccccccc}
			\hline
			Parameters & Mean & SD & 2.5\% & 25\% & 50\% & 75\% & 97.5\% \\ 
			\hline
			$\alpha$ & 1.9640 & 0.3522 & 1.2860 & 1.7262 & 1.9625 & 2.1981 & 2.6628 \\ 
			$\phi$ & 0.4264 & 0.0717 & 0.2808 & 0.3805 & 0.4279 & 0.4757 & 0.5607 \\ 
			$\beta_{1}$ & -0.0131 & 0.0020 & -0.0170 & -0.0144 & -0.0131 & -0.0118 & -0.0093 \\ 
			$\beta_{2}$ & 0.2662 & 0.0469 & 0.1784 & 0.2357 & 0.2644 & 0.2960 & 0.3639 \\ 
			$\beta_{3}$ & 0.7939 & 0.5889 & 0.0370 & 0.3248 & 0.6677 & 1.1586 & 2.1934 \\ 
			$\beta_{4}$ & 0.9264 & 0.6624 & 0.0363 & 0.4000 & 0.8006 & 1.3310 & 2.4865 \\ 
			$\beta_{5}$ & 1.1099 & 0.7127 & 0.0648 & 0.5528 & 1.0309 & 1.5600 & 2.7247 \\ 
			$\beta_{6}$ & 0.9406 & 0.6619 & 0.0418 & 0.4071 & 0.8330 & 1.3559 & 2.4601 \\ 
			$\sigma_{E}$ & 0.0977 & 0.0157 & 0.0722 & 0.0868 & 0.0959 & 0.1067 & 0.1336 \\ 
			\hline
		\end{tabular}
	\end{table}

	\begin{figure}[!h]
		\centering
		\begin{subfigure}[t]{0.495\textwidth}
			\centering
			\includegraphics[width=\textwidth]{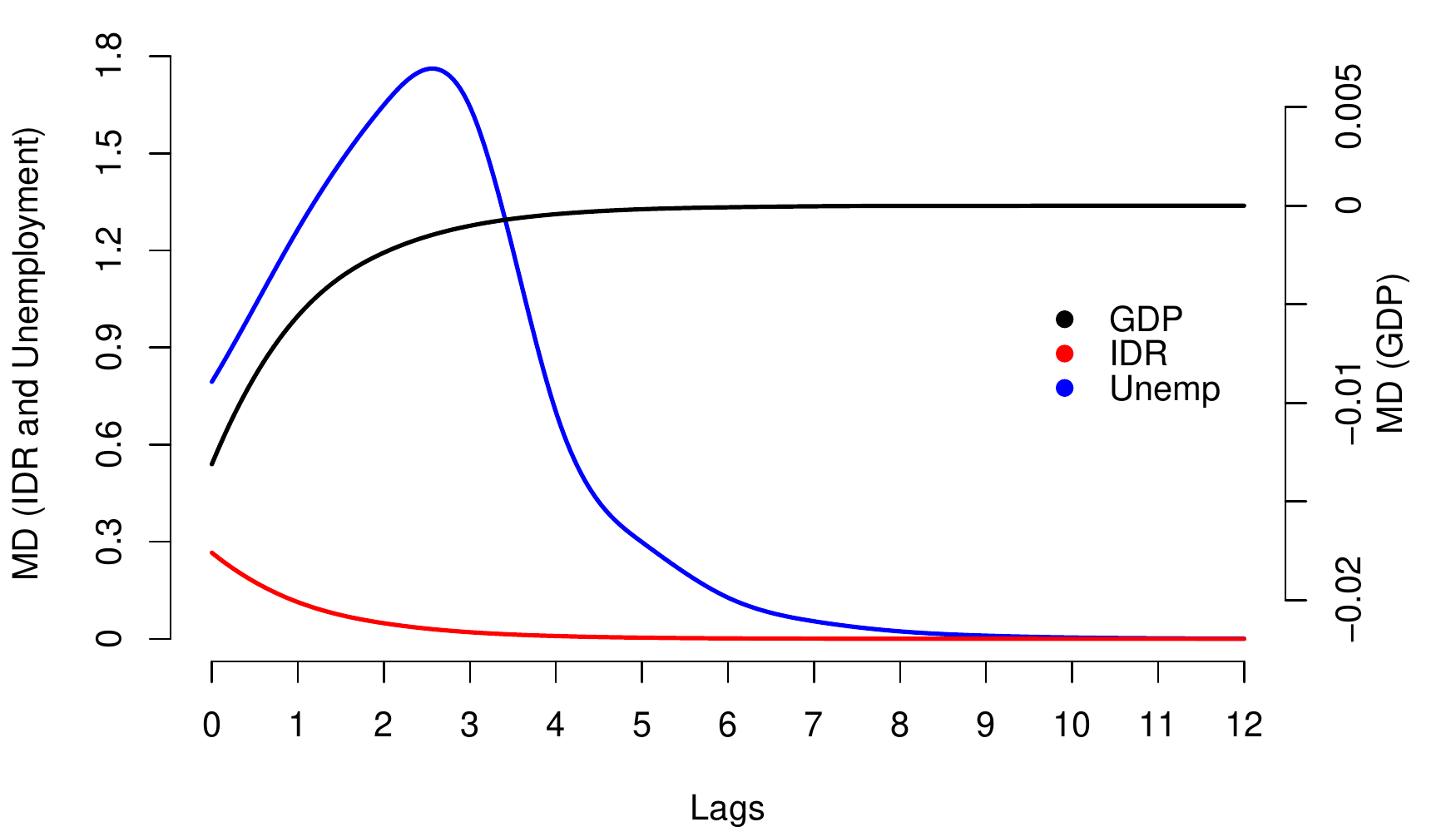}  
			\caption{Dynamic Multiplier of the model variables}
		\end{subfigure}
		\hfill
		\begin{subfigure}[t]{0.495\textwidth}
			\centering
			\includegraphics[width=\textwidth]{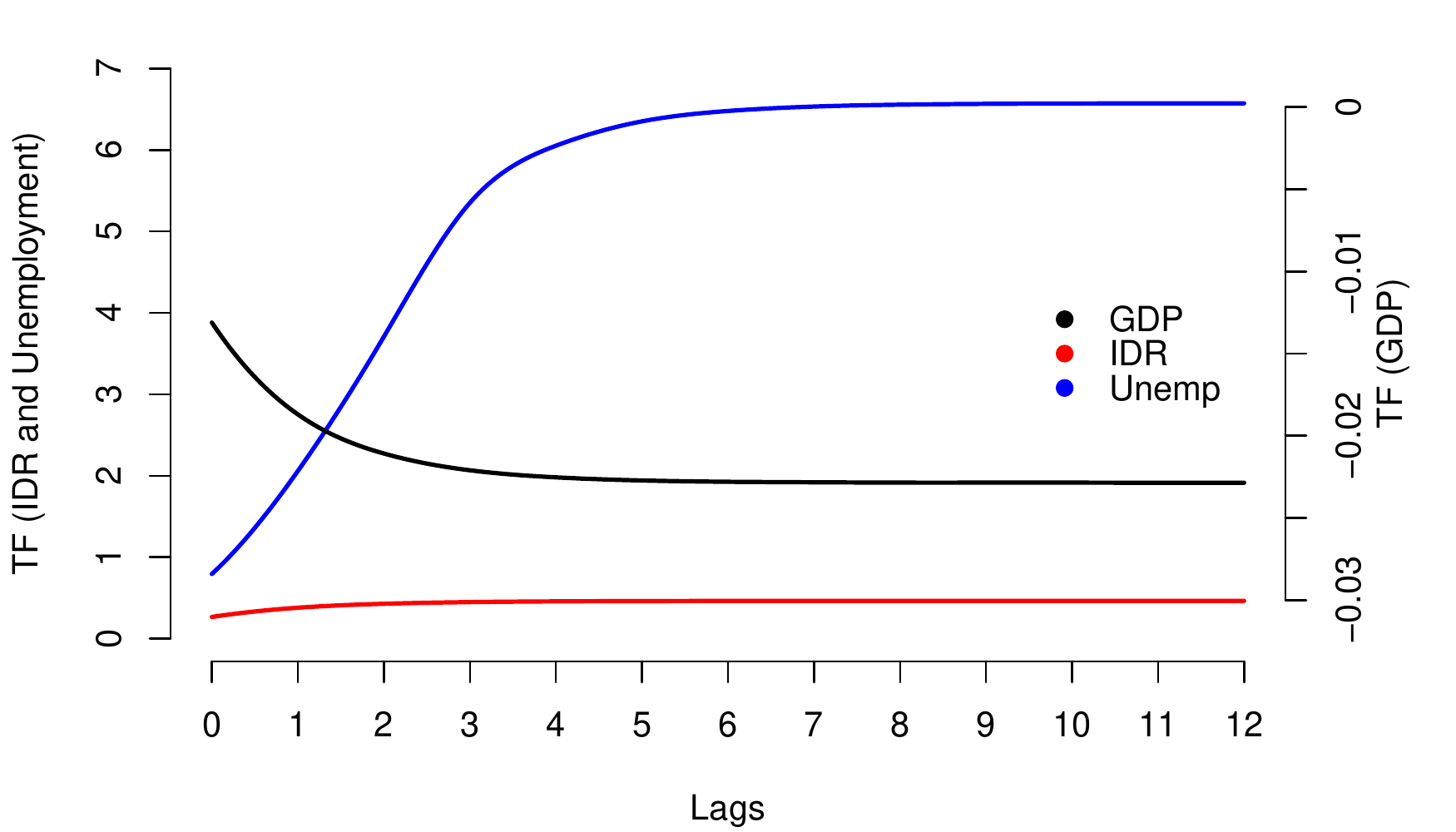}  
			\caption{Transfer function of the model variables}
		\end{subfigure}
		
%		\begin{subfigure}{\textwidth}
%			\centering
%			\includegraphics[width=0.8\linewidth]{images/Projection_Scenarios_Mod1}  
%			\caption{Projection of the stress scenarios}
%		\end{subfigure}
		\caption{Theoretical transmission of model I}
	\end{figure}

	\begin{figure}[h!]
		\centering
		\includegraphics[scale=0.8]{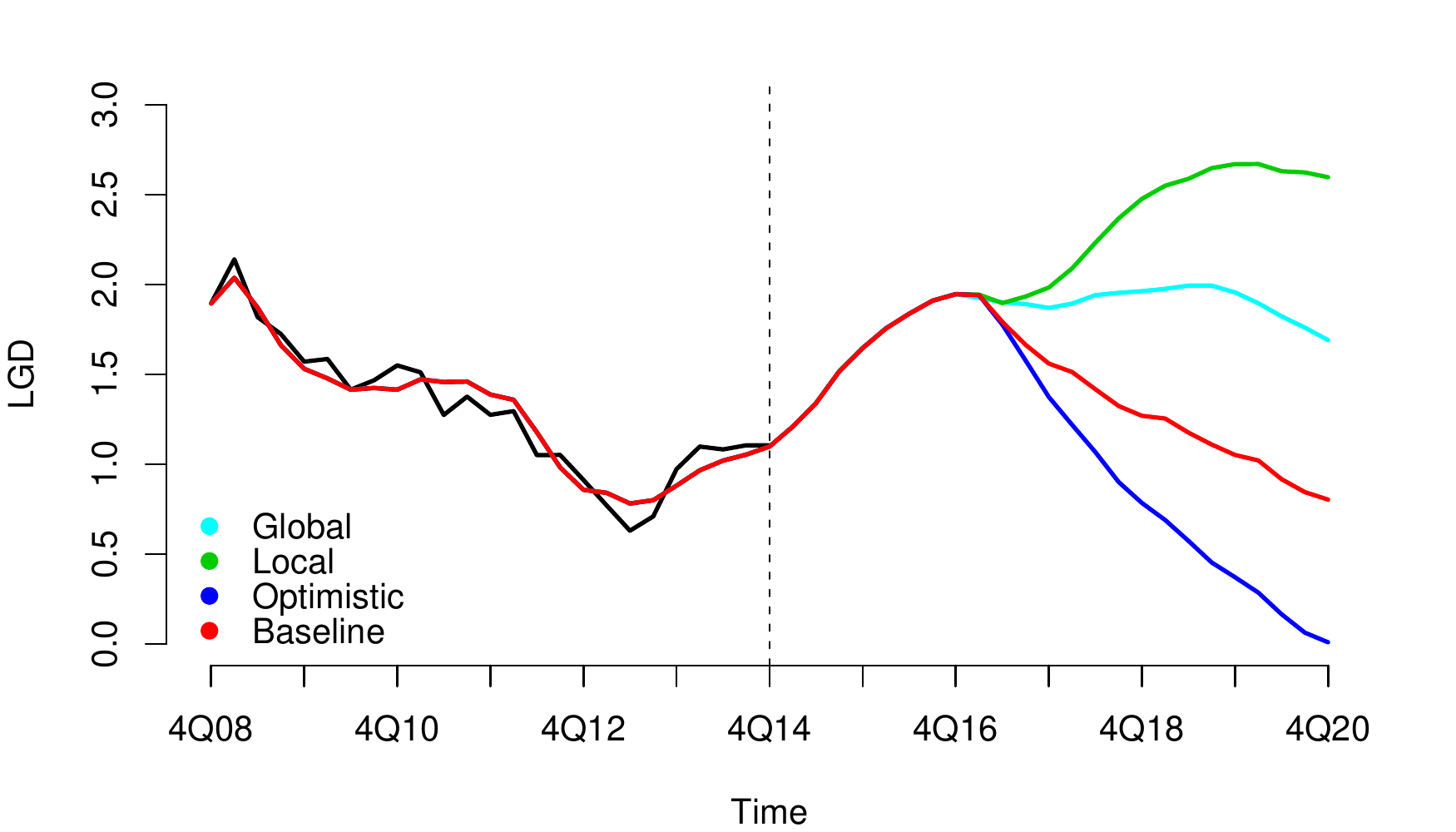} 
		\caption{Projection of the stress scenarios of the model I} 
	\end{figure}

	\subsection{Model II}
	
	The Model II can be represented as,
	
	\begin{align}
	Y_{t} &=E_{t} + v_{t}, \hspace{4.2cm} v_{t}\sim\mathcal{N}(0,\sigma_{Y}^{2}) \nonumber
	\\ 
	E_{t} &=\phi E_{t-1} + X_{t}^{\top}\beta+\partial E_{t}, \hspace{2cm} \partial E_{t}\sim\mathcal{N}(0,\sigma_{E}^{2}),
	\\
	%\nonumber
	%\\	
	\textrm{Initial State:} \hspace{2cm} E_{1} &\sim \mathcal{N}(Y_{1}, \sigma_{E}^{2}), \nonumber
\end{align}
where $X_{t}^{\top}=(1, GDP_{t}, IDR_{t}, Unemp_{t}, Unemp_{t-1}, Unemp_{t-2}, Unemp_{t-3} )$ and $\beta=(\alpha, \beta_{1}, \ldots, \beta_{6})^{\top}$. We defined the following prior distributions,
	\begin{align*}
	\alpha&\sim \mathcal{N}(1.5, 0.5), \hspace{0.5cm} \textrm{with} \hspace{0.5cm} \alpha \in \mathbb{R} \\
	\beta_{j}&\sim\textrm{Half-Normal}(0, 1), \hspace{0.5cm} \textrm{with} \hspace{0.5cm} \beta_{j} \in (-\infty,0] \,\, \textrm{and} \,\, j = 1           \\
	\beta_{j}&\sim\textrm{Half-Normal}(0, 1), \hspace{0.5cm} \textrm{with} \hspace{0.5cm} \beta_{j} \in [0,\infty) \,\, \textrm{and} \,\, j \in \{2,3,4,5,6\}           \\
	\phi&\sim\mathcal{B}eta(2,2), \hspace{0.5cm} \textrm{with} \hspace{0.5cm} \phi \in (0,1) \\
	\sigma_{E}&\sim \textrm{Inv-Gamma}(2, 0.1), \hspace{0.5cm} \textrm{with} \hspace{0.5cm} \sigma_{E} \in [0,\infty) \\
	\eta&\sim \textrm{Gamma}(10, 1), \hspace{0.5cm} \textrm{with} \hspace{0.5cm} \eta \in [0,\infty).
	\end{align*} 
The results of this model are presented below,

	\begin{table}[h!]
		\centering
		\caption{Posterior measures of the parameters of model II}
		\vspace{-0.2cm}
		\begin{tabular}{cccccccc}
			\hline
			Parameters & Mean & SD & 2.5\% & 25\% & 50\% & 75\% & 97.5\% \\ 
			\hline
			$\alpha$ & 1.8739 & 0.3636 & 1.1594 & 1.6281 & 1.8696 & 2.1240 & 2.5704 \\ 
			$\phi$ & 0.4423 & 0.0746 & 0.2888 & 0.3937 & 0.4447 & 0.4938 & 0.5823 \\ 
			$\beta_{1}$ & -0.0126 & 0.0020 & -0.0166 & -0.0140 & -0.0125 & -0.0112 & -0.0086 \\ 
			$\beta_{2}$ & 0.2625 & 0.0487 & 0.1680 & 0.2305 & 0.2619 & 0.2938 & 0.3610 \\ 
			$\beta_{3}$ & 0.7889 & 0.5994 & 0.0276 & 0.3058 & 0.6770 & 1.1423 & 2.1915 \\ 
			$\beta_{4}$ & 0.9183 & 0.6525 & 0.0466 & 0.3966 & 0.7970 & 1.3128 & 2.4412 \\ 
			$\beta_{5}$ & 1.0226 & 0.6931 & 0.0447 & 0.4696 & 0.9275 & 1.4599 & 2.6046 \\ 
			$\beta_{6}$ & 0.9111 & 0.6379 & 0.0458 & 0.4011 & 0.8010 & 1.3052 & 2.3608 \\ 
			$\sigma_{E}$ & 0.0270 & 0.0098 & 0.0143 & 0.0199 & 0.0249 & 0.0317 & 0.0522 \\ 
			$\eta$ & 4.1648 & 1.4817 & 1.6372 & 3.0765 & 4.0338 & 5.1467 & 7.3270 \\ 
			\hline
		\end{tabular}
	\end{table}
	
	\begin{figure}[!h]
		\centering
		\begin{subfigure}[t]{0.495\textwidth}
			\centering
			\includegraphics[width=\textwidth]{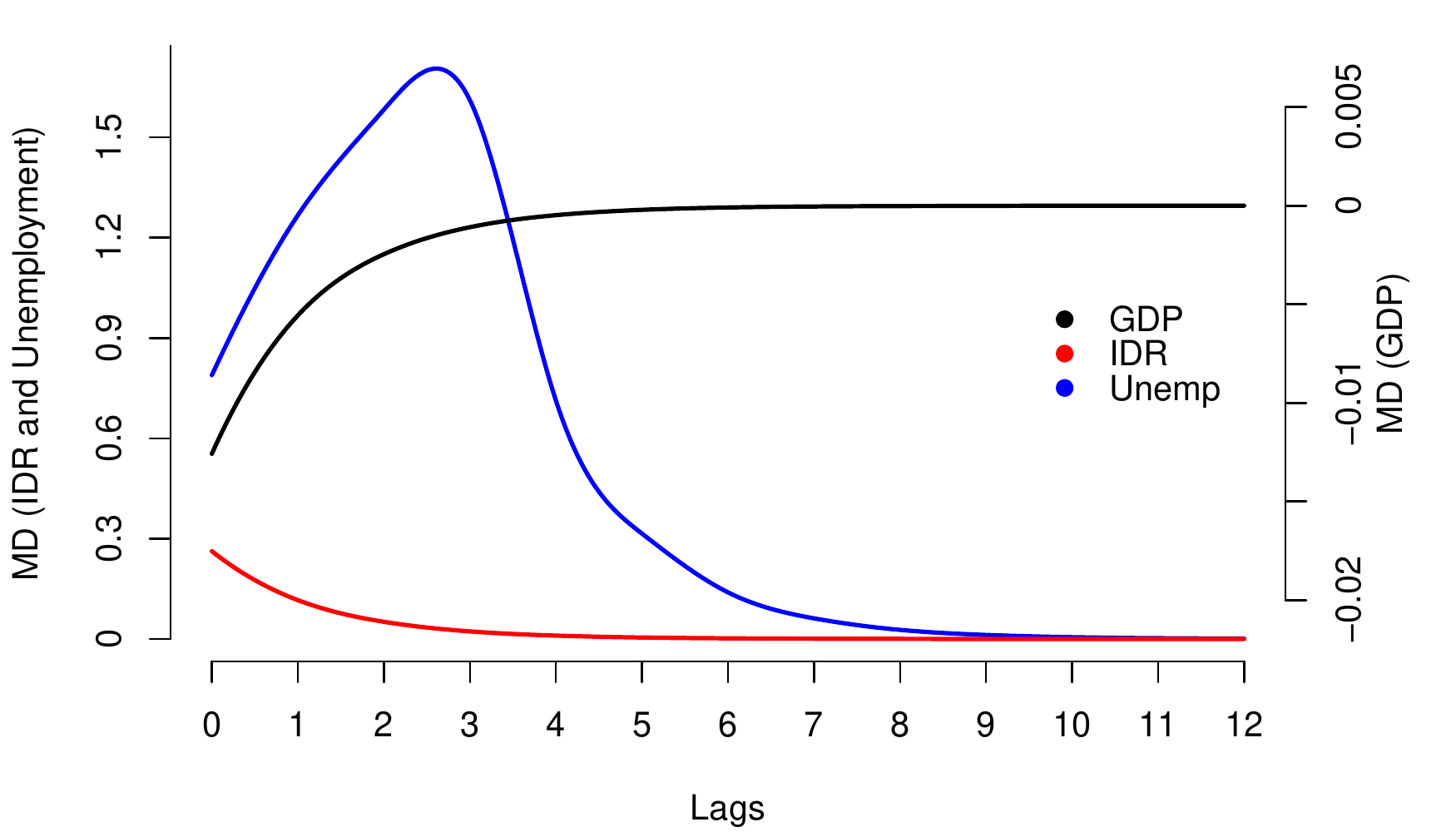}  
			\caption{Dynamic Multiplier of the model variables}
		\end{subfigure}
		\hfill
		\begin{subfigure}[t]{0.495\textwidth}
			\centering
			\includegraphics[width=\textwidth]{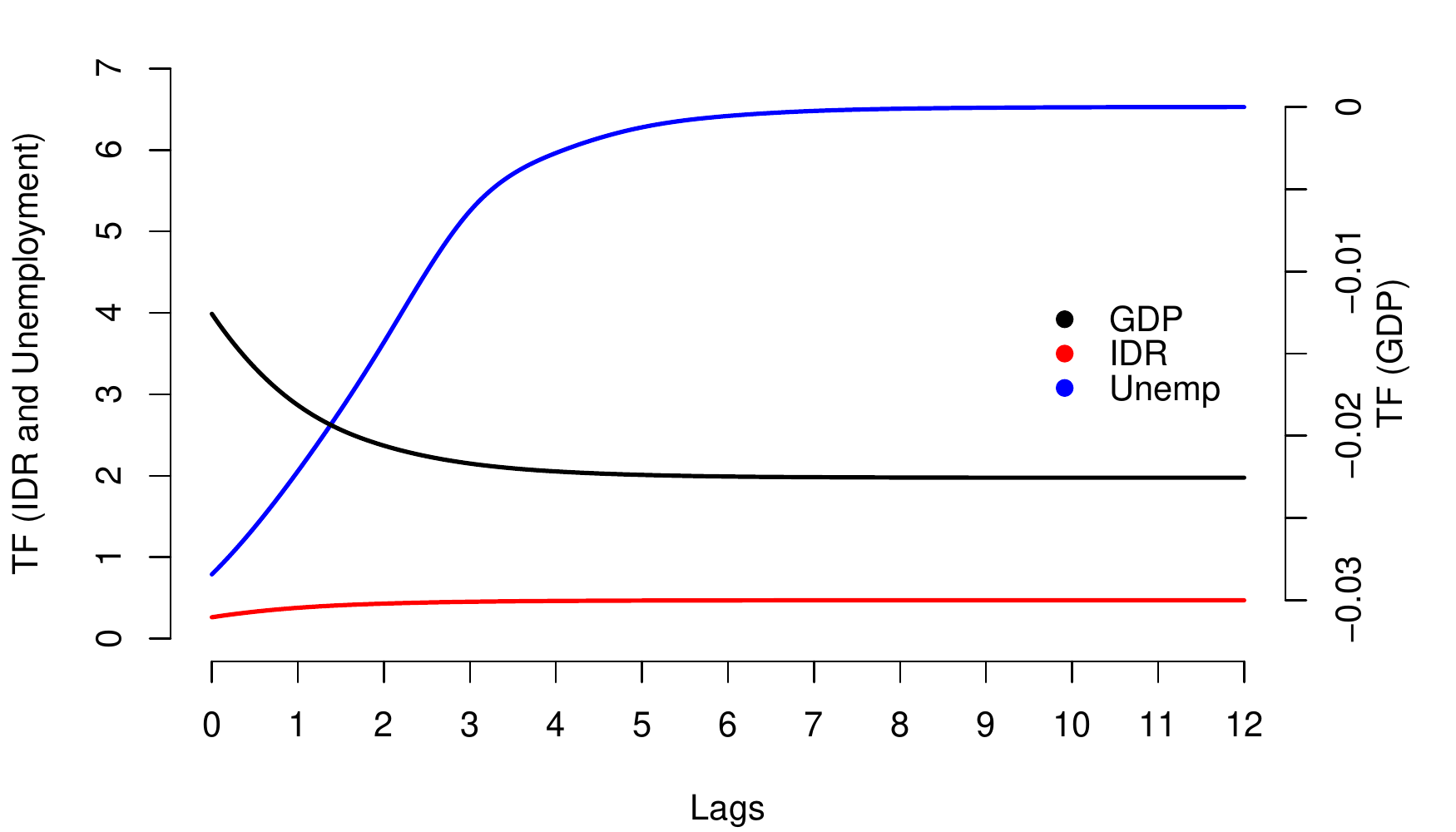}  
			\caption{Transfer function of the model variables}
		\end{subfigure}
		
%		\begin{subfigure}{\textwidth}
%			\centering
%			\includegraphics[width=0.8\linewidth]{images/Projection_Scenarios_Mod2}  
%			\caption{Projection of the stress scenarios}
%		\end{subfigure}
		\caption{Theoretical transmission of model II}
	\end{figure}

	\begin{figure}
		\centering
		\includegraphics[scale=0.8]{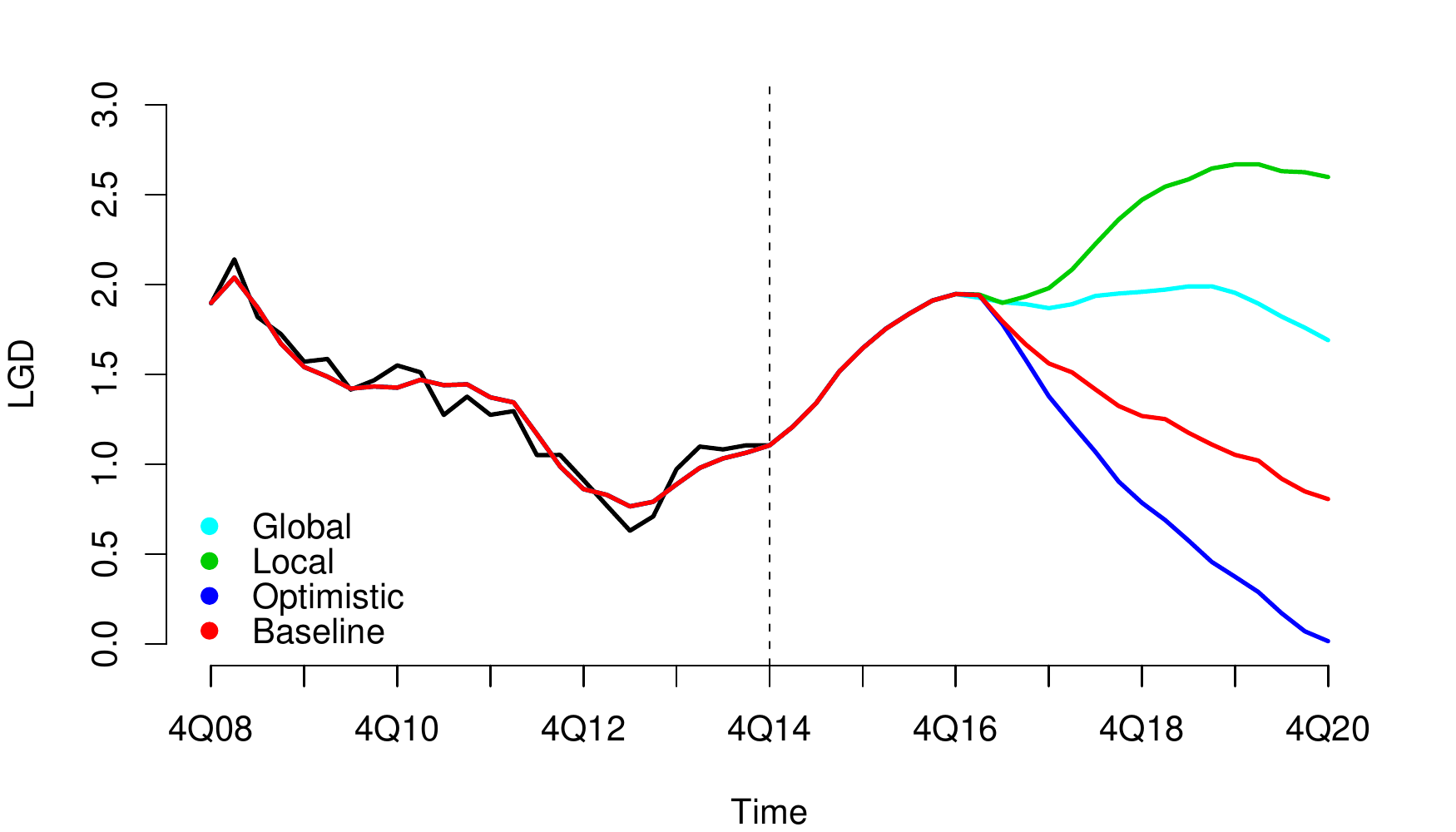} 
		\caption{Projection of the stress scenarios of the model II} 
	\end{figure}

	\clearpage
	\subsection{Model III}
	The Model III can be represented as,
	
	\begin{align}
	Y_{t} &=E_{t} + v_{t}, \hspace{4.35cm} v_{t}\sim\mathcal{N}(0,\sigma_{Y}^{2}) \nonumber
	\\ 
	E_{t} &=\phi_{t} E_{t-1} + X_{t}^{\top}\beta+\partial E_{t}, \hspace{2cm} \partial E_{t}\sim\mathcal{N}(0,\sigma_{E}^{2}) \nonumber 
	\\
	\phi_{t} &= \phi_{t-1} + \partial\phi_{t}, \hspace{3.55cm} \partial\phi_{t}\sim\mathcal{N}(0,\sigma_{\phi}^{2})  , 
	\\
	%\nonumber
	%\\
	\textrm{Initial State:} \hspace{2cm} E_{1} &\sim \mathcal{N}(Y_{1}, \sigma_{E}^{2}), \nonumber
	\\
	\phi_{0} &\sim \mathcal{B}eta(2, 2), \nonumber
	\end{align}
where, $\eta=0$, $X_{t}^{\top}=(1, GDP_{t}, IDR_{t}, Unemp_{t}, Unemp_{t-1}, Unemp_{t-2}, Unemp_{t-3} )$ and $\beta=(\alpha, \beta_{1}, \ldots, \beta_{6})^{\top}$. Note that the resilience parameter varies over time according to random walk process; generating stochastic transmissions. We defined the following prior distributions,
	\begin{align*}
	\alpha&\sim \mathcal{N}(1.5, 0.5), \hspace{0.5cm} \textrm{with} \hspace{0.5cm} \alpha \in \mathbb{R} \\
	\beta_{j}&\sim\textrm{Half-Normal}(0, 1), \hspace{0.5cm} \textrm{with} \hspace{0.5cm} \beta_{j} \in (-\infty,0] \,\, \textrm{and} \,\, j = 1           \\
	\beta_{j}&\sim\textrm{Half-Normal}(0, 1), \hspace{0.5cm} \textrm{with} \hspace{0.5cm} \beta_{j} \in [0,\infty) \,\, \textrm{and} \,\, j \in \{2,3,4,5,6\}           \\
	\sigma_{\phi}&\sim \textrm{Inv-Gamma}(2, 0.1), \hspace{0.5cm} \textrm{with} \hspace{0.5cm} \sigma_{\phi} \in [0,\infty) \\
	\sigma_{E}&\sim \textrm{Inv-Gamma}(2, 0.1), \hspace{0.5cm} \textrm{with} \hspace{0.5cm} \sigma_{E} \in [0,\infty).
	\end{align*} 
The results of this model are presented below,
	
	\begin{table}[ht]
		\centering
		\caption{Posterior measures of the parameters of model III}
		\vspace{-0.2cm}
		\begin{tabular}{cccccccc}
			\hline
			Parameters & Mean & SD & 2.5\% & 25\% & 50\% & 75\% & 97.5\% \\ 
			\hline
			$\alpha$ & 1.6331 & 0.3717 & 0.9173 & 1.3819 & 1.6316 & 1.8777 & 2.3807 \\ 
			$\phi_{1}$ & 0.3996 & 0.1050 & 0.1955 & 0.3282 & 0.3982 & 0.4683 & 0.6105 \\ 
			$\beta_{1}$ & -0.0114 & 0.0023 & -0.0159 & -0.0129 & -0.0114 & -0.0099 & -0.0070 \\ 
			$\beta_{2}$ & 0.3489 & 0.0829 & 0.1985 & 0.2921 & 0.3443 & 0.4020 & 0.5192 \\ 
			$\beta_{3}$ & 0.6499 & 0.5143 & 0.0211 & 0.2448 & 0.5399 & 0.9315 & 1.8955 \\ 
			$\beta_{4}$ & 0.8037 & 0.6091 & 0.0286 & 0.3215 & 0.6829 & 1.1553 & 2.2805 \\ 
			$\beta_{5}$ & 1.0154 & 0.6870 & 0.0422 & 0.4674 & 0.9124 & 1.4499 & 2.5593 \\ 
			$\beta_{6}$ & 0.8513 & 0.6114 & 0.0472 & 0.3721 & 0.7312 & 1.2186 & 2.3146 \\ 
			$\sigma_{E}$ & 0.0621 & 0.0194 & 0.0289 & 0.0475 & 0.0615 & 0.0750 & 0.1028 \\ 
			$\sigma_{\phi}$ & 0.0522 & 0.0216 & 0.0188 & 0.0350 & 0.0499 & 0.0668 & 0.0986 \\ 
			\hline
		\end{tabular}
	\end{table}

	\begin{figure}[h!]
		\centering
		\begin{subfigure}[t]{0.495\textwidth}
			\centering
			\includegraphics[width=\textwidth]{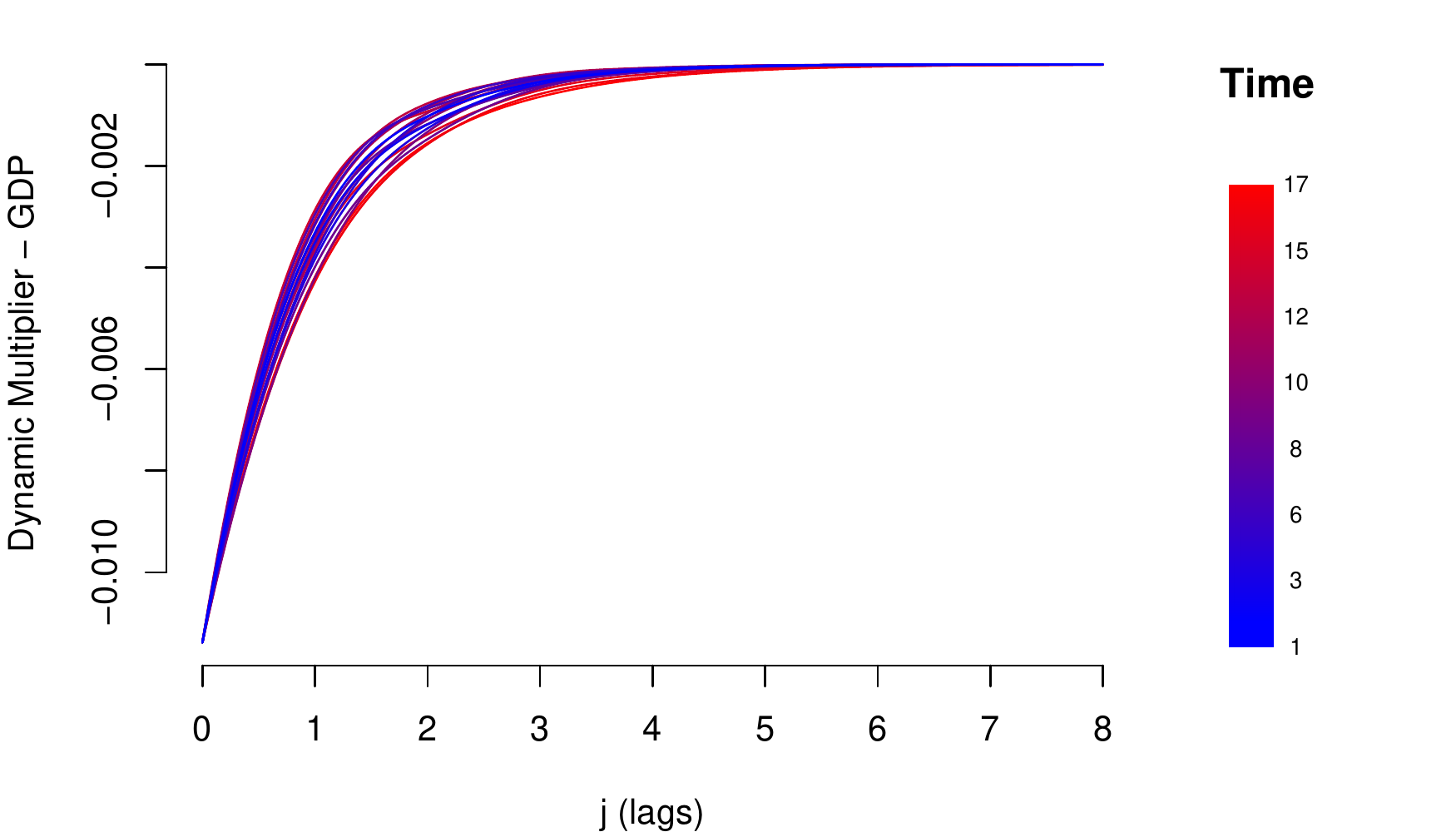}  
%			\caption{Dynamic Multiplier of the GDP}
		\end{subfigure}
		\begin{subfigure}[t]{0.495\textwidth}
			\centering
			\includegraphics[width=\textwidth]{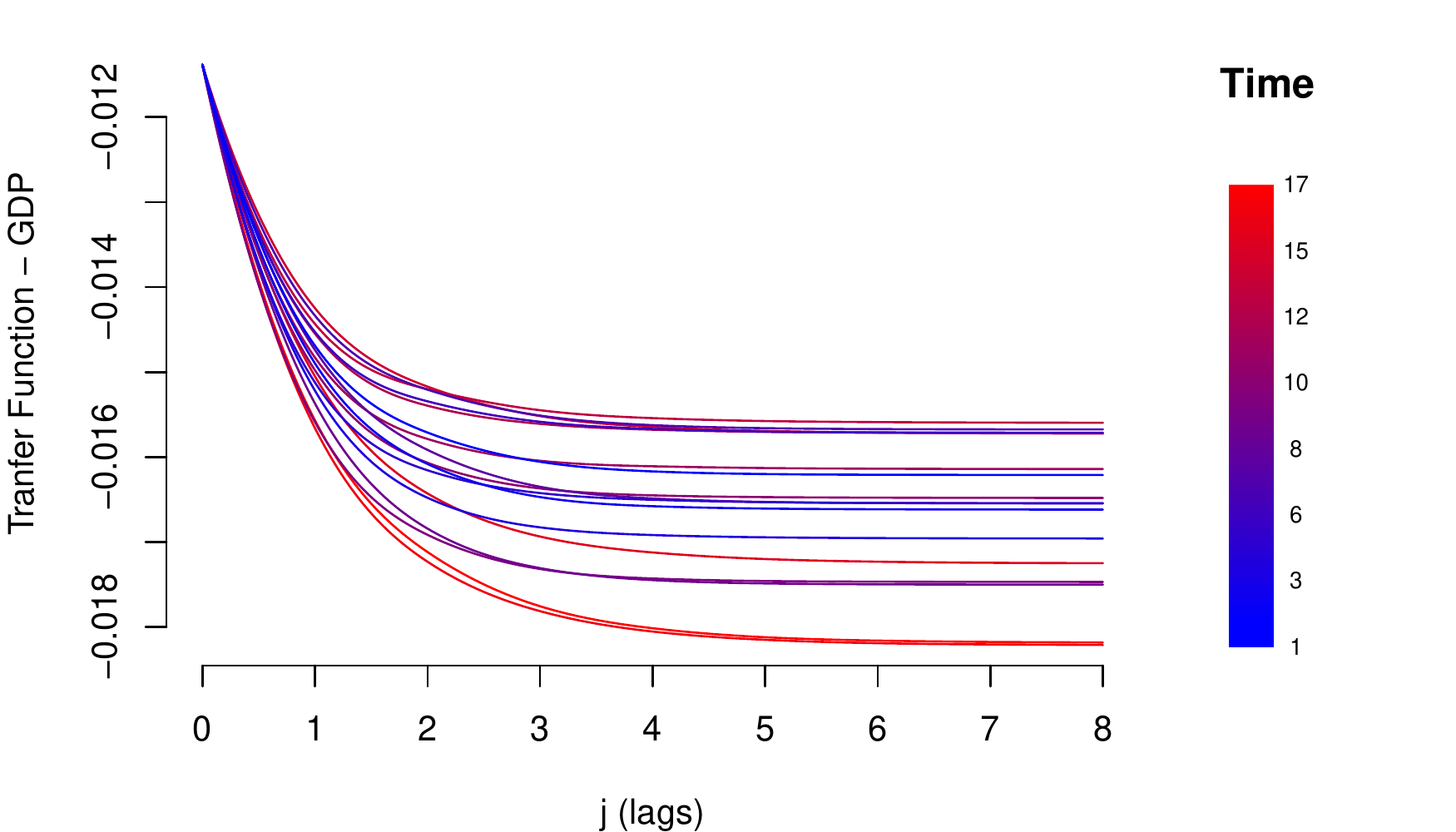}  
%			\caption{Transfer Function of the GDP}
		\end{subfigure}		
		
		\begin{subfigure}[t]{0.495\textwidth}
			\centering
			\includegraphics[width=\textwidth]{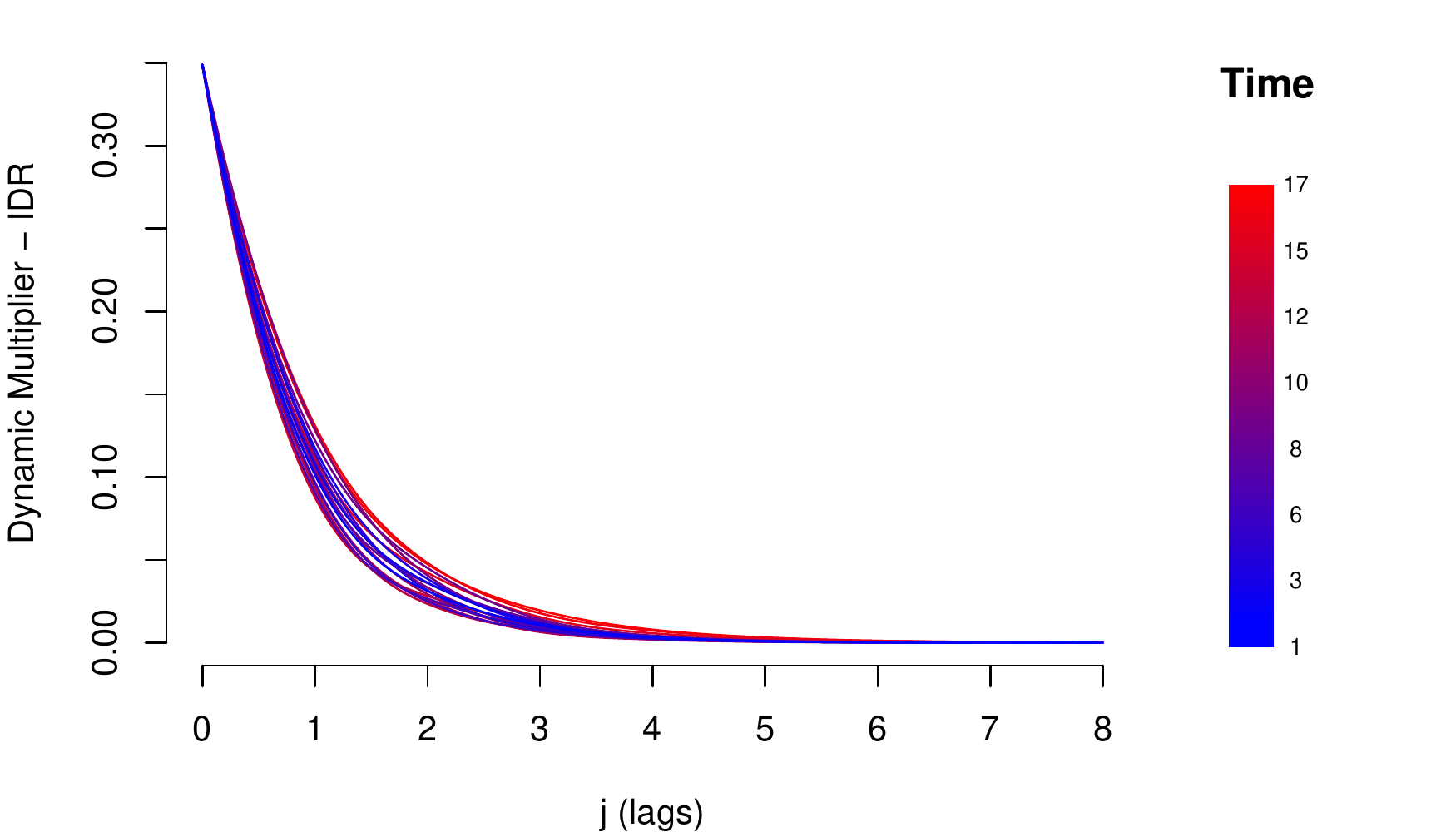}  
%			\caption{Dynamic Multiplier of the IDR}
		\end{subfigure}
		\begin{subfigure}[t]{0.495\textwidth}
			\centering
			\includegraphics[width=\textwidth]{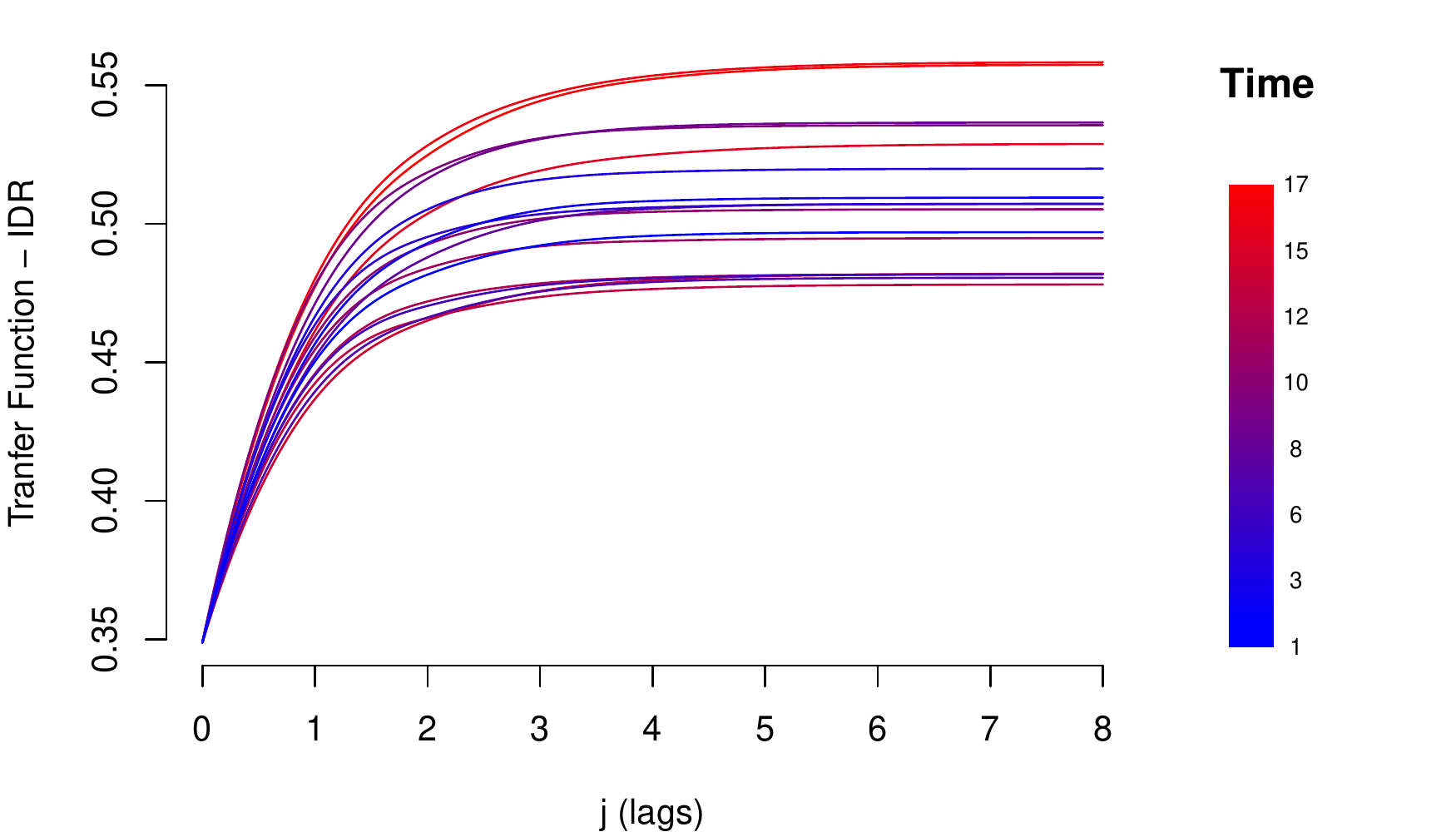}  
%			\caption{Transfer Function of the IDR}
		\end{subfigure}	
		
		\begin{subfigure}[t]{0.495\textwidth}
			\centering
			\includegraphics[width=\textwidth]{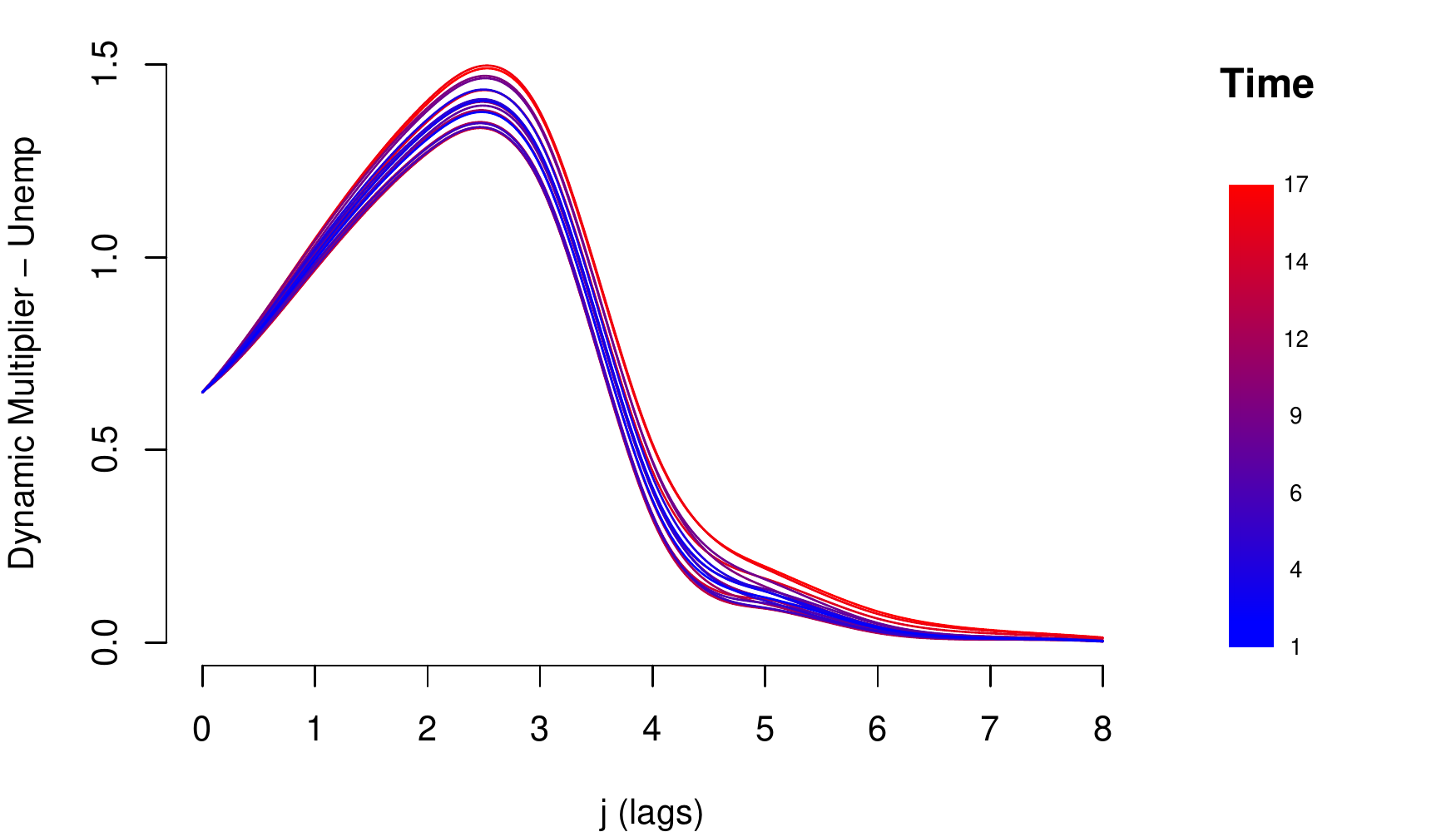}  
%			\caption{Dynamic Multiplier of the Unemp}
		\end{subfigure}
		\begin{subfigure}[t]{0.495\textwidth}
			\centering
			\includegraphics[width=\textwidth]{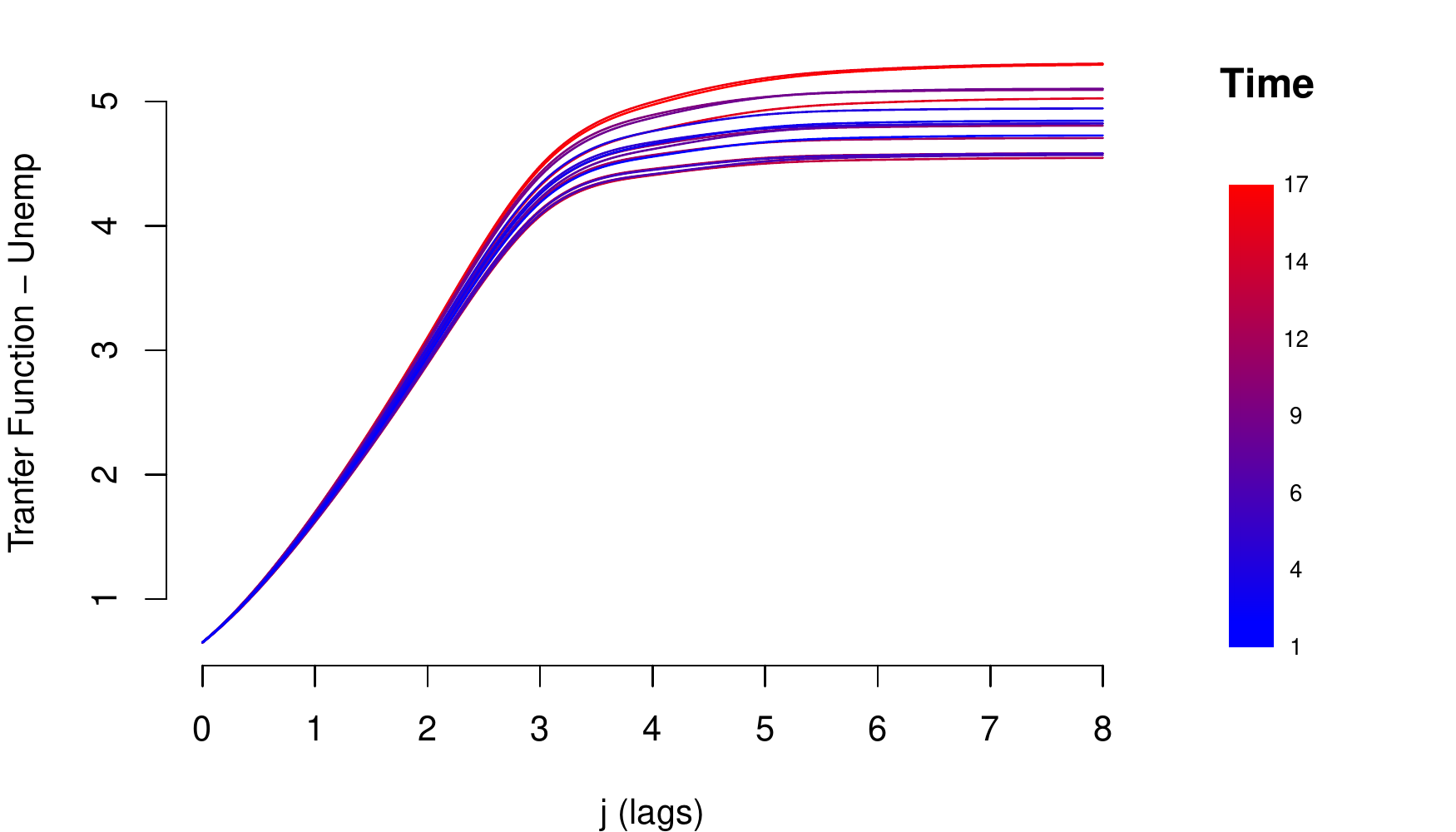}  
%			\caption{Transfer Function of the Unemp}
		\end{subfigure}	
		\caption{Stochastic Dynamic Multiplier and Stochastic Transfer Function of the model III}
	\end{figure}

	\begin{figure}[h!]
		\centering
		\includegraphics[scale=0.595]{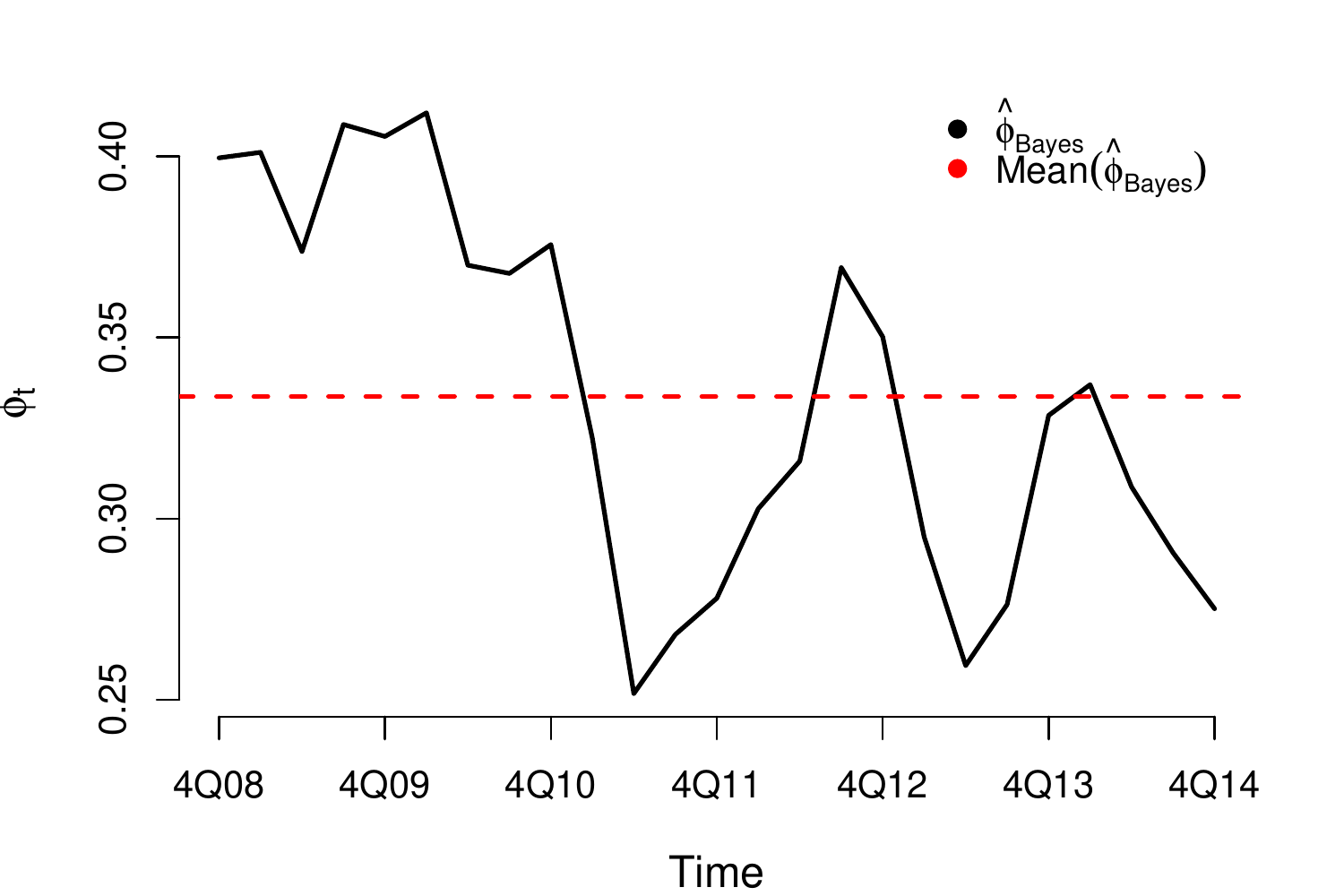} 
		\caption{Evolution of the resilience parameter of the model III} \label{erp3}
	\end{figure}	
	
\clearpage
	
	\begin{figure}[h!]
		\centering
		\includegraphics[scale=0.8]{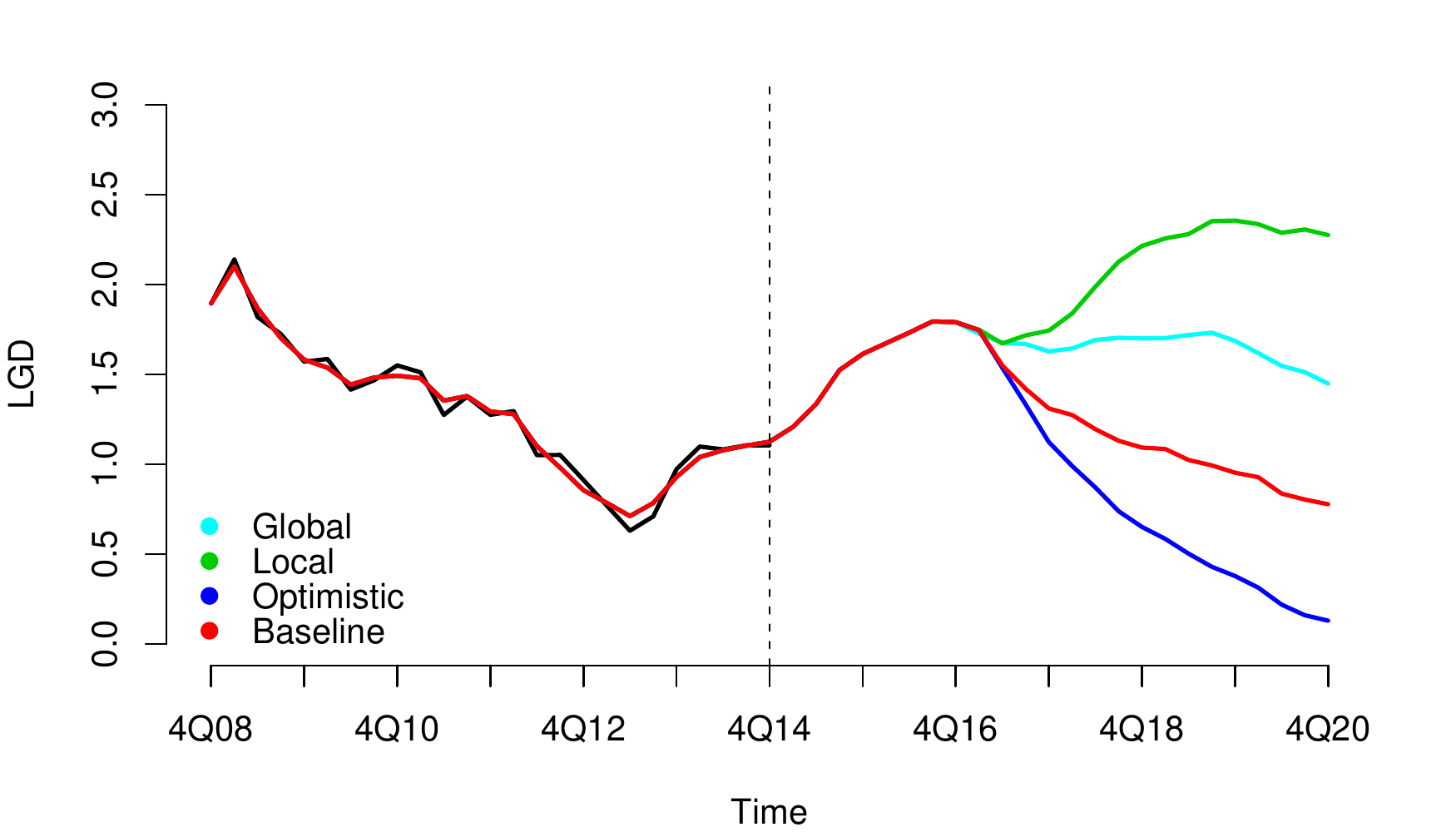} 
		\caption{Projection of the stress scenarios of the model III} 
	\end{figure}
	
%	\begin{figure}[!h]
%		\centering
%		\begin{subfigure}[t]{0.65\textwidth}
%			\centering
%			\includegraphics[width=\textwidth]{images/Evolution_phi_Mod3}  
%			\caption{Evolution of the resilience parameter}
%		\end{subfigure}
%		
%		\begin{subfigure}{\textwidth}
%			\centering
%			\includegraphics[width=0.8\linewidth]{images/Projection_Scenarios_Mod3}  
%			\caption{Projection of the stress scenarios}
%		\end{subfigure}
%		\caption{Results from model 3}
%	\end{figure}

	\subsection{Model IV}
	The Model IV can be represented as,
	
	\begin{align}
	Y_{t} &=E_{t} + v_{t}, \hspace{4.35cm} v_{t}\sim\mathcal{N}(0,\sigma_{Y}^{2}) \nonumber
	\\ 
	E_{t} &=\phi_{t} E_{t-1} + X_{t}^{\top}\beta+\partial E_{t}, \hspace{2cm} \partial E_{t}\sim\mathcal{N}(0,\sigma_{E}^{2}) \nonumber 
	\\
	\phi_{t} &= \phi_{t-1} + \partial\phi_{t}, \hspace{3.55cm} \partial\phi_{t}\sim\mathcal{N}(0,\sigma_{\phi}^{2}),
	\\
	%\nonumber
	%\\
	\textrm{Initial State:} \hspace{2cm} E_{1} &\sim \mathcal{N}(Y_{1}, \sigma_{E}^{2}), \nonumber
	\\
	\phi_{0} &\sim \mathcal{B}eta(2, 2), \nonumber
	\end{align}
where $X_{t}^{\top}=(1, GDP_{t}, IDR_{t}, Unemp_{t}, Unemp_{t-1}, Unemp_{t-2}, Unemp_{t-3} )$ and $\beta=(\alpha, \beta_{1}, \ldots, \beta_{6})^{\top}$. Note that the resilience parameter varies over time according to random walk process; generating stochastic transmissions. We defined the following prior distributions,
	\begin{align*}
	\alpha&\sim \mathcal{N}(1.5, 0.5), \hspace{0.5cm} \textrm{with} \hspace{0.5cm} \alpha \in \mathbb{R} \\
	\beta_{j}&\sim\textrm{Half-Normal}(0, 1), \hspace{0.5cm} \textrm{with} \hspace{0.5cm} \beta_{j} \in (-\infty,0] \,\, \textrm{and} \,\, j = 1           \\
	\beta_{j}&\sim\textrm{Half-Normal}(0, 1), \hspace{0.5cm} \textrm{with} \hspace{0.5cm} \beta_{j} \in [0,\infty) \,\, \textrm{and} \,\, j \in \{2,3,4,5,6\}           \\
	\sigma_{\phi}&\sim \textrm{Inv-Gamma}(2, 0.1), \hspace{0.5cm} \textrm{with} \hspace{0.5cm} \sigma_{\phi} \in [0,\infty) \\
	\sigma_{E}&\sim \textrm{Inv-Gamma}(2, 0.1), \hspace{0.5cm} \textrm{with} \hspace{0.5cm} \sigma_{E} \in [0,\infty) \\
	\eta&\sim \textrm{Gamma}(10, 1), \hspace{0.5cm} \textrm{with} \hspace{0.5cm} \eta \in [0,\infty).
	\end{align*} 
The results of this model are presented below,
	
	\begin{table}[ht]
		\centering
		\caption{Posterior measures of the parameters of model IV}
		\vspace{-0.2cm}
		\begin{tabular}{cccccccc}
			\hline
			Parameters & Mean & SD & 2.5\% & 25\% & 50\% & 75\% & 97.5\% \\ 
			\hline
			$\alpha$ & 1.6645 & 0.3834 & 0.9186 & 1.4080 & 1.6579 & 1.9279 & 2.4081 \\ 
			$\phi_{1}$ & 0.3944 & 0.1140 & 0.1489 & 0.3227 & 0.4009 & 0.4730 & 0.6020 \\ 
			$\beta_{1}$ & -0.0111 & 0.0024 & -0.0157 & -0.0127 & -0.0111 & -0.0095 & -0.0063 \\ 
			$\beta_{2}$ & 0.3201 & 0.0797 & 0.1747 & 0.2677 & 0.3156 & 0.3675 & 0.4935 \\ 
			$\beta_{3}$ & 0.7082 & 0.5655 & 0.0257 & 0.2682 & 0.5760 & 1.0267 & 2.0621 \\ 
			$\beta_{4}$ & 0.8128 & 0.6098 & 0.0338 & 0.3249 & 0.6885 & 1.1847 & 2.2774 \\ 
			$\beta_{5}$ & 0.9609 & 0.6830 & 0.0401 & 0.4221 & 0.8319 & 1.3791 & 2.5442 \\ 
			$\beta_{6}$ & 0.8443 & 0.6226 & 0.0326 & 0.3511 & 0.7248 & 1.2087 & 2.3297 \\ 
			$\sigma_{E}$ & 0.0170 & 0.0053 & 0.0096 & 0.0132 & 0.0160 & 0.0196 & 0.0306 \\ 
			$\sigma_{\phi}$ & 0.0387 & 0.0165 & 0.0153 & 0.0266 & 0.0357 & 0.0478 & 0.0789 \\ 
			$\eta$ & 5.6366 & 1.7542 & 2.7083 & 4.3663 & 5.4529 & 6.7124 & 9.6173 \\ 
			\hline
		\end{tabular}
	\end{table}
	
	\begin{figure}[!h]
		\centering
		\begin{subfigure}[t]{0.495\textwidth}
			\centering
			\includegraphics[width=\textwidth]{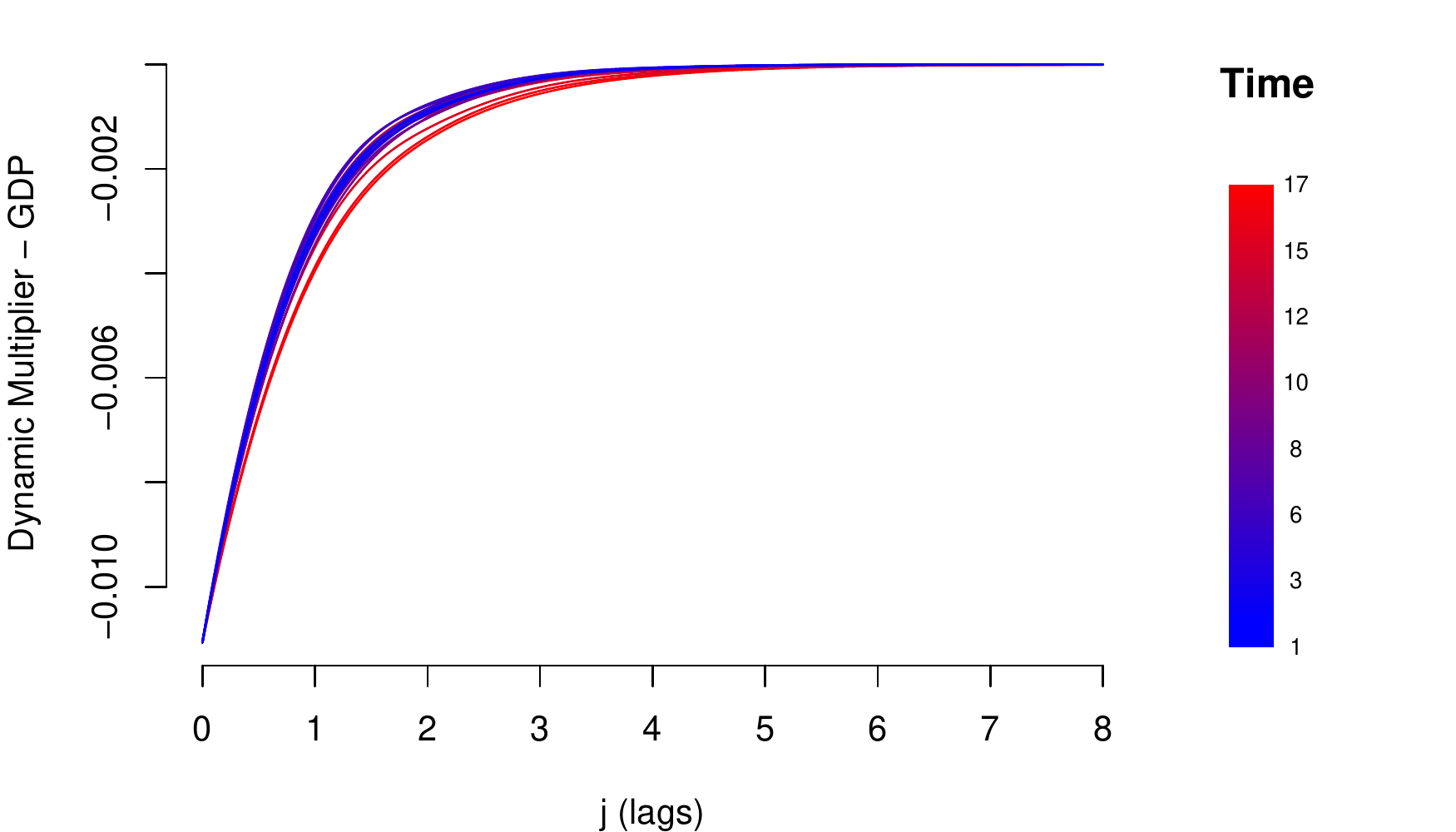}  
%			\caption{Dynamic Multiplier of the GDP}
		\end{subfigure}
		\begin{subfigure}[t]{0.495\textwidth}
			\centering
			\includegraphics[width=\textwidth]{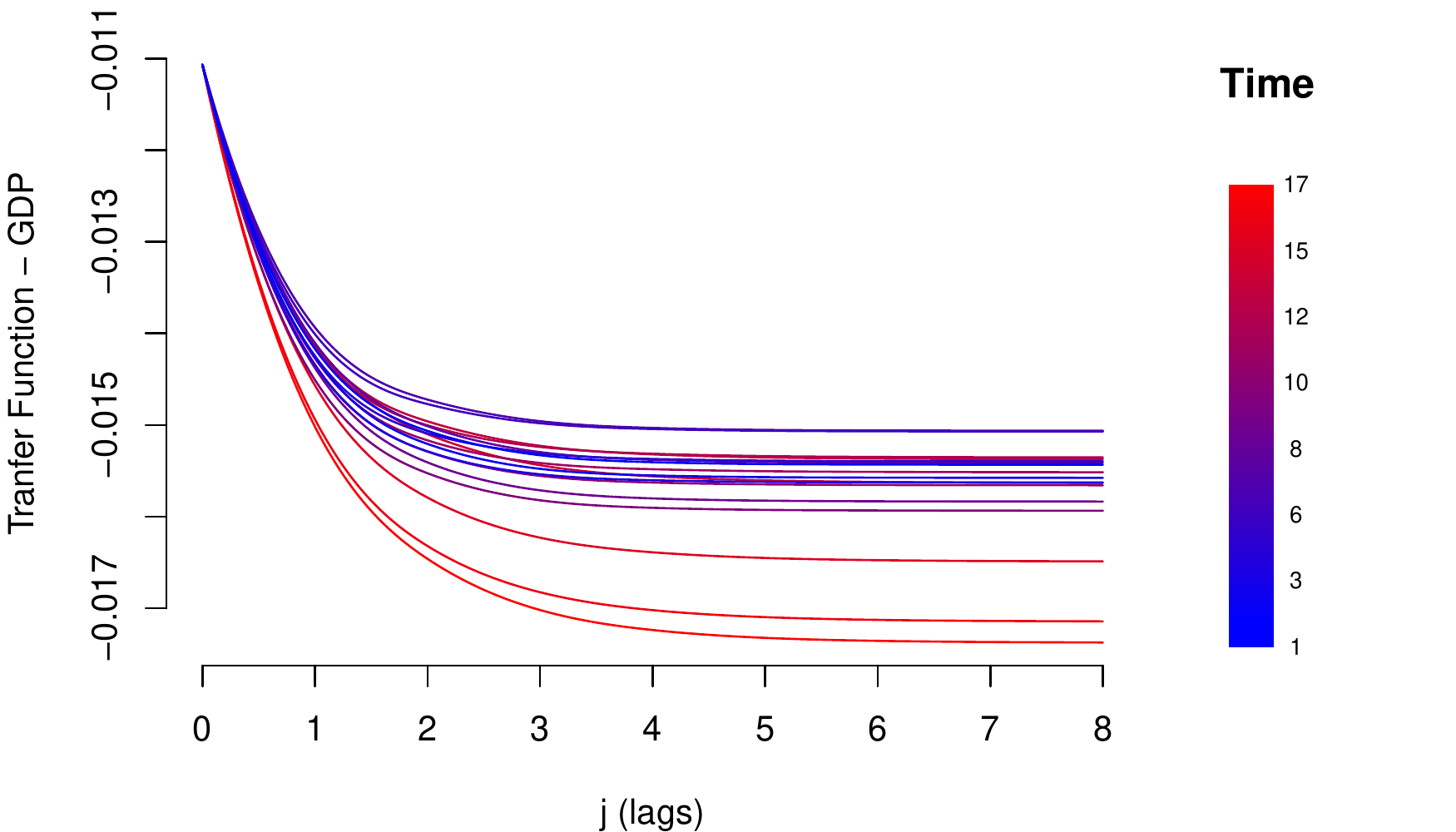}  
%			\caption{Transfer Function of the GDP}
		\end{subfigure}		
		
		\begin{subfigure}[t]{0.495\textwidth}
			\centering
			\includegraphics[width=\textwidth]{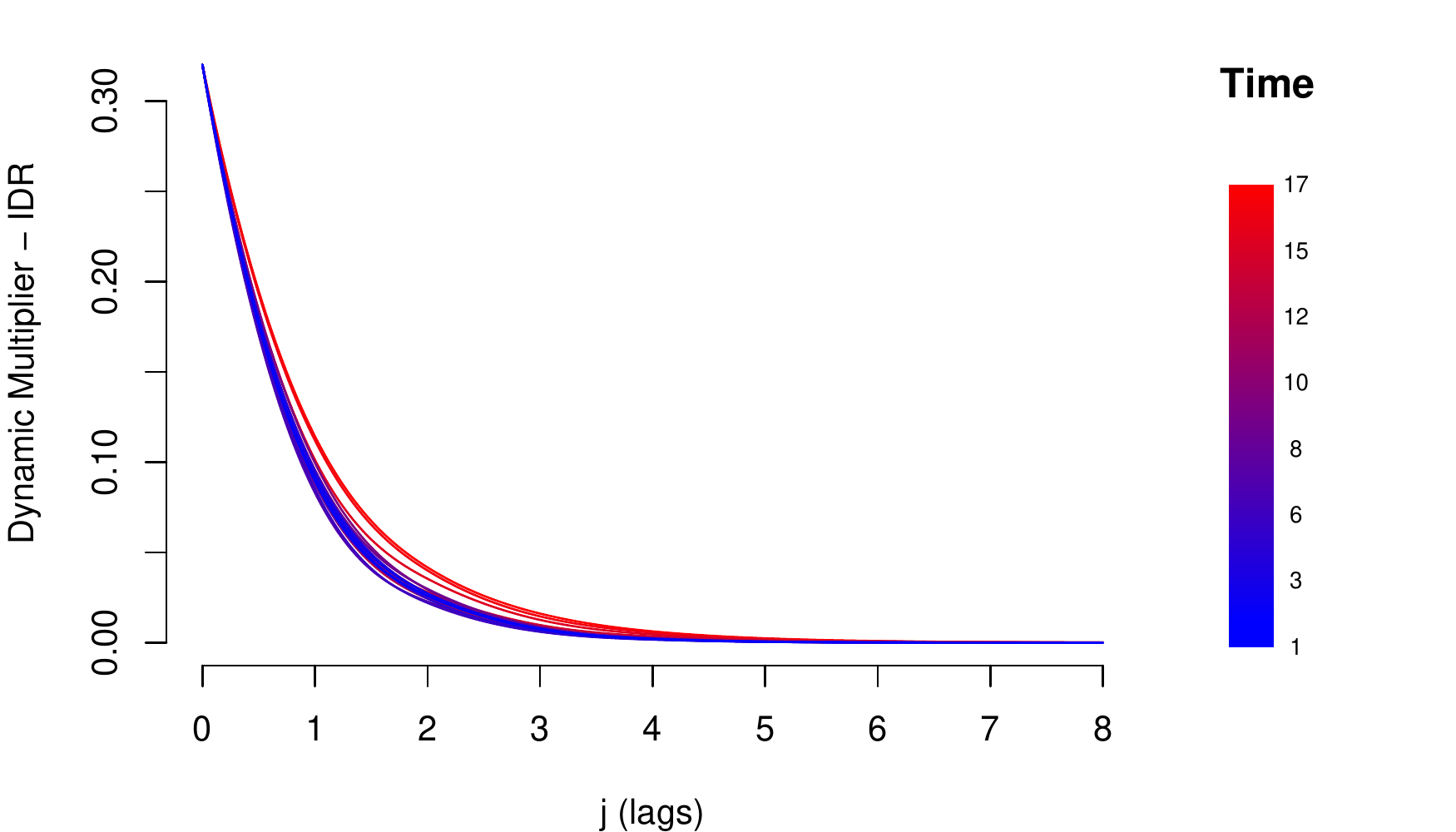}  
%			\caption{Dynamic Multiplier of the IDR}
		\end{subfigure}
		\begin{subfigure}[t]{0.495\textwidth}
			\centering
			\includegraphics[width=\textwidth]{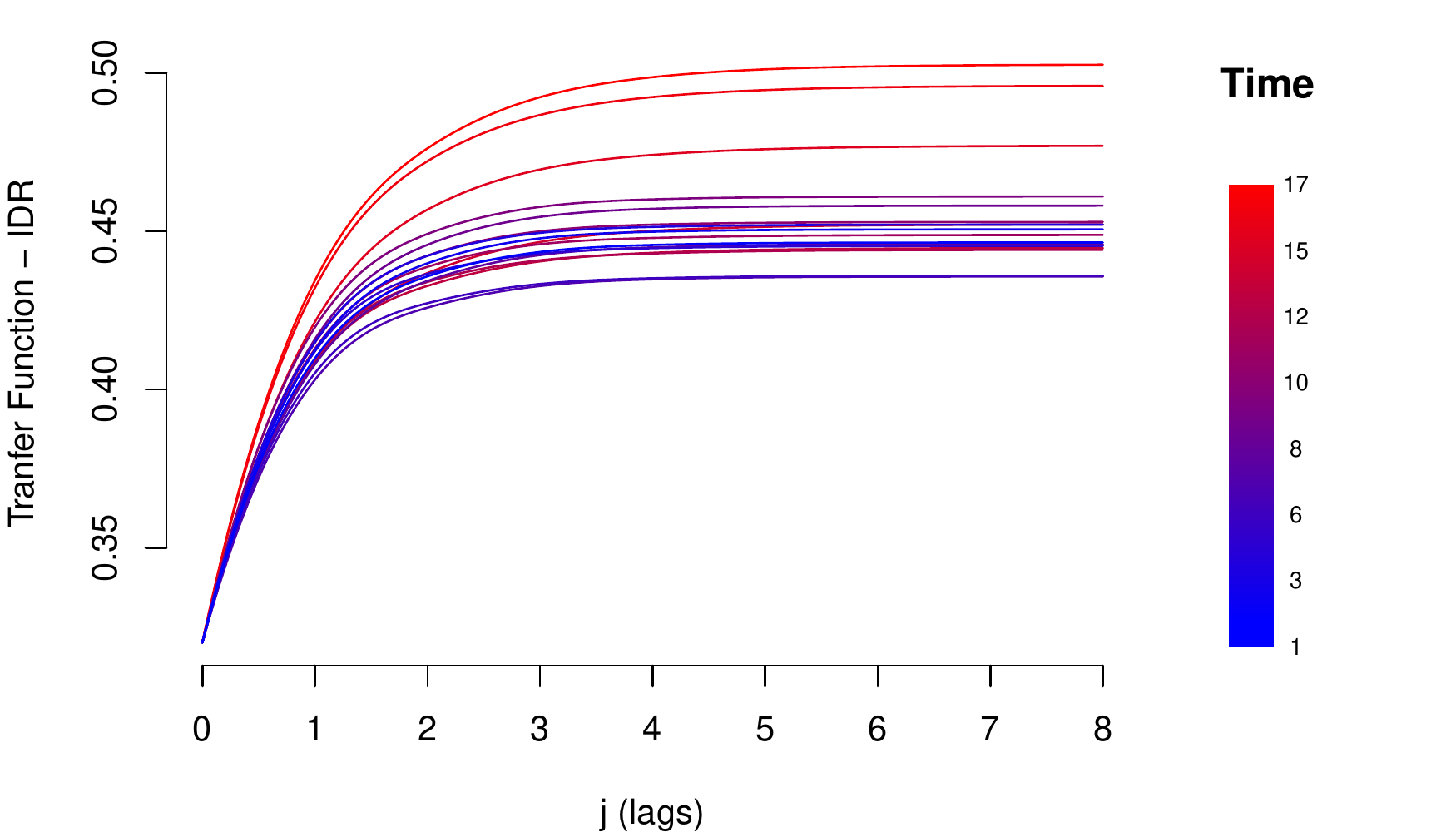}  
%			\caption{Transfer Function of the IDR}
		\end{subfigure}	
		
		\begin{subfigure}[t]{0.495\textwidth}
			\centering
			\includegraphics[width=\textwidth]{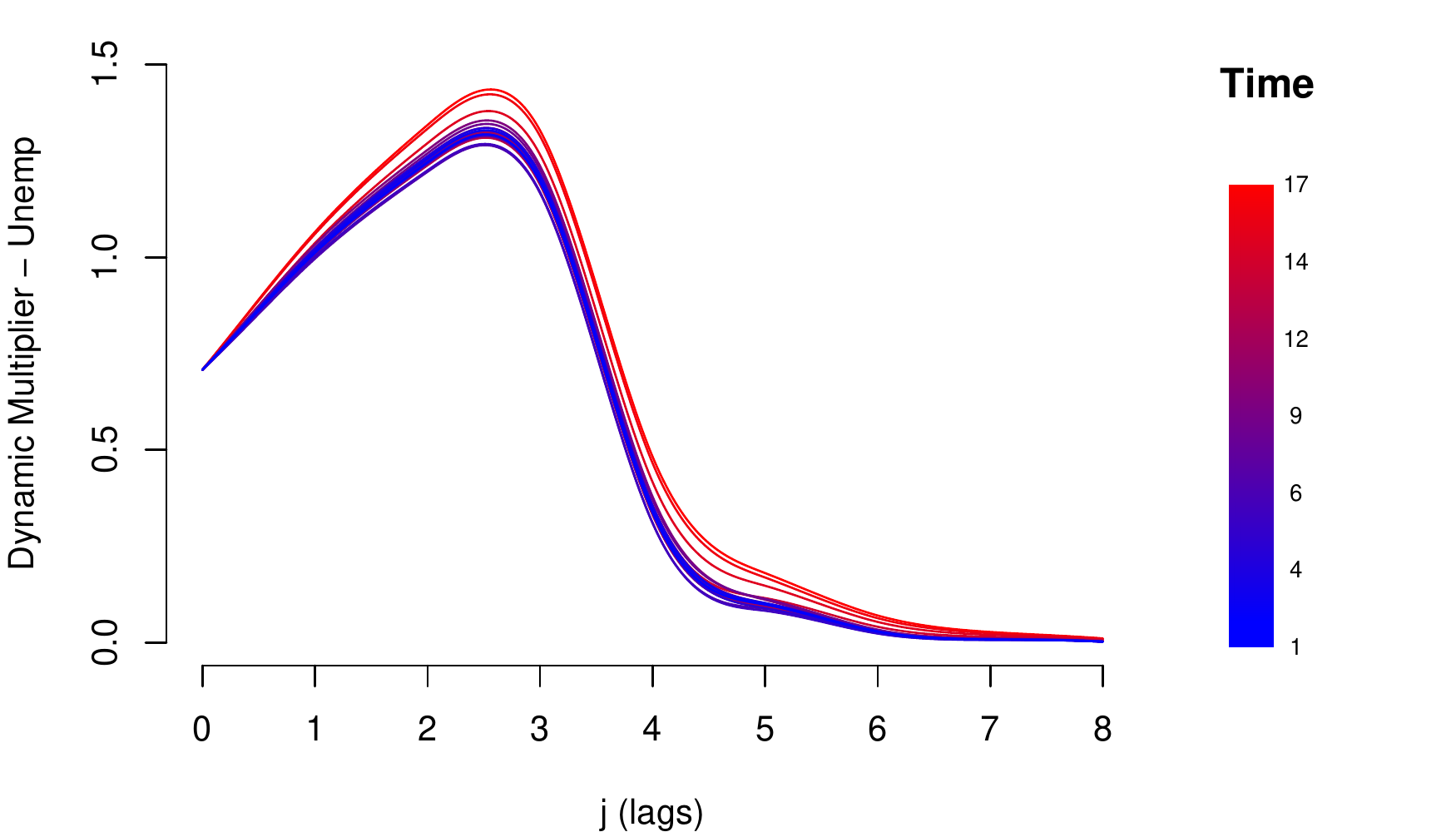}  
%			\caption{Dynamic Multiplier of the Unemp}
		\end{subfigure}
		\begin{subfigure}[t]{0.495\textwidth}
			\centering
			\includegraphics[width=\textwidth]{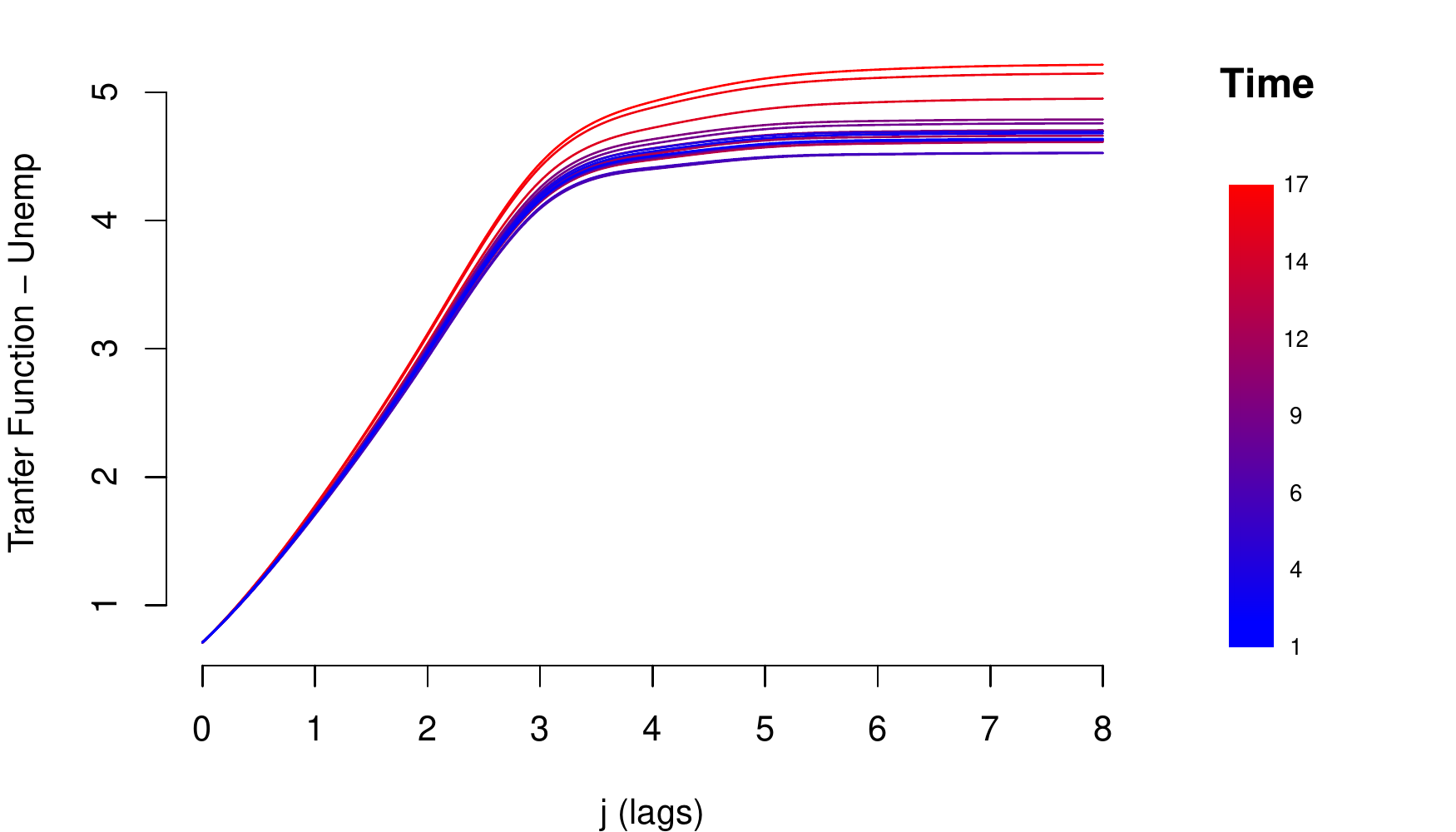}  
%			\caption{Transfer Function of the Unemp}
		\end{subfigure}	
		\caption{Stochastic Dynamic Multiplier and Stochastic Transfer Function of the model IV}
	\end{figure}

	\begin{figure}[ht!] 
		\centering
		\includegraphics[scale=0.595]{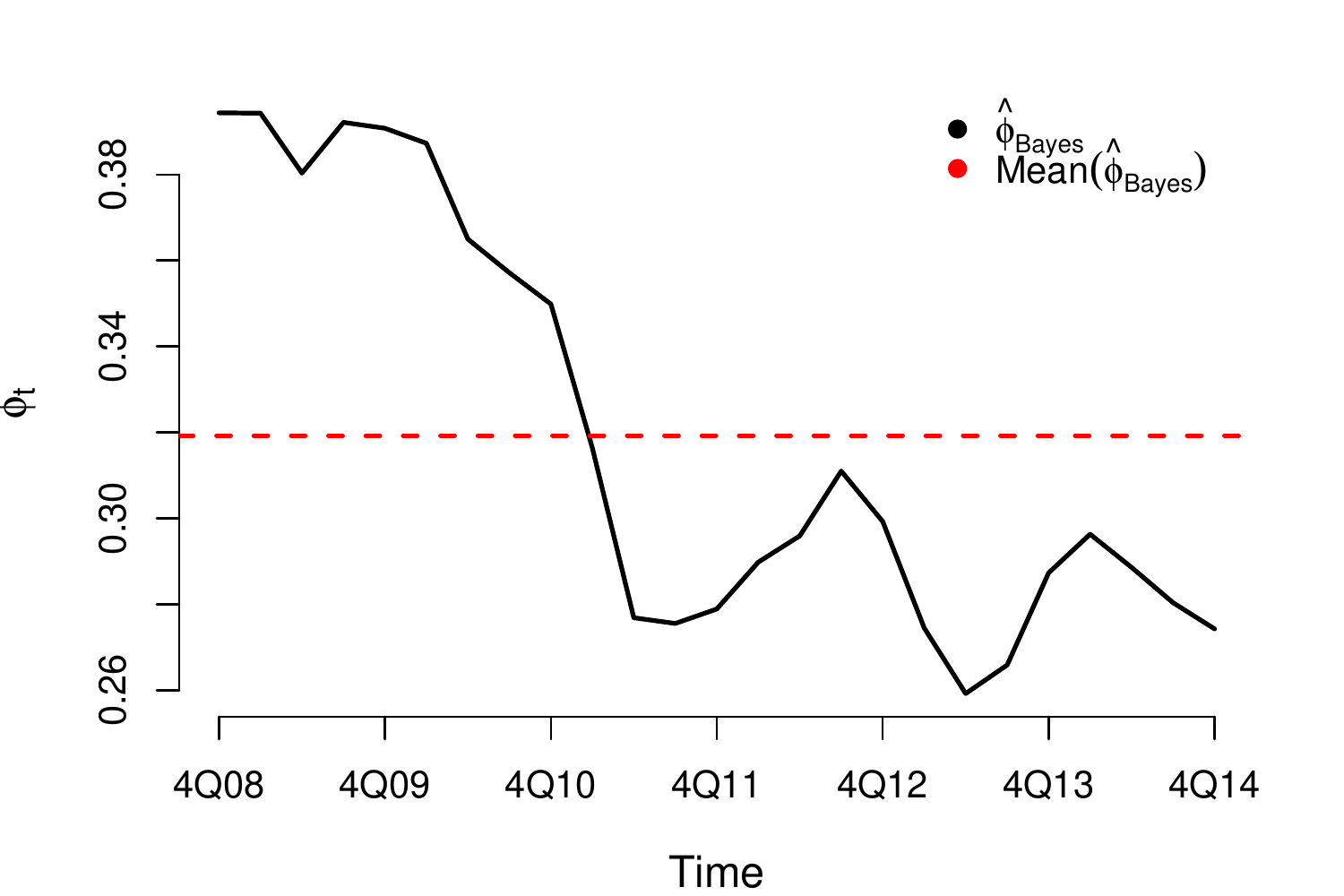} 
		\caption{Evolution of the resilience parameter of the model IV} \label{erp4}
	\end{figure}	
	
	\begin{figure}[h!]
		\centering
		\includegraphics[scale=0.8]{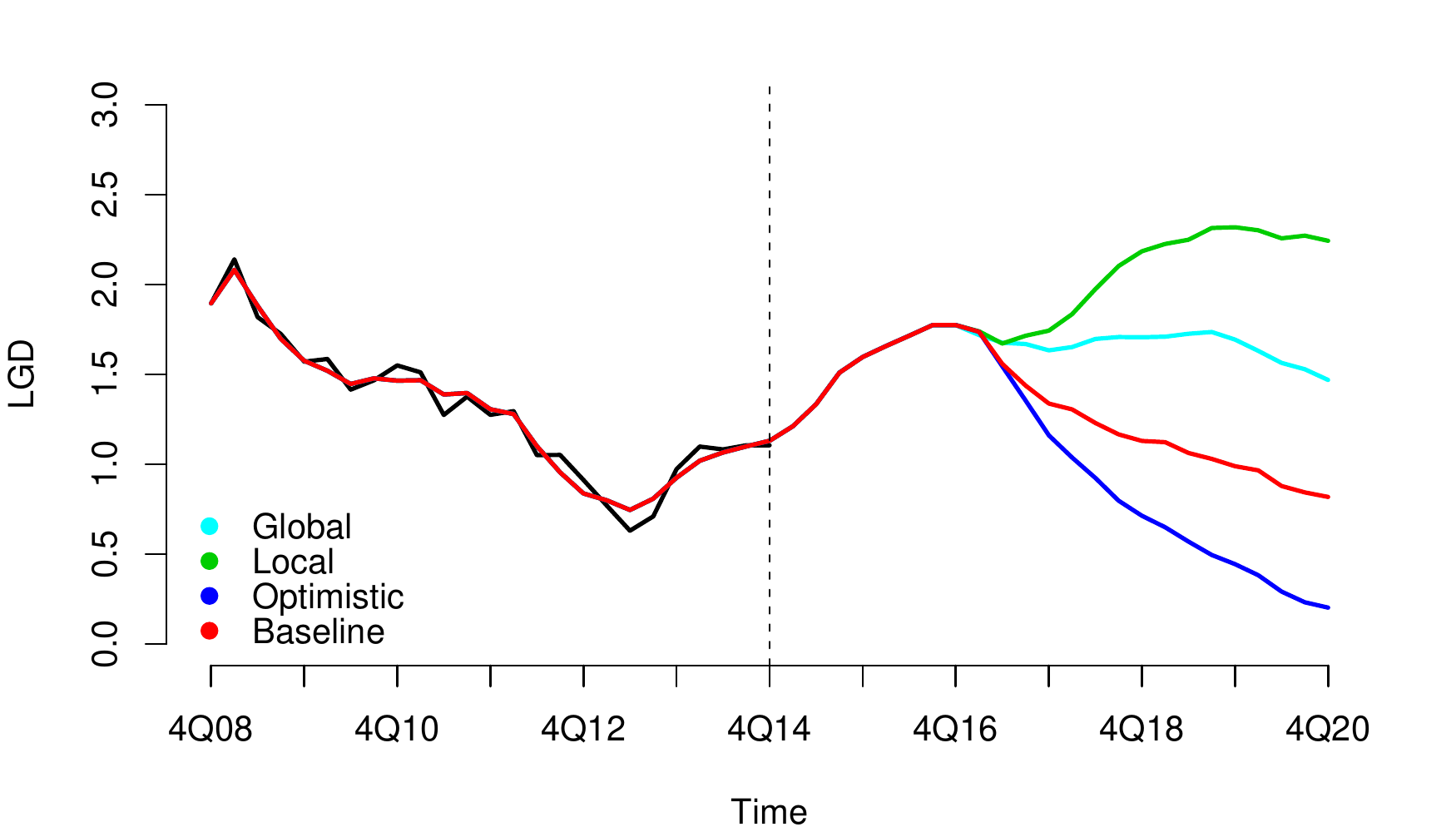} 
		\caption{Projection of the stress scenarios of the model IV} 
	\end{figure}

%	\begin{figure}[!h]
%		\centering
%		\begin{subfigure}[t]{0.65\textwidth}
%			\centering
%			\includegraphics[width=\textwidth]{images/Evolution_phi_Mod4}  
%			\caption{Evolution of the resilience parameter}
%		\end{subfigure}
%		
%		\begin{subfigure}{\textwidth}
%			\centering
%			\includegraphics[width=0.8\linewidth]{images/Projection_Scenarios_Mod4}  
%			\caption{Projection of the stress scenarios}
%		\end{subfigure}
%		\caption{Results from model 4}
%	\end{figure}

%	\clearpage
	\subsection{Model Comparison Results}
	
	The four models proposed for the credit risk data condense the empirical propagation of the shocks, described by the impact measurements in section (\ref{iim}). It is important to highlight that they also reproduce satisfactorily the persistence nature of the shocks of the variable Unemployment. That is, all the proposed models summarize well the transmission process of macroeconomic shocks underlying the data. As an effect of condensing well the transmissions of empirical shocks, the models present good adjustment capacity in the response variable (see the Table \ref{mgf}). In addition, stationarity and all other assumptions of the models are attended. For a correct interpretation of adjustment capacity measures see Appendix \ref{AnexoMeasuresFit}.
	
	\begin{table}[h!]
		\centering
		\caption{Measures of Goodness of Fit}
		\begin{tabular}{cccc}
			\hline
			Model &  MASE  & MSE    & $R^2$   \\  \hline
			I     & 0.6608 & 0.0081 & 0.9414  \\ 
			II    & 0.5901 & 0.0065 & 0.9527  \\ 
			III   & 0.3074 & 0.0019 & 0.9860  \\ 
			IV    & 0.4142 & 0.0035 & 0.9744  \\ \hline
		\end{tabular}\label{mgf}
	\end{table}

	\begin{table}[h!]
		\centering
		\caption{Information Criteria}
		\begin{tabular}{cccccc}
			\hline
			Model &  DIC   & WAIC   & SE$_{\textrm{WAIC}}$ & LOOIC   & SE$_{\textrm{LOOIC}}$   \\  \hline
			I     & -40.7 &	-40.6 &	5.5	& -40.4 & 5.5 \\ 
			II    & -44.6 &	-42.4 &	4.5	& -42.0	& 4.6 \\ 
			III   & -60.9 & -56.9 &	4.4	& -52.8	& 4.8 \\ 
			IV    & -49.0 &	-47.6 &	3.8	& -45.8	& 4.1 \\ \hline
		\end{tabular}\label{ic}
	\end{table}
	
Although the four models are empirically consistent and with good performance, models III and IV are more complete because they consider a flexible structure for the resilience parameter. In both cases, the resilience of the risk parameter improves over time (see Figures \ref{erp3} and \ref{erp4}). That is, during the observation period, the risk parameter improved its response and adaptation to shocks. Given that in the credit risk data there is a significant learning process of adaptation to shocks, the information criteria tend to choose models III and IV (see Table \ref{ic}). 

The four models generated coherent forecasts (non-overlapping scenarios forecast). An advantage of the last two models is that by incorporating the learning process by means of a flexible structure for the resilience parameter, they generated more optimistic projections for all scenarios as an effect of the improvement in the underlying resilience in the data. Note that if there was a deterioration on resilience, the effect on the projections would be the opposite. And, of course, an advantage of models I and II is that with a simpler structure they manage to reproduce the transmission of shocks and generate coherent projections in the scenarios.

%%--------------------------------------------------

	\section{Conclusions and Future work}

%We have shown that in inefficient economic environments, the transmission of macroeconomic shocks to the parameters of risk is characterized by a persistent propagation of the impacts. To identify the intensity of the persistence of shocks, we have proposed impact measures that describe the dynamic relationship between the risk parameter and macroeconomic shocks. We have also proposed the use of a family of models, called General Transfer Function Models. Finally, we have proposed to use the impact measures to propose specific models that satisfactorily summarize the relevant characteristics on the transmission of the collisions.

%With empirical data from a credit risk portfolio, we have shown that these models, with an adequate identification of the transfer functions through the impact measures, satisfactorily describe the transmission process underlying the data and at the same time generate coherent projections in the the stress scenarios.

%For portfolios where correlation exists between their risk parameters, these models can be extended to include the correlation between the risk parameters, for example LGD and PD, and allow simultaneous modeling. It is also possible to extend these models to include a dependence between the resilience and severity of the stress scenario, that is, the response of the parameter would depend on the intensity of the shocks corresponding to the scenario. We look forward to exploring such extensions in a future work.

We showed that in inefficient economic environments, the transmission of macroeconomic shocks to risk parameters, in general, is characterized by a persistent propagation of impacts. To identify the intensity of this persistence, we proposed the use of impact measures that describe the dynamic relationship between the risk parameter and the macroeconomic series. We presented a family of models called General Transfer Function Models, and considered the use of impact measures to propose specific models, that satisfactorily synthesize the relevant characteristics about the transmission process.

We present a simulation study to evaluate the parameter estimation process and the functional form of the Response Function decays. Moreover, from the empirical data of a credit risk portfolio, we show that these models, with an adequate identification of the transfer functions through the impact measures, satisfactorily describe the underlying data transmission process and at the same time generate coherent projections in the scenarios of stress.

These models can be extended to portfolios where there is a correlation between their risk parameters, eg LGD and PD, including this correlation and allowing simultaneous modeling. It is also possible to extend these models to include a dependence between the resilience and the severity of the stress scenario, that is, the response of the parameter would depend on the intensity of the shocks corresponding to the scenario. We intend to explore these extensions in future work.

%%--------------------------------------------------
	\section*{Acknowledgments}
	The authors thank Luciana Guardia, Tatiana Yamanouchi and Rafael Veiga Pocai for the fruitful discussions and contributions to this research. This research was supported by Santander Brazil.
	
	The vision presented in this article is entirely attributed to the authors, not being the responsibility of Santander S.A.

%%--------------------------------------------------
	%------------------------------- Bibliografia
	
	%\bibliographystyle{plainnat}
	\bibliographystyle{apa}
	\bibliography{references}

\clearpage
%%--------------------------------------------------
\appendix
\section{Appendix}
	
\subsection{Stress Scenarios}\label{AnexoProjectionScenarios}
	\begin{figure}[h!]
		\centering
		\begin{subfigure}[b]{0.65\linewidth}
			\includegraphics[width=\linewidth]{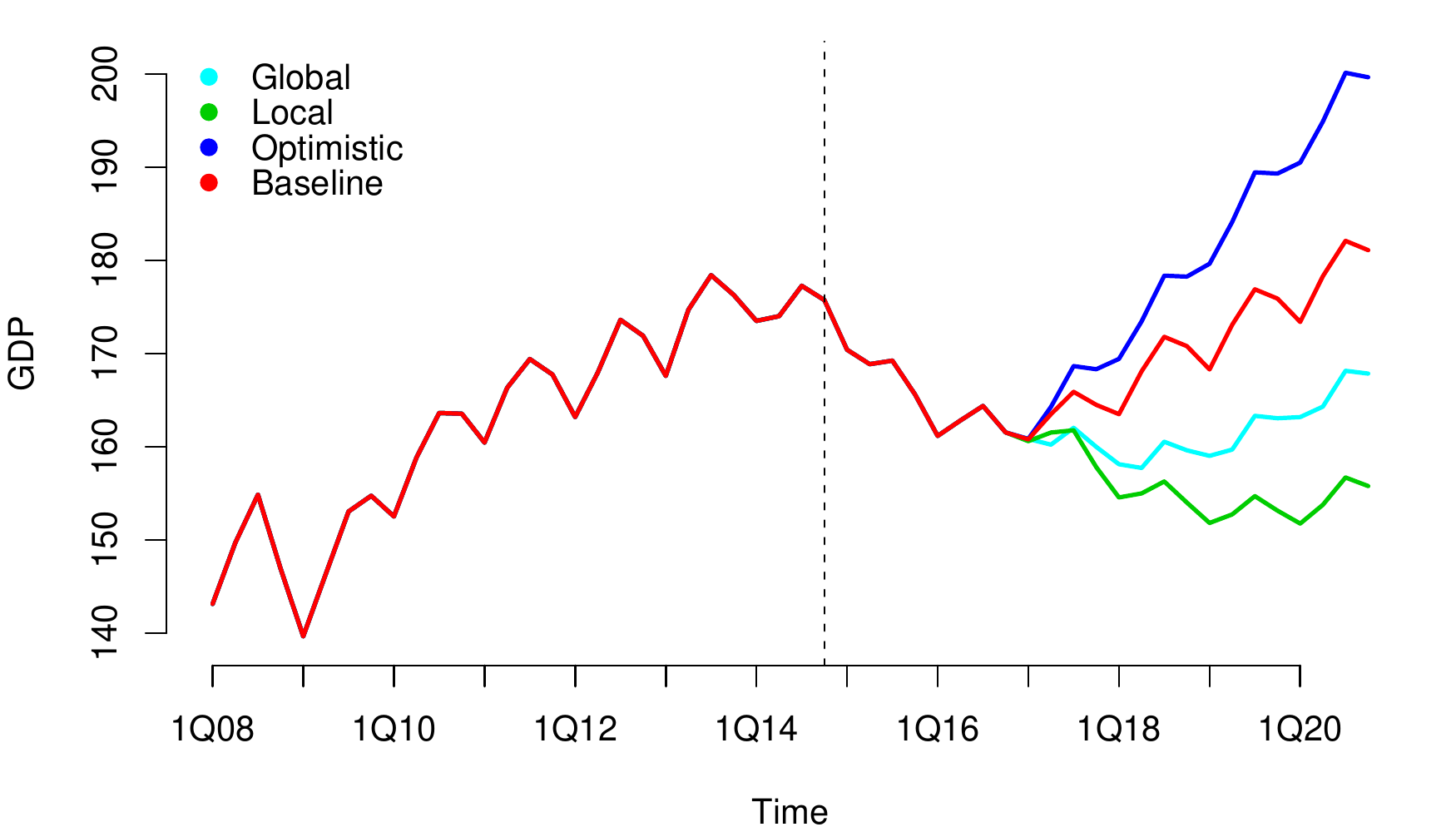}
		\end{subfigure}
		\hfill		
		\begin{subfigure}[b]{0.65\linewidth}
			\includegraphics[width=\linewidth]{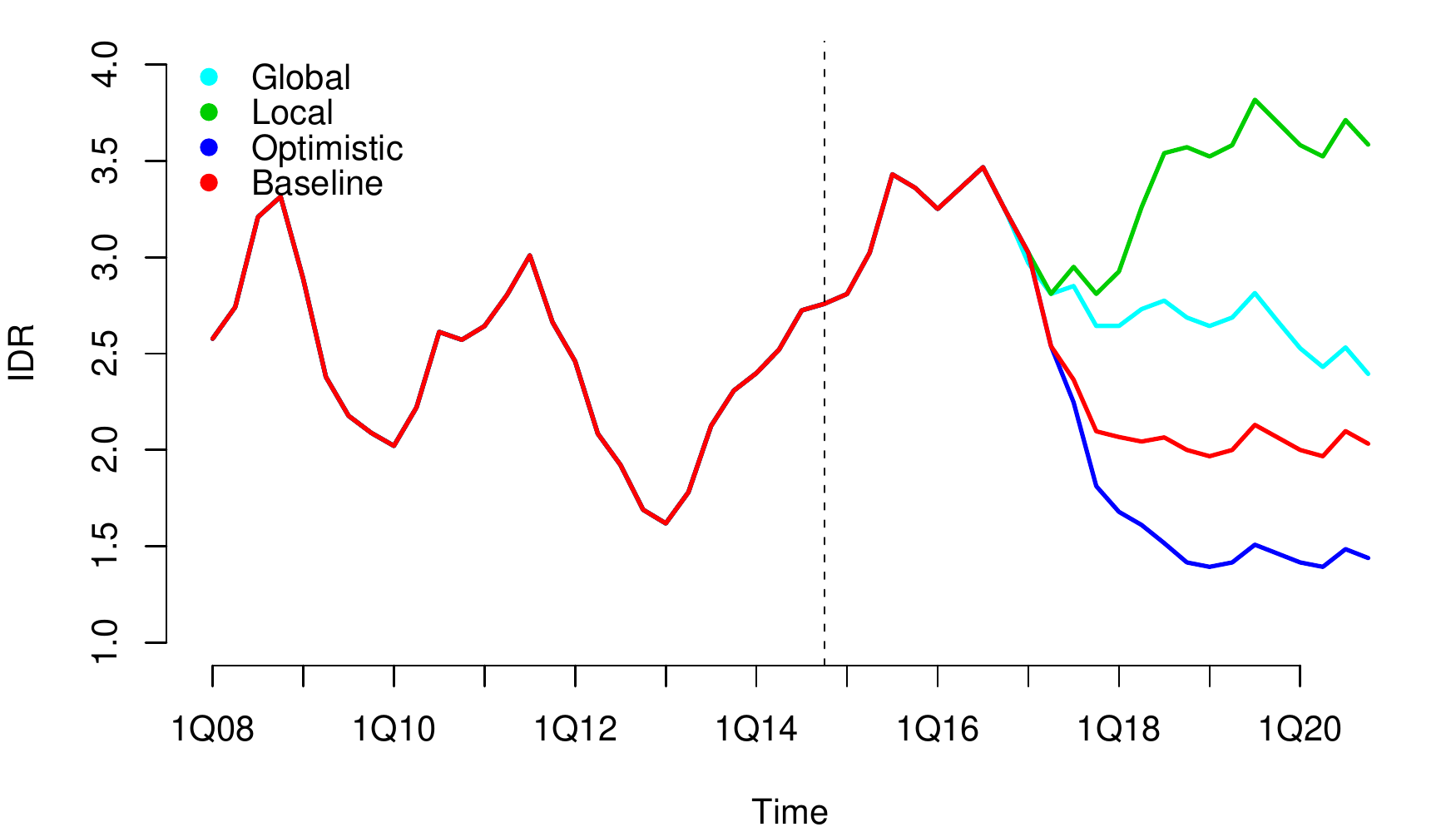}
		\end{subfigure}
		
		\begin{subfigure}[b]{0.65\linewidth}
			\includegraphics[width=\linewidth]{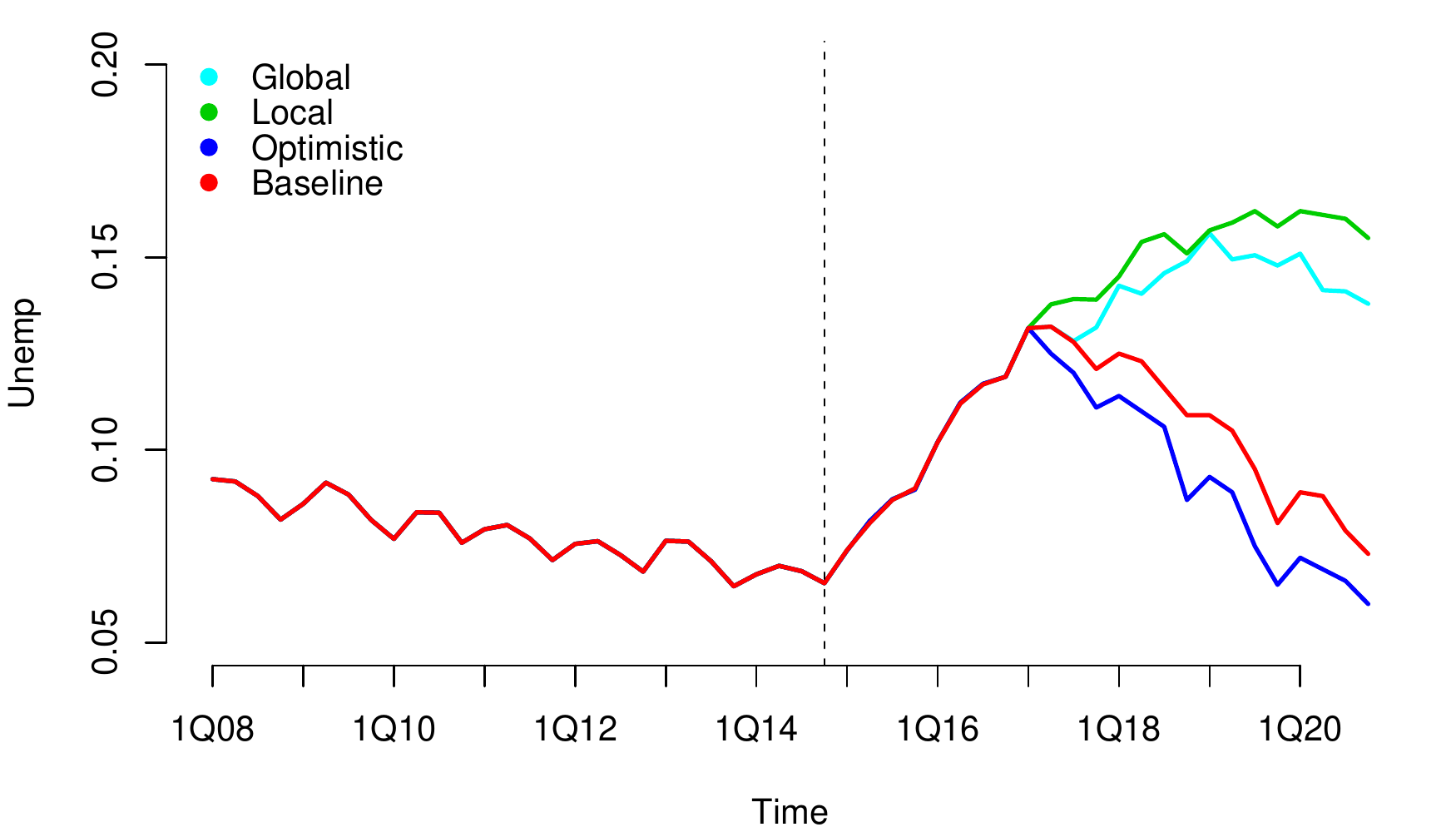}
		\end{subfigure}
		\caption{Scenarios ordered from lowest to highest severity: Optimistic, Baseline, Global, Local. The projections for the periods 1Q2015 to 4Q2016 are common for the four scenarios.}
		\label{series}
	\end{figure}

\clearpage
%-------------------------------------------------------
	\subsection{Definition of the Measures of Goodness of Fit and Some Statistical Tests}\label{AnexoMeasuresFit}
	Let $y_{t}$ denote the observation at time $t$ and $f_{t}$ denote the forecast of $y_{t}$. Then, define the forecast error $e_{t} = y_{t} - f_{t}$. We used some metrics to evaluate the fit of the models as shown below.
	
	\subsubsection*{Mean Absolute Scaled Error - MASE}
	The MASE was proposed by \cite{hyndman2006another} as a generally applicable measurement of forecast accuracy. It is defined as,
	\begin{equation}
	q_{t} = \frac{e_{t}}{\frac{1}{T-1}\sum_{t=2}^{T}|y_{t} - y_{t-1}|}.
	\end{equation}
	
	Then $\textrm{MASE} = mean(|q_{t}|)$. According to \cite{hyndman2006another}, when $\textrm{MASE}<1$, the proposed method gives, on average, lower errors than the one-step errors from the naïve method. 
	
	\subsubsection*{Mean Square Error - MSE}
	A metric widely used to assess whether the model is well adjusted, is the mean square error, which is defined as,
	\begin{equation}
	\textrm{MSE} = mean(e_{t}^{2}).
	\end{equation}
	
	\subsubsection*{Coefficient of Determination - $R^{2}$}
	In statistics, the coefficient of determination or R squared, is the proportion of the variance in the dependent variable that is predictable or explained from the independent variable(s). The most general definition of the coefficient of determination is,
	\begin{equation}
	R^{2} = 1 - \frac{SS_{Res}}{SS_{Tot}},
	\end{equation}
	where $SS_{Res} = \sum_{t=1}^{T}e_{t}^{2}$ and $SS_{Tot} = \sum_{t=1}^{T}(y_{t} - \bar{y})^{2}$.

	\subsubsection*{Ljung-Box Test}
This is a Statistical test to evaluate whether any of a group of autocorrelation of a time series is different from zero. The hypothesis of the test are:

\begin{center}
\begin{tabular}{lcl}
$H_{0}$: There is no serial correlation in the residuals & $vs$ & $H_{1}$: There is serial correlation in the residuals
\end{tabular}
\end{center}
%or mathematically,
%\begin{center}
%\begin{tabular}{lcl}
%$H_{0}$: $\rho_{1}=\rho_{2}=\ldots=\rho_{T}=0$ & $vs$ & $H_{1}$: $\rho_{t}\neq 0$ for some $t \in \{1,\ldots,T\}$ 
%\end{tabular}
%\end{center}

More details in \cite{ljung1978measure}.
	
	\subsubsection*{Augmented Dickey Fuller Test}
The Augmented Dickey Fuller Test (ADF) is a unit root test for stationarity. The hypothesis for the test are:

\begin{center}
\begin{tabular}{lcl}
$H_{0}$: There is a unit root & $vs$ & $H_{1}$: There is no a unit root (or the time series is trend-stationary),
\end{tabular}
\end{center}

More details in \cite{said1984testing}.

%	\subsubsection*{Augmented Dickey Fuller Test}
%The Augmented Dickey Fuller Test (ADF) is unit root test for stationarity. The Hypothesis for the test:
%
%\begin{center}
%\begin{tabular}{lcl}
%$H_{0}$: There is a unit root & $vs$ & $H_{1}$: There is no a unit root (or the time series is trend-stationary),
%\end{tabular}
%\end{center}
%
%More details in Said and Dickey (1984).

\clearpage	
\section{Appendix}
%-------------------------------------------------------
\subsection{Linear Regression with Stochastic Regressors}\label{AnexoRegressionModel}

    \begin{equation*}\label{eq:lm}
	Y_{t}= X_{t}^{\top} \beta+ \xi_{t},
	\end{equation*}
	where the scalar series $Y_{t}$ is the risk parameter LGD, the vector $X_{t}$ contains the three macroeconomic series and the intercept. The vector $\beta$ is the vector of coefficients. The term $\xi_{t}$ is a white noise process with zero mean and variance $\sigma^{2}$. The vector $X_{t}$ is stochastic and independent of $\xi_{s}$ for all $t$ and $s$, which corresponds to the assumption of strict exogeneity . The coefficients of the model are estimated through ordinary least squares (see \cite{hamilton1994time}). 
	
	\begin{table}[h!]
		\centering
		\caption{Summary of Regression Model}
		\begin{tabular}{ccccc}
			\hline
			& Estimate & Std. Error & t value &Pr$(>|t|)$ \\
			\hline
			(Intercept) & 5.91 & 1.25 & 4.71 & 0.00 \\
			GDP & -0.03 & 0.00 & -6.26 & 0.00 \\
			IDR & 0.30 & 0.06 & 4.75 & 0.00 \\
			Unemp & -6.79 & 6.41 & -1.06 & 0.30 \\
			\hline
		\end{tabular}
	    \label{tab:reg_model}
	\end{table}
	
	\begin{figure}[h!]
		\centering
		\includegraphics[scale=0.8]{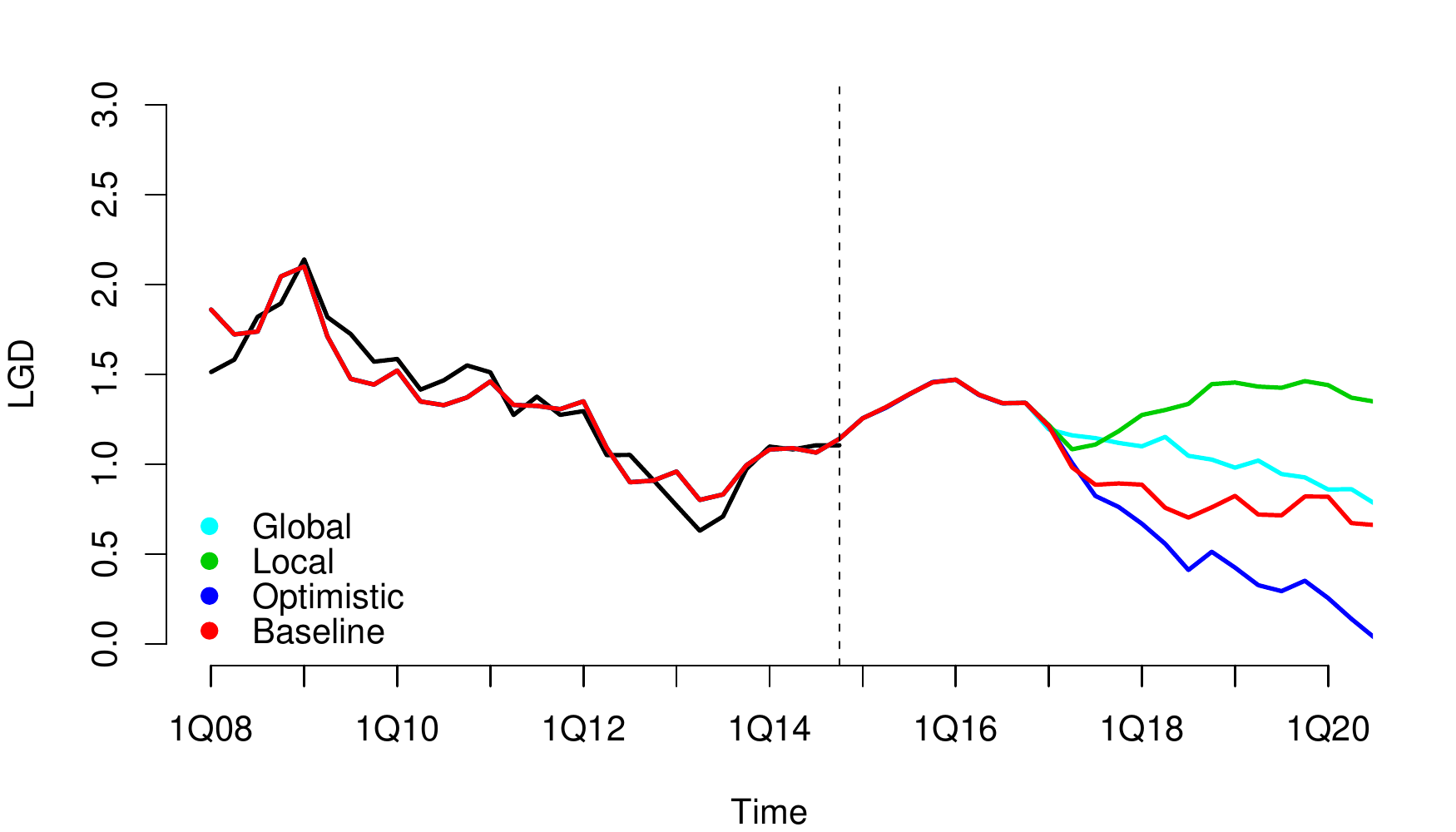} 
		\caption{Projection of the regression model. The dashed line represents the estimation interval of the model (1Q2008 to 4Q2014), the period from 1Q2015 to 4Q2020 refers to the stressed macroeconomic scenarios.} \label{fig:ProjectionRegModel}
	\end{figure}

  The model seems to have a good fit, but presents serial correlation in the residuals. Apparently, the variable Unemployment is not significant, see Table \ref{tab:reg_model}. Note that the Unemployment variable is important in crisis contexts. In addition, is the estimation of a coefficient with a negative sign for the variable Unemployment contradicts the expected economic relationship between Unemployment and LGD, this is reflected in non-coherent forecasting of scenarios (overlapping scenarios forecasts), see Figure \ref{fig:ProjectionRegModel}.
	
	\begin{figure}[!h]
		\centering
		\includegraphics[width=0.6\linewidth]{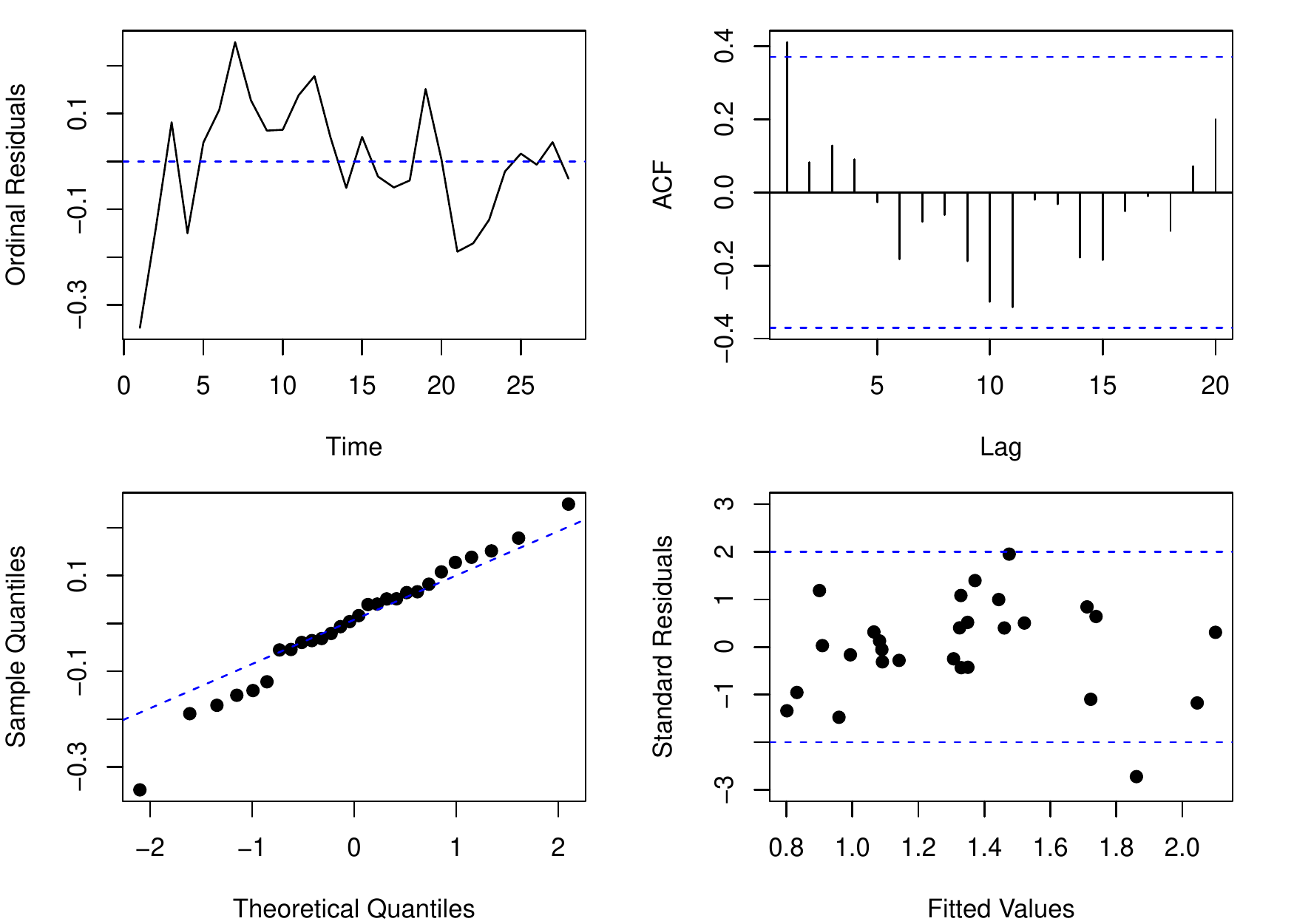}  
		\vspace{-0.1cm} 
		\caption{Residuals graphs of Regression Model}
	\end{figure}
	
	\begin{table}[ht!]
		\centering
		\caption{Normality Test}
		\vspace{-0.3cm}
		\begin{tabular}{ccc}
			\hline
			Tests & Statistics & p.values \\ 
			\hline
			Shapiro-Wilk & 0.9750 & 0.7192 \\ 
  			Kolmogorov-Smirnov & 0.1177 & 0.4107 \\ 
  			Cramer-von Mises & 0.0421 & 0.6297 \\ 
  			Anderson-Darling & 0.2638 & 0.6723 \\ 
			\hline
		\end{tabular}
	\end{table}

\begin{table}[ht!]
\centering
\begin{minipage}{.55\linewidth}
\centering
\caption{Ljung-Box Test}
\vspace{-0.3cm}
		\begin{tabular}{ccc}
			\hline
			Lag & Statistic & p.value \\ 
			\hline
			1 & 5.2500 & 0.0219 \\ 
  			2 & 5.4694 & 0.0649 \\ 
  			3 & 6.0211 & 0.1106 \\ 
  			4 & 6.3076 & 0.1773 \\ 
  			5 & 6.3345 & 0.2750 \\ 
  			6 & 7.6111 & 0.2680 \\ 
  			7 & 7.8700 & 0.3442 \\ 
  			8 & 8.0302 & 0.4305 \\ 
  			9 & 9.5907 & 0.3846 \\ 
  			10 & 13.7686 & 0.1838 \\ 
  			11 & 18.6371 & 0.0679 \\ 
  			12 & 18.6583 & 0.0971 \\ 
			\hline
		\end{tabular}	
\end{minipage}
\begin{minipage}{.4\linewidth}
\centering
	\caption{Measures of Goodness of Fit}
	\vspace{-0.3cm}
		\begin{tabular}{ccc}
			\hline
			 MASE & MSE & $R^{2}$ \\ 
			\hline
		    0.8264 & 0.0158 & 0.8841 \\ 
			\hline
		\end{tabular}	
\vspace{1cm}
\centering
	\caption{Durbin-Watson and Augmented Dickey-Fuller Test}
	\vspace{-0.3cm}
		\begin{tabular}{cccc}
			\hline
			Test & Statistical & p-value  \\ 
			\hline
			DW  & 0.9011 & 0.0002  \\ 
			ADF & 2.7674 & 0.0115  \\ 
			\hline
		\end{tabular}	
\end{minipage}
\end{table}

\end{document}